\documentclass[reprint,superscriptaddress,twocolumn,amsmath,amssymb,amsthm, dsfont,prb]{revtex4-1}
\usepackage[linktocpage=true]{hyperref}
\usepackage{hyperref}
\usepackage{graphicx}
\usepackage{mathrsfs}
\usepackage{dcolumn}
\usepackage{bm}
\usepackage{tikz}
\usepackage{hhline}
\usepackage{tabularx}
\usepackage{multirow} 
\usepackage{float}
\usetikzlibrary{decorations.markings}

\begin{document}
\tikzset{
    vertex/.style={fill,circle,draw,scale=0.3},
    ->/.style={decoration={markings,mark=at position 0.5 with {\fill (2pt,0)--(-2pt,2.31pt)--(-2pt,-2.31pt)--cycle;}},postaction={decorate}},
    ->2/.style={decoration={markings,mark=at position 1 with {\fill (2pt,0)--(-2pt,2.31pt)--(-2pt,-2.31pt)--cycle;}},postaction={decorate}},
}
\definecolor{lblue}{rgb}{0.4,0.4,1} 
\definecolor{lred}{rgb}{1,0.4,0.4} 

\title{Dimensional Reduction and Topological Invariants of Symmetry-Protected Topological Phases}

\author{Nathanan Tantivasadakarn}
\affiliation{Perimeter Institute for Theoretical Physics, Waterloo, ON N2L 2Y5, Canada}

\date{\today}

\begin{abstract}
We review the dimensional reduction procedure in the group cohomology classification of bosonic SPT phases with finite abelian unitary symmetry group. We then extend this to include general reductions of arbitrary dimensions and also extend the procedure to fermionic SPT phases described by the Gu-Wen super-cohomology model. We then show that we can define topological invariants as partition functions on certain closed orientable/spin manifolds equipped with a flat connection. The invariants are able to distinguish all phases described within the respective models. Finally, we establish a connection to invariants obtained from braiding statistics of the corresponding gauged theories.
\end{abstract}

\maketitle

\tableofcontents

\section{Introduction}\label{intro}

In the past years, Symmetry-Protected Topological (SPT) order has emerged as a new type of order in classifying gapped phases of matter with symmetry \cite{GuWen2009,ChenGuWen2011,ChenGuWen2011-2,ChenGuWen2011-3,FidkowskiKitaev2011}. There are currently two different ways to understand SPT phases. In one point of view, they are quantum phases that satisfies the following properties:
\begin{enumerate}
\itemsep0em 
\item Bulk excitations have a finite energy gap.
\item The Hamiltonian is invariant under some set of (on site) symmetries, none of which are spontaneously broken.
\item The ground state cannot be continuously connected with the trivial (product) state without closing the energy gap or breaking the symmetry. However, it can always be connected in the absence of symmetry.
\end{enumerate}
Alternatively, in the low energy limit, their effective theories are known to be Topological Quantum Field Theories (TQFTs). So far, both understandings give rise to similar predictions, but it is not well understood how these two notions are connected in general.

One of the most important questions of understanding SPT phases is their classification. That is, given a symmetry group $G$, what are all the possible SPT phases in $n$ space-time dimensions. This classification has been extensively studied using various methods, including group cohomology \cite{Chen2013,GuWen2014,Wen2015,Chengetal2015,WangGu2017}, cobodism theory \cite{Kapustin2014,Kapustinetal2015,KapustinThorngren2017}, and invertible TQFTs \cite{Freed2014,Freed2016,KapustinTurzillo2017}. A natural followup question is then, how can we distinguish different SPT phases from each other? One of the possibilities is that if the symmetry is on-site and unitary, then one can gauge the symmetry group and study the braiding statistics of the excitations of the gauged theory. Inequivalent SPT phases would then give different braiding statistics \cite{LevinGu2012,WangLevin2014,JiangMesarosRan2014,WangLevin2015,WangWen2015,LinLevin2015,Wang2016,WangLinGu2017,ChanYeRyu2017,PutrovWangYau2017,ChengTantiWang2017}. In this method, the braiding statistics are topological invariants of the theory, and have shown evidence that they can distinguish all possible types of SPT phases.

In this paper, we propose an alternative method to distinguish SPT phases with finite abelian unitary symmetry group. By studying lattice theories that represent different SPT phases, we define topological invariants as partition functions of the theory on certain closed manifolds. The manifolds are equipped with certain flat connections and must be oriented or spin for bosonic or fermionic SPT phases respectively. We will focus our study on two lattice models. For bosonic SPT phases, we study the group cohomology model\cite{Chen2013}, which is believed to characterize all distinct phases in space-time dimensions $n \le 4$ by the cohomology group $\mathcal H^n(G,U(1))$. For fermionic SPT phases, we will study the special group supercohomology model by Gu \& Wen, which provides a partial classification \cite{GuWen2014}.

We would like to point out that these models assume a TQFT description, and therefore do not encompass all possible SPT phases. However, we find that the invariants we defined are related to invariants obtained from braiding statistics, which are topological invariants that are well defined for all gauged SPT phases. Thus, studying the invariants from partition functions could lead to a better understanding of invariants from braiding statistics and vice versa.

\subsection{Summary of the Main Results}
\begin{table*}
\caption{\label{tab:resultsbosons} We define partition functions ($Z$) on the closed orientable manifolds shown below with certain flat connections. The flat connections are defined by placing the generators of the group $G$ to the generators $\pi_1(\mathcal M)$ of the manifold. The set of partition functions given are sufficient to distinguish all inequivalent SPT phases in the group cohomology classification for finite abelian unitary symmetry group $G$. Here, we use the notation $N^{ijk...} = \text{lcm}(N_i,N_j,N_k,...)$ and $N_{ijk...} = \text{gcd}(N_i,N_j,N_k,...)$ to denote the least common multiple and greatest common divisor respectively. There are two invariants, $\mathcal I_{ij}$ and $\mathcal I_{ijk}$, for which we do not know the corresponding $5$-manifolds.}
\begin{tabular}{|c| c| c |c|c|}
\hline
Dimensions & Invariant & Manifold ($\mathcal M$) & $\pi_1(\mathcal M)$ & Generators\\
\hline
1+1D & $Z^{ij}$  & $T^2$ & $\mathbb Z^2$   & $e_i$, $e_j$\\
\hline
\multirow{3}{*}{2+1D} & $Z_{i}$  &$L(N_i;1)$  & $\mathbb Z_{N_{i}}$   & $e_i$ \\
\hhline{~----}
& $Z_{ij}$ &$L(N^{ij};1)$  &$\mathbb Z_{N^{ij}}$ & $e_i+e_j$  \\
\hhline{~----}
& $Z_{ijk}$&$T^3$  & $\mathbb Z^3$ & $e_i$, $e_j$, $e_k$ \\
\hline
\multirow{3}{*}{3+1D} & $Z_{i,l}$  &$L(N_i;1) \times S^1$  & $\mathbb Z_{N_{i}} \times \mathbb Z$   & $e_i$, $e_l$ \\
\hhline{~----}
& $Z_{ij,l}$& $L(N^{ij};1)\times S^1$  &$\mathbb Z_{N^{ij}} \times \mathbb Z$ & $e_i+e_j$, $e_l$  \\
\hhline{~----}
& $Z_{ijk,l}$ &$T^4$  & $\mathbb Z^4$ & $e_i$, $e_j$, $e_k$, $e_l$ \\
\hline
\multirow{6}{*}{4+1D} &$\mathcal I_i$  &$L(N_i;1,1,1)$  & $\mathbb Z_{N_{i}}$   & $e_i$\\
\hhline{~----}
& $\mathcal I_{ij}$& \multirow{2}{*}{} & \multirow{2}{*}{} & \multirow{2}{*}{}\\
\cline{2-2}
& $\mathcal I_{ijk}$ & & & \\
\hhline{~----}
& $Z_{i,l,m}$ &  $L(N_i;1) \times T^2$  & $\mathbb Z_{N_{i}} \times \mathbb Z^2$   & $e_i$, $e_l$, $e_m$  \\
\hhline{~----}
& $Z_{ij,l,m}$& $L(N^{ij};1)\times T^2$  &$\mathbb Z_{N^{ij}} \times \mathbb Z^2$ & $e_i+e_j$, $e_l$, $e_m$  \\
\hhline{~----}
& $Z_{ijk,l,m}$&$T^5$  & $\mathbb Z^5$ & $e_i$, $e_j$, $e_k$, $e_l$, $e_m$ \\
\hline
\end{tabular}
\end{table*}
\begin{table*}
\caption{\label{tab:resultsfermions} We define Gu-Wen partition functions ($\mathcal Z$) on the closed spin manifolds shown below with certain flat connections. For the dimensions we consider, they are essentially the same manifolds as those in the bosonic case, but also equipped with a spin structure. The set of partition functions given are sufficient to distinguish all inequivalent SPT phases in the Gu-Wen group super-cohomology model for finite abelian unitary group $G$. Here, $N^{0i}=\text{lcm}(2,N_i).$}
\begin{tabular}{|c| c| c |c|c|}
\hline
Dimensions & Invariant & Manifold ($\mathcal M$) & $\pi_1(\mathcal M)$ & Generators\\
\hline
\multirow{2}{*}{1+1D} & $\mathcal Z^{ij}(0,0)$  & $T^2$ & $\mathbb Z^2$   & $e_i$, $e_j$\\
\hhline{~----}
&  $\mathcal Z^{ij}(0,1)$  & $T^2$ & $\mathbb Z^2$   & $e_i$, $e_j$\\
\hline
\multirow{5}{*}{2+1D} & $Z_{i}(0)$  &$L(N_i;1)$  & $\mathbb Z_{N_{i}}$   & $e_i$ \\
\hhline{~----}
& $\mathcal Z_{0i}(1)$  &$L(N^{0i};1)$  & $\mathbb Z_{N^{0i}}$   & $e_i$ \\
\hhline{~----}
& $\mathcal Z_{ij}(0)$ &$L(N^{ij};1)$  &$\mathbb Z_{N^{ij}}$ & $e_i+e_j$  \\
\hhline{~----}
& $\mathcal Z_{ijk}(0,0,0)$&$T^3$  & $\mathbb Z^3$ & $e_i$, $e_j$, $e_k$ \\
\hhline{~----}
& $\mathcal Z_{ijk}(0,0,1)$&$T^3$  & $\mathbb Z^3$ & $e_i$, $e_j$, $e_k$ \\
\hline
\multirow{6}{*}{3+1D} & $\mathcal Z_{i,l}(0,0)$  &$L(N_i;1) \times S^1$  & $\mathbb Z_{N_{i}} \times \mathbb Z$   & $e_i$, $e_l$ \\
\hhline{~----}
& $\mathcal Z_{i,l}(0,1)$  &$L(N_i;1) \times S^1$  & $\mathbb Z_{N_{i}} \times \mathbb Z$   & $e_i$, $e_l$ \\
\hhline{~----}
& $\mathcal Z_{0i,l}(1,0)$  &$L(N^{0i};1) \times S^1$  & $\mathbb Z_{N^{0i}} \times \mathbb Z$   & $e_i$, $e_l$ \\
\hhline{~----}
& $\mathcal Z_{ij,l}(0,0)$& $L(N^{ij};1)\times S^1$  &$\mathbb Z_{N^{ij}} \times \mathbb Z$ & $e_i+e_j$, $e_l$  \\
\hhline{~----}
& $\mathcal Z_{ijk,l}(0,0,0,0)$ &$T^4$  & $\mathbb Z^4$ & $e_i$, $e_j$, $e_k$, $e_l$ \\
\hhline{~----}
& $\mathcal Z_{ijk,l}(0,0,0,1)$ &$T^4$  & $\mathbb Z^4$ & $e_i$, $e_j$, $e_k$, $e_l$ \\
\hline
\end{tabular}
\end{table*}
In this paper, we define topological invariants that are able to distinguish all phases within the group cohomology models for finite abelian unitary symmetry group $G$. In the bosonic case, we will represent our group as
\begin{equation}
G=\prod_{i=1}^K \mathbb{Z}_{N_i}
\end{equation}
with addition as the group operation. The generator of the subgroup $\mathbb{Z}_{N_i}$ is denoted $e_i$. We now define the topological invariants as the partition functions $Z$ on the orientable manifolds listed in Table \ref{tab:resultsbosons}. Each manifold $\mathcal M$ is equipped with a flat connection, which is chosen such that we place certain generators of the group $G$ as holonomies around non-contractible loops that are generators of the fundamental group $\pi_1(\mathcal M)$. We then show that these invariants are enough to distinguish all SPT phases labeled by the cohomology group $\mathcal H^n(G,U(1))$ in $n$ space-time dimensions.

An important finding we have made is the classification of SPT phases in 4+1 dimensions using the cohomology group $\mathcal H^5(G,U(1))$. There, we had to develop a general dimensional reduction beyond compactification over a circle, which we introduce in section \ref{dimredbosons}. The new invariants we obtain, called $\mathcal I_{ij}$ and $\mathcal I_{ijk}$, are not possible to obtain using compactification over a circle and are necessary in order to obtain a complete set of invariants. 

For the fermionic case, the group $G$ is extended by the fermionic parity operator $\mathbb Z_2^f$. For the Gu-Wen super-cohomology model, the extension is trivial and we have the total symmetry group
\begin{equation}
G^f=\mathbb Z_2^f \times G.
\end{equation}
Similarly, we define topological invariants $\mathcal Z$ as the Gu-Wen partition function on the spin manifolds listed in Table \ref{tab:resultsfermions}. The partition function has a spin structure dependence\cite{GaiottoKapustin2016}, which is noted as 0 or 1 for trivial or non-trivial spin structure along each non-contractible direction. For example, the partition function $\mathcal Z^{ij}(0,1)$ defined on $T^2$ has a trivial spin structure on the circle with holonomy $e_i$, and non-trivial spin structure on the other circle with holonomy $e_j$. In this case, using different spin structures, we are able to extract topological invariants that can distinguish all the Gu-Wen fermionic SPT phases, which is described by an extension of the obstruction-free subgroup $B\mathcal H^{n-1}(G,\mathbb Z_2)$ by $\mathcal H^n(G,U(1))$.

As an aside, we establish a relation between the invariants defined with the invariants from braiding statistics in the gauged theory. The correspondence is exact in the bosonic case. In the fermionic case, we conjecture a relation between the two invariants and show explicitly that the correspondence is exact for some example groups, namely $\mathbb Z_2 \times \mathbb Z_4$, $\mathbb Z_2 \times \mathbb Z_6$, and $\mathbb Z_2 \times \mathbb Z_2 \times \mathbb Z_2$. Ultimately, we believe that the correspondence is true in general and the equivalence is ultimately related to surgery on links in manifolds.

\subsection{Organization of the Paper}
 In section \ref{prelim}, we review the group cohomology model classification of bosonic SPT phases and the Gu-Wen special super-cohomology model for fermionic SPT phases. The remaining of this paper is then organized into two main parts. The first part will deal with dimensional reduction, a process of compactifying our manifold over a circle, so that we can study the phases in a lower dimension. In section \ref{dimredbosons}, we will review the well known dimensional reduction procedure for bosonic SPT phases over a circle. We will then show that under certain conditions, we can also perform compactification on higher dimensional manifolds as well. In section \ref{dimredfermions}, we will derive a dimensional reduction procedure over a circle for Gu-Wen fermionic phases. For the second part, we will define invariants that can completely distinguish all SPT phases described within the studied models. This will be done in sections \ref{invariantsbosons} and \ref{invariantsfermions} for bosonic and fermionic phases respectively. Moreover, we will show in section \ref{invariantsbosons} that the general dimensional reduction is crucial for obtaining all the invariants that distinguishes phases beyond four space-time dimensions, which we show explicitly in 4+1D. Finally, in section \ref{braidingrelation}, we will show that the invariants we defined are equivalent to those from braiding statistics and we conjecture some possible connections.

\section{Preliminaries}\label{prelim}
\subsection{The Group Cohomology Model for Bosonic SPT Phases}\label{groupcohomology}
The group cohomology model for SPT phases is a lattice model proposed in Ref. \onlinecite{Chen2013}. In $n$ (Euclidean) space-time dimensions, the input of this theory is a symmetry group $G$ and a certain $U(1)$-valued function $\nu_n$, which takes in $n+1$ group elements. The model then constructs a partition function on an $n$-dimensional space-time manifold as follows. First, we triangulate the manifold as a simplicial complex, with group elements living on the vertices. We then take the product of $\nu_n$ evaluated on all simplices, taking in elements from the $n+1$ vertices of the simplex in a certain order. The partition function then sums over all possible group element configurations, and divides by the order of the group to the power of the number of vertices $N_v$:
\begin{equation}
\label{equ:partitionfunctionnu}
Z= \frac{1}{|G|^{N_v}}\sum_{g_i,g_j,...,g_k \in G} \prod_{ij...k} \nu_{n}^{s_{ij...k}}(g_i,g_j,...,g_k).
\end{equation}
Note that there are two possible parities for each simplex depending on its orientation. A simplex with positive orientation $(+)$ has $s_{ijk...}=1$, while a simplex with negative orientation $(-)$ has $s_{ijk...}=-1$. The orientation of a simplex is defined in Table \ref{tab:nu_n}.

In addition to being $U(1)$-valued, the function $\nu_n$ must also satisfy two extra conditions. First, it must be invariant under a translation $g_i \rightarrow gg_i$, where $g\in G$:
\begin{equation}
\label{equ:leftinvariant}
\nu_n(g_0,g_1,g_2,...,g_n) = \nu_n(gg_0,gg_1,gg_2,...,gg_n).
 \end{equation}
A function with such a property is said to be \underline{\smash{homogeneous}}. Second, it must satisfy the so-called \underline{\smash{cocycle condition}}:
 \begin{equation}
\prod_{i=0}^{n+1} \nu_n^{(-1)^i}(g_0,...,\hat{g_i},...,g_{n+1})=1,
 \end{equation}
where $\hat{g_i}$ means skipping over the element $g_i$. The cocycle condition is needed for the partition function to be well defined, as we will explain shortly.

\begin{figure}[H]
\centering
\includegraphics{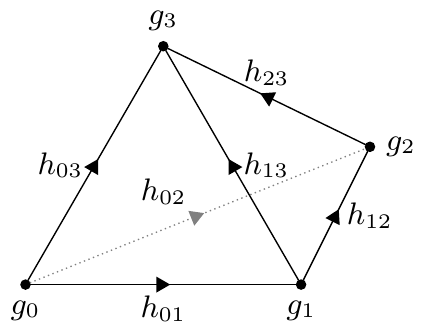}
\caption{Visualization of $\omega_3(g_0^{-1}h_{01}g_1,g_1^{-1}h_{12}g_2,g_2^{-1}h_{23}g_3)$}
\label{fig:omegawithlinks}
\end{figure}

\begin{figure}[h!]
\centering
\includegraphics{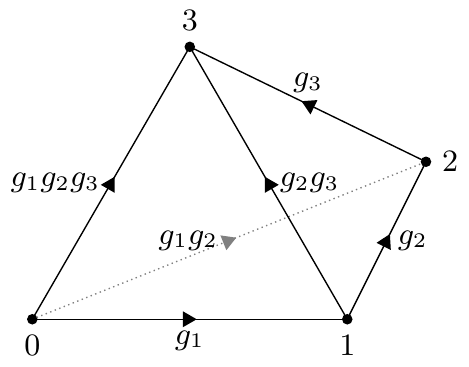}
\caption{Visualization of $\omega_3(g_1,g_2,g_3)$. The numbers on the vertices denote the their local ordering.}
\label{fig:omega3}
\end{figure}

 One can visualize $\nu_n(g_0,...,g_n)$ as an $n$-simplex with the group elements placed on the vertices, as shown in Table \ref{tab:nu_n}. There is a local ordering of vertices, or branching structure, realized by the arrows on the links. The group element $g_i$ is placed on the vertex with $i$ arrows pointing in. The arrows also give the orientation of the simplex. The simplices in the two rows in Table \ref{tab:nu_n} have opposite orientation, and thus opposite parity.\\
 Ref. \onlinecite{Chen2013} has shown that the partition function \eqref{equ:partitionfunctionnu} for two functions $\nu_n$ and $\nu_n'$ describes the same phase if one can be written as another times
 \begin{equation}
\prod_{i=0}^{n} \mu_{n-1}^{(-1)^i}(g_0,...,\hat{g_i},...,g_{n}),
 \end{equation}
for some homogeneous function $\mu_{n-1}$. Distinct SPT phases are thus described by inequivalent $\nu_n$'s which form an abelian group called the cohomology group $\mathcal H^n(G,U(1))$. Physically, the group operation corresponds to stacking two SPT phases on top of each other.\\
\begin{table}[h!]
\caption{Visualization of the cochains $\nu_n$ in various dimensions. The first row has orientation $(+)$, while the second row has orientation $(-)$. For 1+1D, the orientation is determined by following the direction of the vertices in ascending order from 0 to 2 (shown in blue). For 2+1D, the orientation of the tetrahedron is the orientation of the triangle (012) when viewed from vertex 3.}
\begin{center}
\begin{tabular}{|c|c|c|}
\hline
\raisebox{-.5\height}{\begin{tikzpicture}
\node[vertex][label=below:{ $g_0$}] (1) at (0,0) {};
\node[vertex][label=below:{$g_1$}] (2) at (4/2,0) {};
\draw[->] (1) -- (2) node[midway, below] {};
\end{tikzpicture}}&
\raisebox{-.5\height}{\begin{tikzpicture}
\node[vertex][label=below:{$g_0$}] (1) at (0,0) {};
\node[vertex][label=below:{$g_1$}] (2) at (4/2,0) {};
\node[vertex][label=above:{$g_2$}] (3) at (2/2,3.46/2) {};
\draw[->] (1) -- (2) node[midway, below] {};
\draw[->] (2) -- (3) node[midway, right] {};
\draw[->] (1) -- (3) node[midway, left] {};
\draw[->2,blue] (1,0.277) arc (-90:180:0.3) ;
\end{tikzpicture}}
&
\raisebox{-.5\height}{\begin{tikzpicture}
\node[vertex][label=below:{$g_0$}] (1) at (0,0) {};
\node[vertex][label=below:{$g_1$}] (2) at (4/2,0) {};
\node[vertex][label=right:{$g_2$}] (3) at (5/2,2/2) {};
\node[vertex][label=above:{$g_3$}] (4) at (2/2,3.46/2) {};
\draw[->] (1) -- (2) node[midway, below] {};
\draw[->] (2) -- (3) node[midway, right] {};
\draw[->] (3) -- (4) node[midway, above] {};
\draw[->,densely dotted,gray] (1) -- (3) node[midway, left] {};
\draw[->] (1) -- (4) node[midway, left] {};
\draw[->] (2) -- (4) node[midway, right] {};
\end{tikzpicture}}\\
$\nu_1(g_0,g_1)$ & $\nu_2(g_0,g_1,g_2)$ & $\nu_3(g_0,g_1,g_2,g_3)$\\
\hline
\raisebox{-.5\height}{\begin{tikzpicture}
\node[vertex][label=below:{$g_0$}] (1)  at (0,0) {};
\node[vertex][label=below:{ $g_1$}] (2) at (-4/2,0) {};
\draw[->] (1) -- (2) node[midway, below] {};
\end{tikzpicture}}&
\raisebox{-.5\height}{\begin{tikzpicture}
\node[vertex][label=below:{$g_0$}] (1) at (0,0) {};
\node[vertex][label=below:{$g_1$}] (2) at (-4/2,0) {};
\node[vertex][label=above:{$g_2$}] (3) at (-2/2,3.46/2) {};
\draw[->] (1) -- (2) node[midway, below] {};
\draw[->] (2) -- (3) node[midway, left] {};
\draw[->] (1) -- (3) node[midway, left] {};
\draw[->2,blue] (-1,0.277) arc (270:0:0.3) ;
\end{tikzpicture}}
&
\raisebox{-.5\height}{\begin{tikzpicture}
\node[vertex][label=below:{$g_0$}] (1) at (0,0) {};
\node[vertex][label=below:{$g_1$}] (2) at (-4/2,0) {};
\node[vertex][label=right:{$g_2$}] (3) at (0.5,2/2) {};
\node[vertex][label=above:{$g_3$}] (4) at (-2/2,3.46/2) {};
\draw[->] (1) -- (2) node[midway, below] {};
\draw[->,densely dotted,gray] (2) -- (3) node[midway, left] {};
\draw[->] (3) -- (4) node[midway, above] {};
\draw[->] (1) -- (3) node[midway, left] {};
\draw[->] (1) -- (4) node[midway, left] {};
\draw[->] (2) -- (4) node[midway, right] {};
\end{tikzpicture}}\\
$\nu_1(g_0,g_1)^{-1}$ & $\nu_2(g_0,g_1,g_2)^{-1}$ & $\nu_3(g_0,g_1,g_2,g_3)^{-1}$\\
\hline
\end{tabular}
\end{center}
\label{tab:nu_n}
\end{table}
The partition function in equation \eqref{equ:partitionfunctionnu} can be generalized by equipping the manifold with a \underline{\smash{flat connection}}. To do so, we first define another function $\omega_n$
\begin{equation}
 \omega_n(g_1,...,g_n)=\nu_n(1,g_1,g_1g_2,...,g_1g_2g_3 \cdots g_n).
 \end{equation}
By definition, $\omega_n$ is also a $U(1)$-valued function. From the cocycle condition of $\nu_n$, one can check that $\omega_n$ must satisfy
\begin{align}
&\omega_n(g_2, ... ,g_{n+1}) \omega_n(g_1, ... ,g_n)^{(-1)^n} & \nonumber\\
&\cdot \prod_{i=1}^{n-1} \omega_n(g_1, ...,g_{i-1}, g_ig_{i+1},g_{i+1},... ,g_n)^{(-1)^{i}}&=1,
 \end{align}
 which we will call the cocycle condition for $\omega_n$. The partition function can now be rewritten as
\begin{equation}
\label{equ:partitionfunctionomega}
Z= \frac{1}{|G|^{N_v}}\sum_{g_i,g_j,...,g_k} \prod_{ij...k}  \omega_n(g_0^{-1}g_1,g_1^{-1}g_2,...,g_{n-1}^{-1}g_n).
\end{equation}
Note that the expression is invariant under $g_i \rightarrow gg_i$  by $g \in G$, so $\omega_n$ does not need to be homogeneous. We can modify the partition function by inserting group elements $h_{ij} \in G$ into $\omega_n$ such that each term is now
\begin{equation}
\omega_n(g_0^{-1}h_{01}g_1,g_1^{-1}h_{12}g_2,...,g_{n-1}^{-1}h_{n-1,n}g_n).
 \end{equation}
The interpretation of these new variables are group elements that live on the links, as shown in Figure \ref{fig:omegawithlinks}. We demand that the product of group elements along the links must vanish locally around any closed contractible loop in the space-time manifold. For example, the constraints for the simplex in Figure \ref{fig:omegawithlinks} are,
\begin{align}
h_{01}h_{12}=h_{02}, && h_{12}h_{23}=h_{13}, \nonumber \\
\label{equ:flatconnectionconstraints}
 h_{01}h_{13}=h_{03}, &&  h_{02}h_{23}=h_{03}.
\end{align}
However, they are allowed to have holonomies when going around non-contractible loops of the manifold.  Different choices of flat connections on the manifold will correspond to different configurations $h_{ij}$. The trivial flat connection can be represented by the configuration where all $h_{ij}$ are the identity element of $G$. More formally, a flat connection is a homomorphism from the fundamental group $\pi_1(\mathcal M)$ to $G$. The homomorphism gives rise to the constraints in equation \eqref{equ:flatconnectionconstraints}.\\
We remark that the elements $h_{ij}$ are not dynamical. That is, the modified partition function does not sum over all their possible values (such partition function describes a Dijkgraaf-Witten model\cite{DijkgraafWitten}). Another way to phrase it is that we do not sum over all flat connections, but fix one. Physically, the choice of $h_{ij}$ (i.e., the flat connection) corresponds to probing the theory by coupling the degrees of freedom on the vertices to a fixed non-dynamical gauge field, which lives on the links\cite{LevinGu2012,Wen2014,HungWen2014}.\\ 
Next, we notice that each term in the partition function is invariant under the transformation 
\begin{align}
g_i \rightarrow \alpha_i g_i, && h_{ij} \rightarrow \alpha_{i}^{-1}h_{ij} \alpha_{j},
\end{align}
where $\alpha_i \in G$. Thus, we can choose $\alpha_i=\ g_i^{-1}$ to ``fix the gauge'' of our partition function. The partition function is invariant under this gauge transformation. Consequently, after ``integrating out'' all the degrees of freedom on the vertices, we are left with only fixed group elements on the links. Renaming $h$ into $g$, the partition function can now be written as
\begin{equation}
\label{equ:partitionfunction}
Z= \prod_{ij...k} \omega_{n}^{s_{ij...k}}(g_i,g_j,...,g_k),
\end{equation}
where the product now runs over the $n$ independent variables on the links in each simplex triangulating the manifold. For example, the visualization of $\omega_3(g_1,g_2,g_3)$ is a 3-simplex shown in Figure \ref{fig:omega3}. There, the constraints have been applied, and there are only 3 independent variables $g_1$, $g_2$, and $g_3$ for a 3-simplex.

We will now explain the basics of group cohomology necessary to understand these models. Let $G$ be the symmetry group of our theory, with group operation $\cdot$, and $M$ be another abelian group\footnote{More generally, $M$ is a module.}. In this paper, we will only study finite abelian unitary\footnote{By unitary, we mean that we do not consider antiunitary symmetries, such as time-reversal symmetry.} $G$ and mostly deal with $M=U(1)$ (represented using $e^{i\theta})$ or $M=\mathbb{Z}_2$ under addition.

Any function $\omega_n$ from $G^n$ to $M$ is called an \underline{\smash{$n$-cochain}}. The space of all such functions forms a group $\mathcal C^n(G,M)$, with group operation being the product of functions endowed from the group operation in $M$. For the following discussion, let us take this operation to be multiplication.\\
Next, let us define the \underline{\smash{coboundary operator}} $d$ as a map from $\mathcal C^n(G,M)$ to $\mathcal C^{n+1}(G,M)$ by
\begin{align}
d\omega_n(g_1, ... ,g_{n+1}) =& \omega_n(g_2, ... ,g_{n+1}) \omega_n(g_1, ... ,g_n)^{(-1)^{n+1}} \nonumber\\
 &\cdot \prod_{i=1}^{n} \omega_n(g_1, ...,g_{i-1}, g_ig_{i+1},g_{i+1},... ,g_n)^{(-1)^{i}}.
\end{align}
In this way, the cocycle condition of $\omega_n$ can be written simply as
\begin{equation}
\label{equ:cocyclecondition}
d\omega_n = 1.
\end{equation}
Furthermore, one can check from this definition that $d^2$ is null. That is, for any $n$-cochain, $d^2\omega_n =1$.\\
Naturally, we now call the cochain an \underline{\smash{$n$-cocycle}} if it satisfies the cocycle condition \eqref{equ:cocyclecondition}. The $n$-cocycles form a subgroup, which is denoted $\mathcal Z^n(G,M)$. Furthermore, an $n$-cocycle is called an \underline{\smash{$n$-coboundary}} if it is a coboundary of some $(n-1)$-cochain $\mu_{n-1}$ (i.e., $ \omega_n = d\mu_{n-1}$). The subgroup of all $n$-coboundaries is denoted $\mathcal B^n(G,M)$. Accordingly,
\begin{equation}
\mathcal B^n(G,M) \subset \mathcal Z^n(G,M) \subset \mathcal C^n(G,M).
\end{equation}
Let us define the equivalence class where two $n$-cocycles are equivalent if one can be written as the other times an $n$-coboundary. The equivalent classes form the \underline{\smash{$n^\textit{th}$ cohomology group}}
\begin{equation}
\mathcal H^n(G,M) = \mathcal Z^n(G,M)/\mathcal B^n(G,M).
\end{equation}
\begin{table}[ht!]
\caption{Visualization of the cochains $\omega_n$ in various dimensions. The first row has orientation $(+)$, while the second row has orientation $(-)$. The group elements of the remaining links are left out as they are determined from the flatness condition.}
\begin{tabular}{|c|c|c|}
\hline
\raisebox{-.5\height}{\begin{tikzpicture}
\node[vertex][label=below:{$0$}] (1) at (0,0) {};
\node[vertex][label=below:{$1$}] (2) at (4/2,0) {};
\draw[->] (1) -- (2) node[midway, below] {$g_1$};
\end{tikzpicture}}&
\raisebox{-.5\height}{\begin{tikzpicture}
\node[vertex][label=below:{$0$}] (1) at (0,0) {};
\node[vertex][label=below:{$1$}] (2) at (4/2,0) {};
\node[vertex][label=above:{$2$}] (3) at (2/2,3.46/2) {};
\draw[->] (1) -- (2) node[midway, below] {$g_1$};
\draw[->] (2) -- (3) node[midway, right] {$g_2$};
\draw[->] (1) -- (3) node[midway, left] {};
\draw[->2,blue] (1,0.277) arc (-90:180:0.3) ;
\end{tikzpicture}}
&
\raisebox{-.5\height}{\begin{tikzpicture}
\node[vertex][label=below:{$0$}] (1) at (0,0) {};
\node[vertex][label=below:{$1$}] (2) at (4/2,0) {};
\node[vertex][label=right:{$2$}] (3) at (5/2,2/2) {};
\node[vertex][label=above:{$3$}] (4) at (2/2,3.46/2) {};
\draw[->] (1) -- (2) node[midway, below] {$g_1$};
\draw[->] (2) -- (3) node[midway, right] {$g_2$};
\draw[->] (3) -- (4) node[midway, above] {$g_3$};
\draw[->,densely dotted,gray] (1) -- (3) node[midway, left] {};
\draw[->] (1) -- (4) node[midway, left] {};
\draw[->] (2) -- (4) node[midway, right] {};
\end{tikzpicture}}\\
$\omega_1(g_1)$ & $\omega_2(g_1,g_2)$ & $\omega_3(g_1,g_2,g_3)$\\
\hline
\raisebox{-.5\height}{\begin{tikzpicture}
\node[vertex][label=below:{$0$}] (1)  at (0,0) {};
\node[vertex][label=below:{$1$}] (2) at (-4/2,0) {};
\draw[->] (1) -- (2) node[midway, below] {$g_1$};
\end{tikzpicture}}&
\raisebox{-.5\height}{\begin{tikzpicture}
\node[vertex][label=below:{$0$}] (1) at (0,0) {};
\node[vertex][label=below:{$1$}] (2) at (-4/2,0) {};
\node[vertex][label=above:{$2$}] (3) at (-2/2,3.46/2) {};
\draw[->] (1) -- (2) node[midway, below] {$g_1$};
\draw[->] (2) -- (3) node[midway, left] {$g_2$};
\draw[->] (1) -- (3) node[midway, left] {};
\draw[->2,blue] (-1,0.277) arc (270:0:0.3) ;
\end{tikzpicture}}
&
\raisebox{-.5\height}{\begin{tikzpicture}
\node[vertex][label=below:{$0$}] (1) at (0,0) {};
\node[vertex][label=below:{$1$}] (2) at (-4/2,0) {};
\node[vertex][label=right:{$2$}] (3) at (0.5,2/2) {};
\node[vertex][label=above:{$3$}] (4) at (-2/2,3.46/2) {};
\draw[->] (1) -- (2) node[midway, below] {$g_1$};
\draw[->,densely dotted,gray] (2) -- (3) node[midway, left] {\color{black} $g_2$};
\draw[->] (3) -- (4) node[midway, above] {$g_3$};
\draw[->] (1) -- (3) node[midway, left] {};
\draw[->] (1) -- (4) node[midway, left] {};
\draw[->] (2) -- (4) node[midway, right] {};
\end{tikzpicture}}\\
$\omega_1(g_1)^{-1}$ & $\omega_2(g_1,g_2)^{-1}$ & $\omega_3(g_1,g_2,g_3)^{-1}$\\
\hline
\end{tabular}
\label{tab:visual}
\end{table}
To make the visualization introduced earlier more concrete, let us introduce the dual of $n$-cochains called \underline{\smash{$n$-chains}}. They are a collection of $n$-simplices with a well defined ordering of vertices, and form a group $\mathcal C_n(G,M)$. Examples of $n$-simplices are illustrated in Table \ref{tab:visual}. There are two ways to label these simplices. One is to read off the ordering of the vertices. For example, the 3-simplex with positive orientation $(+)$ in Figure \ref{fig:omega3} can be labeled $(0123)$. Alternatively, one can read off the group elements on links $(01)$, $(12)$, and $(23)$. In the figure, the simplex would then be denoted $(g_1,g_2,g_3)$. For the negative orientation, we will label the simplex by raising it to the $^{-1}$ power, such as $(g_1,g_2,g_3)^{-1}$.\\
Dual to the coboundary operator is the boundary operator $\partial$, which takes an $n$-chain to its boundary $(n-1)$-chain:
\begin{align}
\partial(g_1, ... ,g_{n}) &= (g_2, ... ,g_n) (g_1, ... ,g_{n-1})^{(-1)^n} \nonumber\\
\label{equ:boundarylinks}
& \cdot \prod_{i=1}^{n-1} (g_1, ...,g_{i-1}, g_ig_{i+1},g_{i+1},... ,g_n)^{(-1)^{i}}
\end{align}
The multiplication is that of the group $\mathcal C_n(G,M)$ and is interpreted as adjoining the simplices. As an example, the boundary of the simplex in Figure \ref{fig:omega3} is $\partial(g_1,g_2,g_3)=(g_2,g_3)(g_1g_2,g_3)^{-1}(g_1,g_2g_3)(g_1,g_2)^{-1}$. In terms of the vertices,
\begin{align}
\label{equ:boundary}
\partial(012...n) &= \prod_{i=0}^{n} \left (012...\hat{i}...(n-1)n \right)^{(-1)^i},
\end{align}
where $\hat{i}$ means skipping over the element $i$. For example, $\partial(0123)=(123)(023)^{-1}(013)(012)^{-1}$. We remark that $\partial^2$ is null in the same way as $d^2$ is.

When an $n$-cochain $\omega_n$ acts on the simplex $ (g_1,...,g_n)$, or $ (g_1,...,g_n)^{-1}$, the value in $M$ is respectively $\omega_n(g_1,...,g_n)$ or  $\omega_n(g_1,...,g_n)^{-1}$. In general, an $n$-chain can consist of more than one $n$-simplex. Acting $\omega_n$ on the $n$-chain is defined as the multiplication (in $M$) of acting $\omega_n$ individually on each $n$-simplex.  Abstractly, the $n$-cochain $\omega_n$ is a map from such a simplex to the group $M$. It is easier to visualize the evaluated value of the $n$-cochain as the $n$-chain itself.

We can now describe the cocycle condition visually. Consider a 2-cochain $\omega_2$. The coboundary of $\omega_2$ is
\begin{align}
d\omega_2(g_1,g_2,g_3) &=\frac{\omega_2(g_2,g_3)\omega_2(g_1,g_2g_3)}{\omega_2(g_1g_2,g_3)\omega_2(g_1,g_2)} \nonumber\\
&= \omega_2(\partial (g_1,g_2,g_3)).
\end{align}
If we compare it to Figure \ref{fig:omega3}, we can see that $d\omega_2$ evaluated on the tetrahedron is exactly $\omega_2$ evaluated on the four triangles which form the boundary of the tetrahedron.  Note that this is analogous to Stokes' theorem for differential forms. Hence, the cocycle condition means that $\omega_2$ evaluated on the surface of this 3-simplex is unity. In general, a cocycle $\omega_n$ evaluated on any closed $n$-manifold (provided that the connection is trivial) will give unity.\\
The cocycle condition also has a nice property in terms of triangulation. The cocycle condition can be arranged to read 
\begin{align}
\frac{\omega_2(g_1,g_2)}{\omega_2(g_2,g_3)} &=\frac{\omega_2(g_1,g_2g_3)}{\omega_2(g_1g_2,g_3)},\\
\omega_2(g_1,g_2)&= \frac{\omega_2(g_2,g_3) \omega_2(g_1,g_2g_3)}{\omega_2(g_1g_2,g_3)},
\end{align}
or visually,
\begin{align}
\raisebox{-.5\height}{\begin{tikzpicture}
\begin{scope}[scale=1.5]
\node[vertex] (1) at (0,0) {};
\node[vertex] (2) at (1.4,0) {};
\node[vertex] (3) at (0,1.4) {};
\node[vertex] (4) at (1.4,1.4) {};
\draw[->] (1) -- (2) node[midway, below] {$g_1$};
\draw[->] (2) -- (4) node[midway, right] {$g_2g_3$};
\draw[->] (3) -- (4) node[midway, above] {$g_3$};
\draw[->] (2) -- (3) node[midway, left] {$g_2$};
\draw[->] (1) -- (3) node[midway, left] {$g_1g_2$};
\draw[->2,blue] (0.47,0.22) arc (-90:180:0.15) ;
\draw[->2,blue] (0.94,0.79) arc (270:0:0.15) ;
\end{scope}
\end{tikzpicture}} 
= &
\raisebox{-.5\height}{\begin{tikzpicture}
\begin{scope}[scale=1.5]
\node[vertex] (1) at (0,0) {};
\node[vertex] (2) at (1.4,0) {};
\node[vertex] (3) at (0,1.4) {};
\node[vertex] (4) at (1.4,1.4) {};
\draw[->] (1) -- (2) node[midway, below] {$g_1$};
\draw[->] (2) -- (4) node[midway, right] {$g_2g_3$};
\draw[->] (3) -- (4) node[midway, above] {$g_3$};
\draw[->] (1) -- (4) node[midway, above left] {};
\draw[->] (1) -- (3) node[midway, left] {$g_1g_2$};
\node at (0.5, 0.9)[anchor=north, scale=1,rotate=45]{$g_1g_2g_3$}; 
\draw[->2,blue] (0.94,0.22) arc (-90:180:0.15) ;
\draw[->2,blue] (0.47,0.79) arc (270:0:0.15) ;
\end{scope}
\end{tikzpicture}},
\end{align}
\begin{align}
\raisebox{-.5\height}{\begin{tikzpicture}
\begin{scope}[scale=1.5]
\node[vertex] (1) at (0,0) {};
\node[vertex] (2) at (1.5*1.4,0) {};
\node[vertex] (3) at (1.5/2*1.4,1.3*1.4) {};
\draw[->] (1) -- (2) node[midway, below] {$g_1$};
\draw[->] (2) -- (3) node[midway, right] {$g_2$};
\draw[->] (1) -- (3) node[midway, left] {$g_1g_2$};
\draw[->2,blue] (1.05,0.306) arc (-90:180:0.3) ;
\end{scope}
\end{tikzpicture}} 
= &
\raisebox{-.5\height}{\begin{tikzpicture}
\begin{scope}[scale=1.5]
\node[vertex] (1) at (0,0) {};
\node[vertex] (2) at (3/2*1.4,0) {};
\node[vertex] (3) at (1.5/2*1.4,2.6/2*1.4) {};
\node[vertex] (4) at (1.5/2*1.4,0.819/2*1.4) {};
\draw[->] (1) -- (2) node[midway, below] {$g_1$};
\draw[->] (2) -- (3) node[midway, right] {$g_2$};
\draw[->] (1) -- (3) node[midway, left] {$g_1g_2$};
\draw[->] (1) -- (4) node[midway, below] {};
\draw[->] (2) -- (4) node[midway, below] {$g_2g_3$};
\draw[->] (3) -- (4) node[midway, left] {$g_3$};
\node at (0.7, 0.3)[anchor=north, scale=1]{$g_1g_2g_3$}; 
\draw[->2,blue] (1.05,0.1) arc (-90:180:0.15);
\draw[->2,blue] (0.7,0.65) arc (270:0:0.15);
\draw[->2,blue] (1.4,0.65) arc (-90:180:0.15);
\end{scope}
\end{tikzpicture}}.
\end{align}
Let us now return to discuss the partition function we defined in equation \eqref{equ:partitionfunction}. We can see that, if $\omega_n$ is an $n$-cocycle, then the evaluation on a given space-time manifold is triangulation independent. Thus, the partition function is a topological quantity of the manifold.  


To summarize, bosonic SPT phases can be modeled via a partition function given by the evaluation of a cocycle on a manifold with a flat connection by equation \eqref{equ:partitionfunction}. To avoid cumbersome notation, we will write the partition function on $\mathcal{M}_n$, an $n$-manifold using the following schematic notation\footnote{Naively, one can think of it in the same spirit of integrating differential forms on a manifold. However it can also be made precise in the following way. A triangulation of the manifold defines a map from $\mathcal M$ to $BG$, the classifying space of the group $G$. We can then pullback the cocycles $\omega_n$ from $BG$ to $\mathcal M$, which we still call $\omega_n$. The partition function is then the evaluation of the cocycle on the manifold, which is represented by the integral.}:
\begin{equation}
\label{equ:intomega}
Z= \int_{\mathcal{M}_n} \omega_n.
\end{equation}

\subsection{The Gu-Wen Group Super-cohomology Model for Fermionic SPT Phases}\label{GuWen}
Fermionic SPT phases are an extension of the bosonic case. There is a generalization of the bosonic SPT phase partition function to fermionic SPT phases by Gu \& Wen, which they call a special group super-cohomology model \cite{GuWen2014}. The symmetry group of the theory is now $G_f = \mathbb Z_2^f \times G$, where $\mathbb Z_2^f$ is generated by the fermionic parity operator. In this presentation, we have repeated the procedure of generalizing the model to include manifolds with non-trivial flat connections.\\
 The model introduces a second layer using a cocycle $\beta_{n-1} \in \mathcal Z^{n-1}(G,\mathbb{Z}_2)$. That is,
\begin{align}
\label{equ:dbeta}
d\beta_{n-1}=0.
\end{align}
Note that we are using addition as the group operation for $\mathbb Z_2$. The function $\omega_n$ is no longer a cocycle. Instead, it is a cochain whose coboundary is
\begin{align}
\label{equ:domega}
d\omega_n &= (-1)^{Sq^2(\beta_{n-1})}.
\end{align}
Here, $Sq^2$ is the second Steenrod square\cite{Steenrod1947}, defined as
\begin{align}
Sq^2(\beta_{n-1}) = \beta_{n-1} \cup_{n-3} \beta_{n-1},
\end{align}
where $\cup_i$ is the cup-$i$ product (see Appendix \ref{app:cup} for a discussion of cup-$i$ products and Steenrod squares).  Roughly speaking, $\beta_{n-1}$ ``twists'' the cocycle condition of $\omega_n$ and will add an extra layer of fermions into the theory. Equation \eqref{equ:domega} is called the \underline{\smash{Gu-Wen equation}}.\\
The Steenrod square is a cohomology operation. That is, $Sq^2(\beta_{n-1})$ is a $\mathbb Z_2$-valued cocycle. However, in order for equation \eqref{equ:domega} to make sense. It is important that $Sq^2(\beta_{n-1})$ becomes a coboundary when we describe it as a $U(1)$-valued function, i.e.,
\begin{align}
\label{equ:obstruction}
(-1)^{Sq^2(\beta_{n-1})} \in \mathcal B^{n+1}(G,U(1)).
\end{align}
This is known as the \underline{\smash{obstruction-free condition}}. For certain cocycles $\beta_{n-1}$, the $U(1)$-valued function $(-1)^{Sq^2(\beta_{n-1})}$ is not a coboundary, and a theory with such input is said to be \underline{\smash{obstructed}}. The equivalence class of $\beta_{n-1}$ that has no obstruction forms an obstruction-free subgroup of $\mathcal H^{n-1}(G,\mathbb Z_2)$ called $B\mathcal H^{n-1}(G,\mathbb Z_2)$. The fermionic SPT phases in this model are defined by a pair $(\omega_n,\beta_{n-1})$ such that equations \eqref{equ:dbeta}, \eqref{equ:domega}, and \eqref{equ:obstruction} are satisfied. The group structure of these phases is a ``super-cohomology'' group $\mathscr  H^n(G^f,U(1))$, which is an extension of $B\mathcal H^{n-1}(G,\mathbb Z_2)$ by $\mathcal H^n(G,U(1))$. That is, we have the short exact sequence
\begin{equation}
0 \rightarrow \mathcal H^n(G,U(1)) \rightarrow \mathscr H^n(G^f,U(1)) \rightarrow B\mathcal H^{n-1}(G,\mathbb Z_2)  \rightarrow 0.
\end{equation}

\begin{figure}[h!]
\centering
\includegraphics{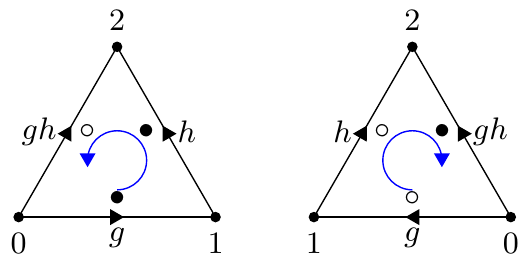}
\caption{Position of Grassmann variables in 1+1D in each simplex. We place $\theta$ at \textbullet \ and $\bar\theta$ at $\circ$ if $\beta_{n-1}$ evaluated on the corresponding face is one. For example, the Grassmann variable on the link $(01)$ of the $(+)$ and $(-)$ triangles are respectively $\theta_{01}^{\beta_1(g)}$ and $\bar\theta_{01}^{\beta_1(g)}$.}
\label{fig:Grassmann}
\end{figure}

Next, we will review the construction of the Gu-Wen partition function. Similarly to the bosonic case, we triangulate our manifold and place group elements on the links.  However, in addition, we place Grassmann variables $\theta$ or $\bar\theta$ inside each simplex located at each face (i.e., codimension-1 subsimplex) only if $\beta_{n-1}$ evaluated on that face is one. For the $(+)$ orientation, the variable placed will be $\theta$ or $\bar\theta$ if the vertex opposite of that face is even or odd respectively, and in the opposite fashion for a $(-)$ simplex. An example is given in Figure \ref{fig:Grassmann} for 1+1D with  \textbullet \ and $\circ$ representing $\theta$ and $\bar\theta$ respectively.

We can now introduce the partition function for the group super-cohomology model. The partition function now has 3 terms:
\begin{align}
\label{equ:GuWenpartitionfunction}
\mathcal Z(\eta)= Z \cdot \sigma(\beta) \cdot (-1)^{\int_S m}.
\end{align}
The first factor is the product of $\omega$ evaluated on all simplices in the manifold $\int_\mathcal{M} \omega$ in the same spirit as equation \eqref{equ:intomega}. We will call this term the \underline{\smash{bosonic integral}}, which we still denote $Z$, but note that it is no longer an invariant under retriangulations in the fermionic case, since $\omega$ is not necessarily a cocycle.\\
The second factor is a \underline{\smash{Grassmann integral}} $\sigma(\beta)$. The integrand is a product of all Grassmann variables in a certain order, where they are grouped according to the simplex they live in. Since $\beta$ is a cocycle, there is always an even number of Grassmann variables in each simplex, and so each group will be Grassmann even. Therefore, we can insert the groups in any order. However, there is a specific ordering within each simplex. For a $(+)$ simplex, we write out all $\theta$'s first in order of the vertex opposite of that face, followed by all $\bar\theta$'s in an identical order. The order is reversed for a $(-)$ simplex. Thus, in the example in Figure \ref{fig:Grassmann}, the $(+)$ simplex has $[\theta_{12}^{\beta(h)}\theta_{01}^{\beta(g)}\bar\theta_{02}^{\beta(gh)}]$ and the $(-)$ simplex has $[\theta_{02}^{\beta(gh)}\bar\theta_{01}^{\beta(g)}\bar\theta_{12}^{\beta(h)}]$. The measure of the integral is a product of all $[d\theta d\bar\theta]$ pairs that are on opposite sides of a face. Note that, each pair is also Grassmann even, so the order of the pairs that enter into the measure does not matter. The only important thing is that $d\theta$ always comes before $d\bar\theta$. Schematically,
\begin{equation}
\sigma(\beta) = \int \prod_{\text{faces}} [d\theta d\bar\theta] \prod_\text{simplices} [\text{variables in each simplex}] 
\end{equation}
Lastly, the third factor is an evaluation of a quantity called $m$ evaluated over a collection of codimension-2 simplices $S$, which is written schematically as $\int_S m$, where $m$ satisfies $dm=\beta$. $m$ is not a physical quantity in the theory\footnote{In fact, since $\beta$ is not necessarily a coboundary, there is no function $m$ that will satisfy this relation. $m$ is only useful for calculation purposes and its interpretation is further discussed in Appendix \ref{app:m}}, but is needed so that the partition function is invariant under Pachner moves i.e. retriangulations. The set $S$ is defined (modulo 2) as the sum of all codimension-2 subsimplices and the subsimplices in the positions given in Table \ref{tab:m} for various dimensions\cite{GuWen2014}. For example, in 2+1D, in additional to all links in the triangulation, for each tetrahedron with local ordering $(0123)$, we add the link $(02)$ to $S$ if that tetrahedron is $(+)$ or $(13)$ if that tetrahedron is $(-)$.
 \begin{table}[h!]
 \caption{Positions of subsimplices in each simplex that need to be included in the set $S$ in addition to all codimension-2 simplices (modulo 2).}
\begin{tabular}{|c|c|c|}
\hline
Dim & $(+)$ simplex & $(-)$ simplex\\
\hline
0+1D & none & none\\
1+1D & none & (1)\\
2+1D & (02) & (13)\\
3+1D & (013),(134),(123) & (024)\\
\hline
\end{tabular}
 \label{tab:m}
\end{table}

It was later proposed in Ref. \onlinecite{GaiottoKapustin2016} that $S$ is the chain representative of the second Stiefel-Whitney class $[w_2]$, which is trivial in the cohomology group $H^2(\mathcal M,\mathbb Z_2)$ on spin manifolds (we refer the reader to Appendix \ref{app:w_2} for more details on the Stiefel-Whitney class). If we let $w_2$ be a coboundary that represents this class, then there exists a cochain $\eta\in C^1(\mathcal{M},\mathbb{Z}_2)$ such that $d\eta = w_2$. The choice of $\eta$ is not unique, and represents the choice of spin structure on the manifold. If we denote $E$ as the chain representative of $\eta$, then $\partial E =S$, and so we can rewrite the integral as $\int_{S} m = \int_{E} \beta$, which depends on $\beta$, a physical quantity. Similarly to $\eta$, $E$ is not unique and depends on the spin structure of the manifold. Note that this implies that the Gu-Wen partition function is only well defined on spin manifolds. We will call this last term in the partition function the \underline{\smash{spin structure term}}, which we will denote $E(\beta)$.

In this paper, we will use an alternative definition of the set $S$, which has been given in general in Ref. \onlinecite{goldstein1976}. This definition has an advantage that it is independent of the orientation of the simplex and simplifies the dimensional reduction procedure in section \ref{dimredfermions}. The definitions of $S$ in different dimensions (mod 2) are
\begin{subequations}
\begin{align}
 \label{equ:S0_1D}
S^{0+1} =&\{ \},\\
 S^{1+1}=&\{\text{all 0-simplices}\} + \{\text{(0) in any 1-simplex}\}\nonumber\\
 \label{equ:S1_1D}
  &+\{\text{(0) in any 2-simplex}\},\\
 S^{2+1}=&\{\text{all 1-simplices}\} + \{\text{(02) in any 2-simplex}\} \nonumber\\
\label{equ:S2_1D}
 &+ \{\text{(03) in any 3-simplex}\},\\
 S^{3+1}=&\{\text{all 2-simplices}\}\nonumber\\
 &+ \{\text{(012),(023) in any 3-simplex}\}\nonumber\\
 \label{equ:S3_1D}
 &+ \{\text{(012),(023),(034) in any 4-simplex}\}.
 \end{align}
 \end{subequations}
We will call the three contributions to $S$ as $S_1$, $S_2$, and $S_3$. Visually, the set $S$ is
\begin{align}
 \label{equ:S1_1Dpic}
 S_1^{1+1}=&
 \raisebox{-.5\height}{\begin{tikzpicture}[scale=0.5]
\node[vertex,blue][label=below:$0$]  (0) at (0,0) {};
\end{tikzpicture}}, &
 S_2^{1+1}=&
\raisebox{-.5\height}{\begin{tikzpicture}[scale=0.5]
\node[vertex,blue][label=below:$0$]  (0) at (0,0) {};
\node[vertex][label=below:$1$]  (1) at (2,0) {};
\draw[->,blue,opacity=0.5] (0) -- (1) node[midway, below] {};
\end{tikzpicture}}, \nonumber\\
S_3^{1+1}=&
   \raisebox{-.5\height}{\begin{tikzpicture}
   \begin{scope}[scale=0.6]
\filldraw[ fill=blue,   fill opacity=0.5,draw=white](0,0)--(2,0)--(1,3.46/2)--cycle;
\node[vertex,blue][label=below:$0$] (0) at (0,0) {};
\node[vertex][label=below:] (1) at (2,0) {};
\node[vertex][label=above:] (2) at (1,3.46/2) {};
\draw[->] (0) -- (1) node[midway, below] {}  node[midway, above] {};
\draw[] (1) -- (2) node[midway, left] {} node[midway, right] {};
\draw[->] (0) -- (2) node[midway, right] {} node[midway, left] {} ;
\end{scope}
\end{tikzpicture}}, & &
\end{align}
\begin{align}
 S_1^{2+1}=&
\raisebox{-.5\height}{\begin{tikzpicture}[scale=0.5]
\node[vertex][label=below:$0$]  (0) at (0,0) {};
\node[vertex][label=below:$1$]  (1) at (2,0) {};
\draw[->,blue] (0) -- (1) node[midway, below] {};
\end{tikzpicture}} , &
S_2^{2+1}=&
   \raisebox{-.5\height}{\begin{tikzpicture}
   \begin{scope}[scale=0.6]
\filldraw[ fill=blue,   fill opacity=0.5,draw=white](0,0)--(2,0)--(1,3.46/2)--cycle;
\node[vertex][label=below:$0$] (0) at (0,0) {};
\node[vertex][label=below:$1$] (1) at (2,0) {};
\node[vertex][label=above:$2$] (2) at (1,3.46/2) {};
\draw[->] (0) -- (1) node[midway, below] {}  node[midway, above] {};
\draw[->] (1) -- (2) node[midway, left] {} node[midway, right] {};
\draw[->,blue] (0) -- (2) node[midway, right] {} node[midway, left] {} ;
\end{scope}
\end{tikzpicture}}, \nonumber\\
\label{equ:S2_1Dpic}
S_3^{2+1}=&
   \raisebox{-.5\height}{\begin{tikzpicture}
   \begin{scope}[scale=0.6]
\filldraw[ fill=blue,   fill opacity=0.5,draw=white](0,0)--(2,0)--(1,3.46/2)--cycle;
\filldraw[ fill=blue,   fill opacity=0.7,draw=white](2.5,1)--(2,0)--(1,3.46/2)--cycle;
\node[vertex][label=below:$0$] (0) at (0,0) {};
\node[vertex][label=below:] (1) at (2,0) {};
\node[vertex][label=above:] (2) at (2.5,1) {};
\node[vertex][label=above:$3$] (3) at (1,3.46/2) {};
\draw[->] (0) -- (1) node[midway, below] {}  node[midway, above] {};
\draw (1) -- (2) node[midway, left] {} node[midway, right] {};
\draw[->,densely dotted,gray] (0) -- (2) node[midway, right] {} node[midway, left] {} ;
\draw[->,blue] (0) -- (3);
\draw[->] (1) -- (3);
\draw[->] (2) -- (3);
\end{scope}
\end{tikzpicture}}, & &
 \end{align}
where the elements of the sets $S^{1+1}$ and $S^{2+1}$ are denoted by the blue vertices and blue links, respectively. Note that in this definition, there is no dependence on the orientation of the simplices. A derivation of these formulas and a proof that they are equivalent to the set $S$ defined in Table \ref{tab:m} are given in Appendix \ref{app:w_2}.

\section{Dimensional Reduction for Bosonic SPT Phases}\label{dimredbosons}
\subsection{Dimensional Reduction over $S^1$}
\begin{figure}[h!]
\centering
\includegraphics{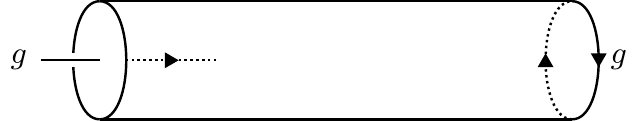}
\caption{A schematic picture of dimensional reduction, with a holonomy $g$ going around the circle. One can think of this as inserting a ``flux'' $g$ in the tube.}
\label{fig:dimred}
\end{figure}
In this section, we will introduce a technique called dimensional reduction to study higher dimensional SPT phases by reducing it to a lower dimension via compactification. This is shown schematically in Figure \ref{fig:dimred}. In doing so, one could also add a ``flux'' corresponding to a group element $g \in G$ going through the hole so that there is a holonomy of $g$ going around the loop. Doing this, we can obtain different lower dimensional theories depending on the choice of $g$.

For bosonic SPT phases described by a cocycle, there is a precise realization of the described process called the slant product $i$. This product pairs a group element $g$ in the subscript with an $(n+1)$-cochain $\omega_{n+1}$ and gives an $n$-cochain $i_{g}\omega_{n+1}$. It is defined as
\begin{align}
\label{equ:slant}
i_{g}\omega_{n+1}(a_1, ..., a_n) = \prod_{i=1}^{n+1} \omega_{n+1}^{(-1)^{n+1+i}}(a_1,...,a_{i-1},g,a_{i},...,a_n).
\end{align}
That is, it is a product of the $\omega_{n+1}$'s with $g$ inserted at all possible positions. There is a parity exponent, which is positive for $g$ inserted at the rightmost slot, and alternates as $g$ moves to the left. As an example, for the 3-cocycle $\omega_3$, we have the following slant product, which can also be presented pictorially as a prism:
\begin{align}
\label{equ:prism}
i_{g}\omega_{3}(a_1,a_2) &= \frac{\omega_3(g,a_1,a_2)\omega_3(a_1,a_2,g)}{\omega_3(a_1,g,a_2)}\nonumber\\
 &=\raisebox{-.5\height}{\begin{tikzpicture}
\node[vertex] (1) at (0,0) {};
\node[vertex] (2) at (1,-0.5) {};
\node[vertex] (3) at (2,0) {};
\node[vertex] (4) at (0,2) {};
\node[vertex] (5) at (1,1.5) {};
\node[vertex] (6) at (2,2) {};
\draw[->,densely dotted,color=gray] (1) -- (3) node[midway, above] {};
\draw[->,densely dotted,color=gray] (1) -- (6)  {};
\draw[->] (1) -- (2) node[midway, below] {$a_1$};
\draw[->] (2) -- (3) node[midway, below] {$a_2$};
\draw[->] (1) -- (4) node[midway, left] {$g$};
\draw[->] (2) -- (5) node[midway, above right] {$g$};
\draw[->] (3) -- (6) node[midway, right] {$g$};
\draw[->] (4) -- (5) node[midway, below] {$a_1$};
\draw[->] (5) -- (6) node[midway, below] {$a_2$};
\draw[->] (4) -- (6) node[midway, above left] {};
\draw[->] (1) -- (5)  {};
\draw[->] (2) -- (6)  {};
\end{tikzpicture}} .
\end{align}
Using the definition in equation \eqref{equ:slant}, one can show that
\begin{equation}
di_g\omega = i_gd\omega.
\end{equation}
That is, the slant product commutes with the coboundary operator. From this, one can show that $i_g$ maps cocycles to cocycles and coboundaries to coboundaries, and hence defines a map from $\mathcal H^{n+1} (G,U(1)) \rightarrow \mathcal H^{n}(G,U(1))$. The physical interpretation of the slant product is that, if $\omega_{n+1}$ defines an SPT phase in $n+1$ dimensions, then $i_g \omega_{n+1}$ describes an SPT phase in $n$ dimensions, corresponding to compactifying the former over a circle with a $g$-flux inserted.
In terms of the partition function, the original one in the higher dimension is an ``integral'' of $\omega$  over  $\mathcal{M} \times S^1$, while the compactified version is now the ``integral'' of $i_g\omega$ over the compactified manifold $\mathcal{M}$. This can be written schematically as
\begin{equation}
\label{equ:dimredbosons}
Z=\int_{\mathcal{M} \times S^1} \omega = \int_{\mathcal{M}} i_g\omega.
\end{equation}
\subsection{General Dimensional Reduction}
Dimensional reduction can also be further done by repeatedly applying the slant product $i_g$ using different group elements. Doing so would allow one to compute partition functions on any $n$-torus $T^n$. However, one can also do dimensional reduction by using the general notion of the slant product. In the general definition, the slant product takes an $n$-cochain $\omega_n$ along with any $m$-chain $(g_1,...,g_m)$. In the case where $m=1$, this is equivalent to $i_g$, the compactification over a circle. For example, in equation \eqref{equ:prism}, the triangulation that represents $i_g\omega_3$ is a Cartesian product of a 2-simplex that triangulates the manifold, and a 1-chain $(g)$ that represents the circle with holonomy $g$. The general slant product can be explicitly defined as a map $i: \mathcal C_m(G,M) \times \mathcal C^{n+m}(G,M) \rightarrow \mathcal C^n(G,M) \nonumber$. For an $m$-chain containing only one simplex $(g_1,..., g_m)$,
\begin{widetext}
\begin{gather}
i_{(g_1,..., g_m)}\omega_{m+n}(a_1,..., a_n) = \prod_{\substack{i_1,...,i_m =1\\i_1<...<i_m}}^{m+n}\omega_{m+n}^{\sigma(i_1,...,i_m)}(a_1,...,a_{i_1-1},g_{1},a_{i_1},...,a_{i_2-2},g_2,a_{i_2-1},...,a_{i_m-m},g_{m},a_{i_m-m+1},...,a_n), \nonumber\\
\text{where \ \ } \sigma(i_1,...,i_m) = (-1)^{\sum_{j=1}^m n+j+i_j}.
\end{gather}
\begin{figure}[H]
\centering
\includegraphics{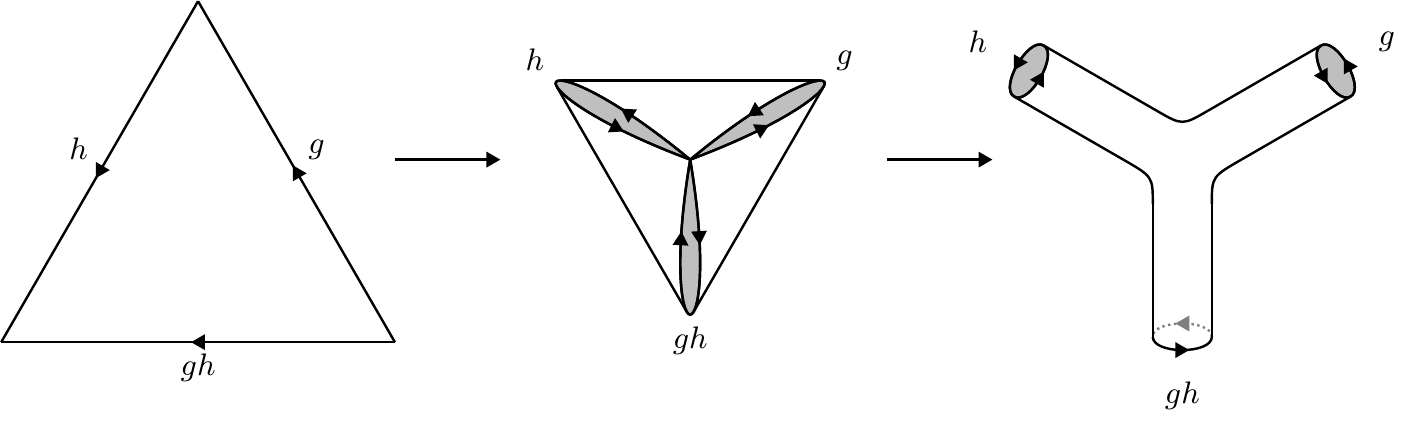}
\caption{How the slant product on the 2-chain $(g,h)$ is equivalent to compactification over a 3-tube configuration. The reduced phase described by the cocycle $\alpha_{(g,h)}$ lives two dimensions lower at the junction of three trivial phases described by cocycles $i_g \omega$, $i_h \omega$, and $1/i_{gh}\omega$.}
\label{fig:3hole}
\end{figure}
As an example, the slant product of a 5-cochain $\omega_5$ and a 2-chain $(g,h)$ is a 3-cochain $i_{(g,h)}\omega_5$ given by
\begin{equation}
\label{equ:slantgh5}
i_{(g,h)}\omega_5(a,b,c)  =  \frac{\omega_5(a,b,c,g,h)\omega_5(a,b,g,h,c)\omega_5(a,g,h,b,c)\omega_5(g,h,a,b,c)\omega_5(a,g,b,c,h)\omega_5(g,a,b,h,c)}   {\omega_5(a,b,g,c,h)\omega_5(a,g,b,h,c)\omega_5(g,a,h,b,c)\omega_5(g,a,b,c,h)}.
\end{equation}
\end{widetext}
Intuitively, $g_1,...,g_m$ are inserted in all possible slots that perserve the ordering $1,...,m$. The variable $i_j$ represents the position where we insert $g_j$ and the sign factor $\sigma(i_1,...,i_m)$ sums the number of times $g_j$ is moved away from position $n+j$ for all $j$. That is, the term $\omega_{m+n} (a_1,a_2,..., a_n,g_1, g_2,..., g_m)$ has $\sigma(i_1,...,i_m)=1$, and all terms are calculated relative to this term.\\
 Geometrically, it is easier to visualize the slant product pictorially as the Cartesian product of an $m$-simplex and an $n$-simplex:\\
 \begin{align}
 i_{(g_1,..., g_m)}\omega_{m+n}(a_1,...,a_n)=
\raisebox{-.5\height}{\begin{tikzpicture}
\begin{scope}[scale=0.55]
\node[vertex] (1) at (0,0) {};
\node[vertex] (2) at (1,0) {};
\node[vertex] (3) at (2,0) {};
\node[vertex] (4) at (3,0) {};
\node[vertex] (5) at (4,0) {};
\node[vertex] (6) at (0,1) {};
\node[vertex] (7) at (1,1) {};
\node[vertex] (8) at (2,1) {};
\node[vertex] (9) at (3,1) {};
\node[vertex] (10) at (4,1) {};
\node[vertex] (11) at (0,2) {};
\node[vertex] (12) at (1,2) {};
\node[vertex] (13) at (2,2) {};
\node[vertex] (14) at (3,2) {};
\node[vertex] (15) at (4,2) {};
\node[vertex] (16) at (0,3) {};
\node[vertex] (17) at (1,3) {};
\node[vertex] (18) at (2,3) {};
\node[vertex] (19) at (3,3) {};
\node[vertex] (20) at (4,3) {};
\node[vertex] (21) at (0,4) {};
\node[vertex] (22) at (1,4) {};
\node[vertex] (23) at (2,4) {};
\node[vertex] (24) at (3,4) {};
\node[vertex] (25) at (4,4) {};
\draw[->] (1) -- (2) node[midway, below] {$a_1$};
\draw[->] (2) -- (3) node[midway, below] {$a_2$};
\draw[->] (3) -- (4) node[midway, below] {$\cdots$};
\draw[->] (4) -- (5) node[midway, below] {$a_n$};
\draw[->] (6) -- (7) node[midway, above] {};
\draw[->] (7) -- (8) node[midway, above] {};
\draw[->] (8) -- (9) node[midway, above] {};
\draw[->] (9) -- (10) node[midway, above] {};
\draw[->] (11) -- (12) node[midway, above] {};
\draw[->] (12) -- (13) node[midway, above] {};
\draw[->] (13) -- (14) node[midway, above] {};
\draw[->] (14) -- (15) node[midway, above] {};
\draw[->] (16) -- (17) node[midway, above] {};
\draw[->] (17) -- (18) node[midway, above] {};
\draw[->] (18) -- (19) node[midway, above] {};
\draw[->] (19) -- (20) node[midway, above] {};
\draw[->] (21) -- (22) node[midway, above] {};
\draw[->] (22) -- (23) node[midway, above] {};
\draw[->] (23) -- (24) node[midway, above] {};
\draw[->] (24) -- (25) node[midway, above] {};
\draw[->] (1) -- (6) node[midway,   left] {$g_1$};
\draw[->] (6) -- (11) node[midway,  left] {$g_2$};
\draw[->] (11) -- (16) node[midway, left] {$\vdots$};
\draw[->] (16) -- (21) node[midway, left] {$g_m$};
\draw[->] (2) -- (7) node[midway, above] {};
\draw[->] (7) -- (12) node[midway, above] {};
\draw[->] (12) -- (17) node[midway, above] {};
\draw[->] (17) -- (22) node[midway, above] {};
\draw[->] (3) -- (8) node[midway, above] {};
\draw[->] (8) -- (13) node[midway, above] {};
\draw[->] (13) -- (18) node[midway, above] {};
\draw[->] (18) -- (23) node[midway, above] {};
\draw[->] (4) -- (9) node[midway, above] {};
\draw[->] (9) -- (14) node[midway, above] {};
\draw[->] (14) -- (19) node[midway, above] {};
\draw[->] (19) -- (24) node[midway, above] {};
\draw[->] (5) -- (10) node[midway, above] {};
\draw[->] (10) -- (15) node[midway, above] {};
\draw[->] (15) -- (20) node[midway, above] {};
\draw[->] (20) -- (25) node[midway, above] {};
\end{scope}
\end{tikzpicture}},
\end{align}
where we calculate the product of all $\omega_{m+n}$'s taking $m+n$ values from all the possible paths going from the bottom left to the top right of the grid. The term $\omega_{m+n} (a_1,a_2,..., a_n,g_1, g_2,..., g_m),$ corresponds to the path going all the way to the right, followed by going all the way to the top. Paths that shift by one square will have opposite parity. For example, the term $\omega_{m+n} (a_1 ,a_2,..., a_{n-1},g_1, a_n,g_2,...,g_m)$ will go in the denominator. Geometrically, that is because two terms whose path differ by a square share a codimension-1 face. Thus, they must have opposite parity.

For a general chain made up of multiple simplices, we use the following properties:
\begin{align}
i_{cc'}\omega &= i_c\omega i_{c'}\omega,\\
i_{c^{-1}}\omega &= i_c\omega^{-1},
\end{align}
where $c$ and $c'$ are chains of the same dimension, and $c^{-1}$ is the chain with opposite orientation.

In general, the slant product does not necessarily commute with $d$. The commutation relation (in the multiplicative sense) is given by \cite{Spanier1994}
\begin{align}
\label{equ:commutator}
\frac{di_{(g_1,...,g_m)}\omega_{n+m}}{i_{(g_1,...,g_m)}d\omega_{n+m}}=   i_{\partial(g_1, ... ,g_m)}\omega_{n+m}^{(-1)^{n+m}}.
\end{align}
We can see that they only commute when $m=1$ because $i_{\partial(g)}$ is null.\\
In the remaining discussion, let us now restrict ourselves to the case where $m=2$:
\begin{equation}
di_{(g,h)}\omega_{n+2}
      = i_{(g,h)}d\omega_{n+2} \cdot \left ( \frac{i_{g}\omega_{n+2} i_{h}\omega_{n+2}}{i_{gh}\omega_{n+2}} \right ) ^{(-1)^{n}}.
\end{equation}
Now, let us consider the case where the dimensional reduction by one dimension is always a coboundary. That is, for any $g$, there is always some cochain $\mu_g$ such that $d\mu_g = i_{g} \omega_{n+2}$. If we insert this relation into the above equation, we find that upon defining the cochain
 \begin{equation}
 \label{equ:cocyclegh}
\alpha_{(g,h)}=i_{(g,h)}\omega_{n+2} \left ( \frac{\mu_{gh}}{\mu_{g} \mu_{h}} \right ) ^{(-1)^{n}},
\end{equation}
then $d\alpha_{(g,h)} = i_{(g,h)}d\omega_{n+2}$. Thus, if $\omega_{n+2}$ is an $(n+2)$-cocycle, then $\alpha_{(g,h)}$ is an $n$-cocycle!

Let us discuss the geometric interpretation of the general slant product. The slant product with a 1-chain $(g)$ corresponds to bending the $1$-chain into a circle with holonomy $g$ as in Figure \ref{fig:dimred}. In Figure \ref{fig:3hole}, we can take the 2-chain $(g,h)$ and fold the vertices over to the center. One can then deform it into a configuration of a junction of three tubes with holonomies $g$, $h$, and $gh$ respectively. The surface given has a boundary, which explains why $i_{g,h}\omega_{n+2}$ is not a cocycle, and requires corrections from $\mu_g$, $\mu_h$ and $\mu_{gh}$. On each leg, the compactification gives an SPT phase living in one dimension lower described by the cocycles $i_g \omega$, $i_h \omega$, and $1/i_{gh}\omega$ \footnote{The last cocycle is not $i_{gh}\omega$ because the holonomy is in the opposite direction}. The SPT phase described by the cocycle in equation \eqref{equ:cocyclegh} lives at the junction of these three systems and only exists when they are all trivial phases. Note that the cochain defined in equation \eqref{equ:cocyclegh} does not define a map from $H^{n+2}(G,U(1))$ to $H^{n}(G,U(1))$. This is because the choice of the cochain $\mu_g$ that satisfies $\mu_g = di_g\omega_{n+2}$ is not unique. Accordingly, $\alpha_{(g,h)}$ is not necessarily a coboundary even if $\omega_{n-2}$ is a coboundary.

In general, we can see that the dimensional reduction using $i_{(g_1,...,g_m)}$ is possible for any $m>2$ provided that $i_{(g_1,...,g_{m-1})}\omega$ is always a coboundary. The general slant product will come in handy in section \ref{invariantsbosons} when we distinguish the bosonic SPT phases in 4+1D classified by $\mathcal H^5(G,U(1))$.

\section{Dimensional Reduction for Gu-Wen Fermionic SPT Phases}\label{dimredfermions}
Now, we would like to repeat the dimensional reduction procedure for fermionic SPT phases in the Gu-Wen model. That is, given a cochain-cocycle pair $(\omega_{n+1},\beta_{n})$ describing a certain SPT phase in $n+1$ dimensions, what is the expression of the pair in $n$ dimensions after compactifying over a circle? For the bosonic case, we saw in the previous section that we could simply apply the slant product because of equation \eqref{equ:dimredbosons}. However, the procedure is now more involved since there are three terms in the Gu-Wen partition function \eqref{equ:GuWenpartitionfunction}. Furthermore, although the partition function is invariant under retriangulations, each of the three terms individually are not. Thus, we must derive the dimensional reduction formula.\\
First, let us outline the procedure. We want to triangulate the manifold $\mathcal{M}_{n} \times S^1$, with a group element $g$ on $S^1$. Since the bosonic integral  $\int_{\mathcal{M}_n \times S^1} \omega$ now depends on the triangulation, we will pick the triangulation we used in the bosonic case, i.e. the triangulation such that equation \eqref{equ:dimredbosons} holds.\\
For the Grassmann integral $\sigma_{n+1}(\beta_n)$, we integrate out the variables on faces that are not along $S^1$. We should then be able to arrange it into the Grassmann integral in the lower dimension $\sigma_{n}(i_g \beta)$ times some leftover sign factor. That is,
\begin{align}
\sigma_{n+1}(\beta_n) = (-1)^{\int_\mathcal{M} \Delta \sigma }\sigma_{n}(i_g \beta_n)
\end{align}
for some cochain $\Delta\sigma$.\\
For the spin structure term, we want to rewrite the integral as 
\begin{align}
\int_{S_{n+1}} m = \int_{S_{n} \times (g)} m + \int_{\Delta S} m,
\end{align}
where $S_{n} \times (g)$ is the schematic notation for the Cartesian product of $S_{n}$ with the 1-chain $(g)$, and $\Delta S$ represents the leftover term. From the property of the slant product, the first term is equal to the spin structure term in the lower dimension
\begin{align}
\label{equ:lift}
\int_{S_{n} \times (g)} m = \int_{S_{n}} i_g m,
\end{align}
while the second term is equal to $\int_{\Delta E}\beta$, for some chain $\Delta E$ such that $\partial( \Delta E)  = \Delta S$. This choice is not unique and depends on the spin structure on $S^1$.\\
Now, we can combine the extra sign factors from the last two terms and call it a cochain $\gamma \in \mathcal C^{n} (G,\mathbb{Z}_2)$. Explicitly,
\begin{align}
\int_{\mathcal{M}}\gamma= \int_{\mathcal{M}} \Delta \sigma + \int_{\Delta E} \beta.
\end{align}
The dimensional reduction procedure (analogous to equation \eqref{equ:dimredbosons} for the bosonic case) can now be written as
\begin{align}
\mathcal Z &= \int_{\mathcal{M} \times S^1} \omega_{n+1} \cdot \sigma_{n+1}(\beta_{n}) \cdot (-1)^{\int_{E_{n+1}} \beta_{n}} \nonumber\\
&= \int_\mathcal{M}(-1)^{\gamma} i_g\omega_{n+1}  \cdot \sigma_{n}(i_g\beta_{n}) \cdot (-1)^{\int_{E_{n}} i_g\beta_{n}}.
\end{align}
This equation tells us that the compactification of the SPT phase described by $(\omega_{n+1},\beta_{n})$ gives us an SPT phase described by $((-1)^\gamma i_g\omega_{n+1},i_g\beta_n)$.\\
Before we proceed, it is still important to check that the reduced system can be described by the Gu-Wen model. That is, it obeys equations \eqref{equ:dbeta}, \eqref{equ:domega}, and \eqref{equ:obstruction} in $n-1$ dimensions:
\begin{align}
\label{equ:reduced1}
di_g\beta&=0,\\
\label{equ:reduced2}
d \left((-1)^{\gamma}i_g\omega \right) &= (-1)^{Sq^2 (i_g\beta)},\\
\label{equ:reduced3}
(-1)^{Sq^2(i_g\beta)} &\in \mathcal B^{n}(G,U(1)).
\end{align}
Equation \eqref{equ:reduced1} is always satisfied by commutativity of $d$ and $i_g$. Inserting the Gu-Wen equation \eqref{equ:domega} into equation \eqref{equ:reduced2} gives us that $\gamma$ must satisfy
\begin{align}
\label{equ:reduced4}
d\gamma  = Sq^2 (i_g\beta) - i_g Sq^2(\beta).
\end{align}
If this is obtained, then the condition \eqref{equ:reduced3} is automatically satisfied from the obstruction-free condition \eqref{equ:obstruction}. Thus, the only equation we need to check is equation \eqref{equ:reduced4}.

In the following subsections, we will show that the explicit expression for $\gamma$ in space-time dimensions up to 3 is
\begin{align}
\label{equ:gamma}
\gamma = Sq^1i_g \beta + i_g Sq^1 \beta + i_g \beta \cup_{n-1} \beta +\eta(S^1)\beta + \epsilon.
\end{align}
Here, $\eta(S^1)$ means $\int_{S^1} \eta$ and is 0 or 1 depending on whether the spin structure corresponds to a periodic or antiperiodic boundary condition respectively, and $\epsilon$ is a specific cochain that is non-zero for $n=3$ (3+1D to 2+1D reduction). In higher dimensions, we also expect $\epsilon$ to be non-zero. This stems from the fact that $i_g$ commutes with the Steenrod square only if the latter is zero or is represented by the cup product. We refer the reader to Appendix \ref{app:cup} for more details.\\
For $n<3$, we can check that the expression above satisfies equation \eqref{equ:reduced4}. $\eta(S^1)\beta$ is a cocycle regardless of $\eta(S^1)$ and the first two terms are also cocycles by the definition of the Steenrod square. Thus, using equation \eqref{equ:dcup} to expand the coboundary of cup-$i$ products,
\begin{equation}
\label{equ:checkgamma}
d \gamma = d(i_g\beta \cup_{n-1} \beta) +d\epsilon =i_g\beta \cup_{n-2} \beta + \beta \cup_{n-2} i_g\beta +d\epsilon.
\end{equation}
Thus, the following equation must be satisfied (mod 2):
\begin{equation}
\label{equ:checkepsilon}
i_g\beta \cup_{n-2} \beta + \beta \cup_{n-2} i_g\beta + Sq^2(i_g \beta) + i_g Sq^2 (\beta) + d\epsilon =0.
\end{equation}
For $n=1$, all terms are zero. For $n=2$, we need to check that $i_g\beta \cup \beta + \beta \cup i_g\beta = i_gSq^2(\beta)$, which is satisfied by the Leibniz rule of $i_g$ on the cup product \eqref{equ:icup}. For $n=3$, we need to check from the explicit expression of $\epsilon$, which will be obtained in the last subsection.
\subsection{1+1D to 0+1D}
\begin{figure}[ht!]
\centering
\includegraphics{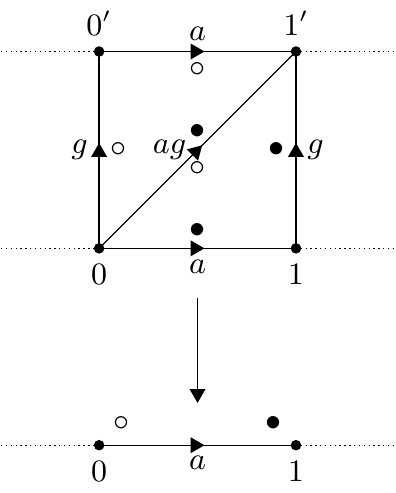}
\caption{Triangulation for 1+1D to 0+1D dimensional reduction. The Grassmann variables on links are reduced to those on vertices.}
\label{fig:1to0D}
\end{figure}

We choose the triangulation shown in Figure \ref{fig:1to0D} for our dimensional reduction so that the bosonic integral matches $i_g\omega$. For each square, the Grassmann integral is
\begin{align}
  \sigma(\beta)&=\int \prod [d\theta d\bar \theta] \ \ \  [\theta_{11'}^{\beta(g)}\theta_{01}^{\beta(a)}\bar\theta_{01'}^{\beta(ag)}] [\theta_{01'}^{\beta(ag)}\bar\theta_{00'}^{\beta(g)}\bar\theta_{0'1'}^{\beta(a)}] \nonumber\\
 &=(-1)^{\beta(a)\beta(g)+\beta(a)} \int \prod [d\theta d\bar \theta] \ \ \ \theta_{11'}^{\beta(g)}\bar\theta_{00'}^{\beta(g)} \nonumber\\
 &=(-1)^{i_g \beta \cup \beta(a) + \beta(a)}  \int \prod [d\theta d\bar \theta] \ \ \  \theta_{1}^{\beta(g)}\bar\theta_{0}^{\beta(g)}.
\end{align}
\begin{table}[h!]
\caption{Reducing $S^{1+1}$ to $S^{0+1}$. The terms in each row are reorganized depending on whether they are supported by one or two vertices in the base manifold. Summing over the columns show that $S^{0+1}$ is empty as desired.}
\centering
\begin{tabular}{|c|c|c|}
\hline
Terms & $S_1^{0+1} \times (g)$ & $S_2^{0+1} \times (g)$ \\
      & (Per vertex) & (Per link) \\
\hline
$S_1^{1+1}$ & 
\raisebox{-.5\height}{\begin{tikzpicture}[scale=0.35]
\node[vertex,blue][label=below:$0$]  (0) at (0,0) {};
\node[opacity=0] at (0,1) {};
\end{tikzpicture}}
& -\\
\hline
$S_2^{1+1}$ & 
\raisebox{-.5\height}{\begin{tikzpicture}[scale=0.35]
\node[vertex,blue][label=below:$0$]  (0) at (0,0) {};
\node[vertex][label=$0'$]  (1) at (0,4) {};
\draw[->,lblue] (0) -- (1) {};
\end{tikzpicture}}
&
\raisebox{-.5\height}{\begin{tikzpicture}[scale=0.35]
\node[vertex,blue][label=below:$0$]  (0) at (0,0) {};
\node[vertex][label=below:$1$]  (1) at (4,0) {};
\node[vertex][label=$0'$]  (2) at (0,4) {};
\node[vertex][label=$1'$]  (3) at (4,4) {};
\draw[->,lblue] (0) -- (1) {};
\draw[->] (0) -- (2) {};
\draw[->] (0) -- (3) {};
\draw[->] (1) -- (3) {};
\draw[->] (2) -- (3) {};
\end{tikzpicture}}
\raisebox{-.5\height}{\begin{tikzpicture}[scale=0.35]
\node[vertex,blue][label=below:$0$]  (0) at (0,0) {};
\node[vertex][label=below:$1$]  (1) at (4,0) {};
\node[vertex][label=$0'$]  (2) at (0,4) {};
\node[vertex][label=$1'$]  (3) at (4,4) {};
\draw[->] (0) -- (1) {};
\draw[->] (0) -- (2) {};
\draw[->,lblue] (0) -- (3) {};
\draw[->] (1) -- (3) {};
\draw[->] (2) -- (3) {};
\end{tikzpicture}}\\
\hline
$S_3^{2+1}$ & - &
\raisebox{-.5\height}{\begin{tikzpicture}[scale=0.35]
\fill[lblue] (0,0)-- (4,0)--(4,4)--cycle;
\node[vertex,blue][label=below:$0$]  (0) at (0,0) {};
\node[vertex][label=below:$1$]  (1) at (4,0) {};
\node[vertex][label=$0'$]  (2) at (0,4) {};
\node[vertex][label=$1'$]  (3) at (4,4) {};
\draw[->] (0) -- (1) {};
\draw[->] (0) -- (2) {};
\draw[->] (0) -- (3) {};
\draw[->] (1) -- (3) {};
\draw[->] (2) -- (3) {};
\end{tikzpicture}}
\raisebox{-.5\height}{\begin{tikzpicture}[scale=0.35]
\fill[lblue] (0,0)-- (0,4)--(4,4)--cycle;
\node[vertex,blue][label=below:$0$]  (0) at (0,0) {};
\node[vertex][label=below:$1$]  (1) at (4,0) {};
\node[vertex][label=$0'$]  (2) at (0,4) {};
\node[vertex][label=$1'$]  (3) at (4,4) {};
\draw[->] (0) -- (1) {};
\draw[->] (0) -- (2) {};
\draw[->] (0) -- (3) {};
\draw[->] (1) -- (3) {};
\draw[->] (2) -- (3) {};
\end{tikzpicture}} \\
\hline
\end{tabular}
\label{tab:S2toS1}
\end{table}
In simplifying the expression, we have used certain rules of moving the Grassmann variables, which are mentioned in Appendix \ref{app:grassmann}. In the second line, we integrated out $\bar\theta_{01'}^{\beta(ag)}\theta_{01'}^{\beta(ag)}$ and moved $\theta_{01}^{\beta(a)}$ to the back with sign $(-1)^{\beta(a)\beta(g)+\beta(a)}$ so that it can be integrated out. In the last line, the indices were renamed by removing the second number and the sign factor was rewritten in terms of the cup product. We now have the reduced form of the Grassmann integral with leftover sign factor
\begin{equation}
\Delta \sigma = i_g \beta \cup \beta + \beta.
\end{equation}
For the spin structure term, we need to show that the set $S^{0+1}$ is empty. This is illustrated in Table \ref{tab:S2toS1}. First, consider all vertices that will contribute to $S_1^{0+1}$. That is, those that are directly above a vertex $(0)$ in the base manifold. These are $(0)$ in $S_1^{1+1}$, and $(0)$ coming from $(00')$ in $S_2^{1+1}$, which cancel out. Next, consider the contributions to $S_2^{0+1}$. That is, those that are directly above a link $(01)$ in the base manifold. From $S_{2}^{1+1}$, we have $(0)$ from $(01)$ and $(0)$ from $(01')$, which cancels. From $S_{3}^{1+1}$, we have $(0)$ from $(011')$ and $(0)$ from $(00'1')$, which also cancels. Thus, $S^{0+1}$ is empty.\\
There are two choices of $E^{0+1}$ whose boundary gives $S^{0+1}$: the empty set, and the entire 0+1D manifold. The two choices of $E^{0+1}$ corresponds to the two spin structures we can assign on the circle we compactify. These two choices give a difference of $\beta$ evaluated on the manifold. Since $\eta(S^1)$ takes value 0 or 1, we can write both cases together as\footnote{The reason we know that the last term is $\eta(S^1)  \beta$ and not $(1+\eta(S^1))  \beta$ is because the antiperiodic boundary condition corresponds to $\eta(S^1)=1$ and upon stacking with itself, must give a phase with periodic boundary conditions, which is $\eta(S^1)=0$.}
\begin{align}
\label{equ:gamma1}
\gamma =i_g\beta \cup \beta+\eta(S^1)  \beta.
\end{align}
This matches equation \eqref{equ:gamma} since $Sq^1 i_g \beta = i_g Sq^1 \beta =0$.
\subsection{2+1D to 1+1D}
\begin{figure}[h!]
\centering
\includegraphics{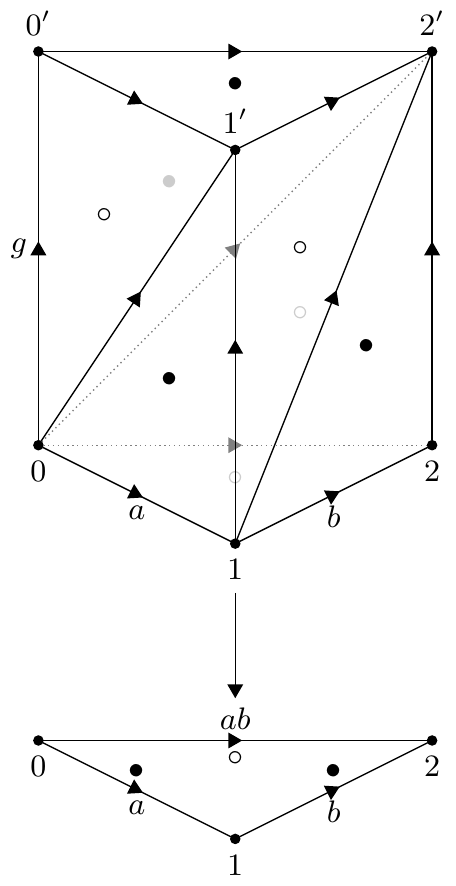}
\caption{Triangulation for 2+1D to 1+1D dimensional reduction. The Grassmann variables on triangles are reduced to those on links.}
\label{fig:2to1D}
\end{figure}

In this calculation, we will not write out the sign factors, but instead collect them along the way and put them all together in the end. We will also omit the measure of the Grassmann integral. The triangulation for the positive orientation is shown in Figure \ref{fig:2to1D}. The integrand is
\begin{widetext}
\begin{align*}
[\theta_{0'1'2'}^{\beta(a,b)}\theta_{00'2'}^{\beta(g,ab)}\bar\theta_{01'2'}^{\beta(ag,b)}\bar\theta_{00'1'}^{\beta(g,a)}][\theta_{011'}^{\beta(a,g)}\theta_{01'2'}^{\beta(ag,b)}\bar\theta_{012'}^{\beta(a,bg)}\bar\theta_{11'2'}^{\beta(g,b)}][\theta_{122'}^{\beta(b,g)}\theta_{012'}^{\beta(a,bg)}\bar\theta_{022'}^{\beta(ab,g)}\bar\theta_{012}^{\beta(a,b)}].
\end{align*}
 First, we swap the order of each group of $\theta$ and $\bar\theta$. One can check that this gives a factor of $(-1)^{i_g(\beta \cup_1 \beta)(a,b)}$ times
\begin{align*}
[\theta_{00'2'}^{\beta(g,ab)}\theta_{0'1'2'}^{\beta(a,b)}\bar\theta_{00'1'}^{\beta(g,a)}\bar\theta_{01'2'}^{\beta(ag,b)}][\theta_{01'2'}^{\beta(ag,b)}\theta_{011'}^{\beta(a,g)}\bar\theta_{11'2'}^{\beta(g,b)}\bar\theta_{012'}^{\beta(a,bg)}][\theta_{012'}^{\beta(a,bg)}\theta_{122'}^{\beta(b,g)}\bar\theta_{012}^{\beta(a,b)}\bar\theta_{022'}^{\beta(ab,g)}].
\end{align*}
\end{widetext}

Now, the variables on the faces inside each prism (i.e., $(012')$ and $(01'2')$) are paired and we can integrate them out for free. Next, we need to pair up variables on the caps of the prism, $\theta_{012}^{\beta(a,b)}$ and $\bar\theta_{0'1'2'}^{\beta(a,b)}$, by cycling one of them to the other and integrating them out. This gives a factor of $(-1)^{\beta(a,b)(\beta(g,ab)+\beta(ab,g))+\beta(a,b)}=(-1)^{i_g \beta \cup_1 \beta (a,b)+\beta(a,b)} $. The remaining integrand is
\begin{align*}
[\theta_{00'2'}^{\beta(g,ab)}\bar\theta_{00'1'}^{\beta(g,a)}\theta_{011'}^{\beta(a,g)}\bar\theta_{11'2'}^{\beta(g,b)}\theta_{122'}^{\beta(b,g)}\bar\theta_{022'}^{\beta(ab,g)}]
\end{align*}
Now, we move $\theta_{00'2'}^{\beta(g,ab)}$ to the back, which gives a factor of $(-1)^{\beta(g,ab)}$, then swap the positions of $\bar\theta_{00'1'}^{\beta(g,a)}\theta_{011'}^{\beta(a,g)}$ and $\bar\theta_{11'2'}^{\beta(g,b)}\theta_{122'}^{\beta(b,g)}$, which gives a factor of $(-1)^{(\beta(g,a)+\beta(a,g))(\beta(g,b)+\beta(b,g))} = (-1)^{i_g\beta \cup i_g \beta (a,b)}$. We are now left with
\begin{align*}
[(\bar\theta_{11'2'}^{\beta(g,b)}\theta_{122'}^{\beta(b,g)})(\bar\theta_{00'1'}^{\beta(g,a)}\theta_{011'}^{\beta(a,g)})(\bar\theta_{022'}^{\beta(ab,g)}\theta_{00'2'}^{\beta(g,ab)})].
\end{align*}
We have now grouped the remaining variables according to the sides of the triangle $(012)$ in the base manifold. The leftover sign so far is
\begin{equation*}
(-1)^{i_g(\beta \cup_1 \beta)(a,b)+i_g\beta \cup_1 \beta (a,b)+i_g\beta \cup i_g \beta (a,b) +  \beta(a,b)+\beta(g,ab)}.
\end{equation*}
Let us separate the last term $\beta(g,ab)$ and write it as $\int_{E_{G3}} \beta $, where $E_{G3}$ is $(00'2')$ of each prism.\\
Next, consider the last pair in the reduced integrand: $\bar\theta_{022'}^{\beta(ab,g)}\theta_{00'2'}^{\beta(g,ab)}$, which is above the link $(02)$. We will show that it can be integrated to $\bar\theta_{02}^{i_g\beta(ab)}$ with a leftover sign factor $(-1)^{\beta(g,ab)}$.
\begin{enumerate}
\itemsep0em 
\item If $\beta(ab,g)=\beta(g,ab)$, then the pair is Grassmann even. Likewise, in the adjacent prism, there will be a term $\bar\theta_{00'2'}^{\beta(g,ab)}\theta_{022'}^{\beta(ab,g)}$. We can move both pairs out and integrate them against $d\theta_{022'}^{\beta(ab,g)}d \bar\theta_{022'}^{\beta(ab,g)}d\theta_{00'2'}^{\beta(g,ab)}d \bar\theta_{00'2'}^{\beta(g,ab)}$ to get $(-1)^{\beta(g,ab)}$. Since $i_g\beta(ab) =0$ modulo 2, we can place $d\theta_{02}^{i_g\beta(ab)}$, $d\bar\theta_{02}^{i_g\beta(ab)}$, $\theta_{02}^{i_g\beta(ab)}$, and $\bar\theta_{02}^{i_g\beta(ab)}$ back into the integral at their respective positions.
\item If $\beta(g,ab)=1$ and $\beta(ab,g)=0$, we can redefine $\theta_{00'2'}^{\beta(g,ab)} \equiv \theta_{02}^{\beta(g,ab)}=\theta_{12}^{i_g\beta(ab)}$ and swap positions with $\bar \theta_{02}^{i_g\beta(b)}$ in the adjacent prism. Since they go through all the same variables plus pass each other, the sign gained is just $(-1)^{\beta(g,ab)}$.
\item If $\beta(ab,g)=1$ and $\beta(g,ab)=0$, we can redefine $\bar\theta_{022'}^{\beta(ab,g)} \equiv \bar\theta_{02}^{\beta(ab,g)}=\bar\theta_{02}^{i_g\beta(b)}$. We can freely multiply by $(-1)^{\beta(g,ab)}$, which is unity.
\end{enumerate}

If we repeat this procedure for all three sides of the triangle $(012)$, we can define $E_{G2}$ as the set of $(00'1')$ from all the vertical rectangles above each link in the base manifold. Doing so, we can see that the sign factor from all the relabeling above is $\int_{E_{G2}} \beta$. The Grassmann integral now has the desired reduced form
\begin{align*}
\int \prod(d\theta d\bar\theta)  \ \ \ [\theta_{12}^{i_g\beta(b)}\theta_{01}^{i_g\beta(a)}\bar\theta_{02}^{i_g\beta(ab)}].
\end{align*}

We will combine $\int_{E_{G2}+E_{G3}}\beta$ to the spin structure term. Since we will be calculating with $m$, let us define $S_{G2}$ and $S_{G3}$ as the boundary of those two sets respectively. We then have $S_{G2}=(00')+(01)+(01')$ from all rectangles and $S_{G3}=(00')+(02')+(02)$ from all prisms. Now, let us combine this with the set $S^{2+1}$ and show that it can be reduced to $S^{1+1}$. To do this, we need to reorganize $S^{2+1}$ into three parts depending on whether they are supported by one, two or three vertices in the base manifold. This is separated into three columns in Table \ref{tab:S3toS2}.
\begin{enumerate}
\itemsep0em 
\item $S_1^{1+1} \times (g)$ come from links that depend on one vertex in the base manifold. These are the vertical links from $S_1^{2+1}$, which can be easily reduced to $S_1^{1+1}$ using the slant product on $m$.\\
\item $S_2^{1+1} \times (g)$ comes from links that depend on two vertices in the base manifold. Let us look at the link $(01)$ as an example. Above it, we have the remaining links from $S_1^{2+1}$ that are not vertical: $(01)$, and $(01')$. From $S_3^{2+1}$, we have two $(01')$ coming from the triangles $(011')$ and $(00'1')$. Finally, the set $S_G^2$, gives us $(00')$, $(01')$, and $(0'1')$. Hence in total, We are left with only $(00')$ defined per link $(01)$ in 1+1D, which is precisely reduced to $S_2^{1+1}$.\\
\item $S_3^{1+1} \times (g)$ comes from links that depend on three vertices in the base manifold. From $S_2^{2+1}$, there is $(02)$ coming from $(012)$ and two $(02')$ coming from $(012')$ and $(01'2')$. From $S_3^{2+1}$, there are three $(02')$ coming from the three tetrahedra that triangulate the prism. Lastly, from $S_{G3}$, we have $(00')$, $(0'2')$, and $(02')$. Adding everything together, the only term remaining is $(00')$ per triangle, which can be reduced to $S_3^{1+1}$.
\end{enumerate}
Thus, we have shown that special terms $S_G$ from the Grassmann integral combined with $S^{2+1}$ can be reduced to $S^{1+1}$.

\begin{table*}
\caption{Reducing $S^{2+1}$ and $S_G$ to $S^{1+1}$. The terms in each row are reorganized depending on whether they depend on one, two, or three vertices in the base manifold. Summing over the columns gives the desired set $(00')$. Since $m$ evaluated on this set is equal to $i_gm$ evaluated on $(0)$, we can reduce the set to $S^{1+1}$, which matches equation \eqref{equ:S1_1Dpic}.}
\begin{tabular}{|c|c|c|c|}
\hline
Terms & $S_1^{1+1} \times (g)$ & $S_2^{1+1} \times (g)$& $S_3^{1+1} \times (g)$ \\
      & (Per vertex) & (Per link) & (Per triangle) \\
\hline
$S_1^{2+1}$ & 
\raisebox{-.5\height}{\begin{tikzpicture}[scale=0.35]
\node[vertex][label=below:$0$]  (0) at (0,0) {};
\node[vertex][label=$0'$]  (1) at (0,4) {};
\draw[->,blue] (0) -- (1) {};
\end{tikzpicture}}
&
\raisebox{-.5\height}{\begin{tikzpicture}[scale=0.35]
\node[vertex][label=below:$0$]  (0) at (0,0) {};
\node[vertex][label=below:$1$]  (1) at (4,0) {};
\node[vertex][label=$0'$]  (2) at (0,4) {};
\node[vertex][label=$1'$]  (3) at (4,4) {};
\draw[->,blue] (0) -- (1) {};
\draw[->] (0) -- (2) {};
\draw[->] (0) -- (3) {};
\draw[->] (1) -- (3) {};
\draw[->] (2) -- (3) {};
\end{tikzpicture}}
\raisebox{-.5\height}{\begin{tikzpicture}[scale=0.35]
\node[vertex][label=below:$0$]  (0) at (0,0) {};
\node[vertex][label=below:$1$]  (1) at (4,0) {};
\node[vertex][label=$0'$]  (2) at (0,4) {};
\node[vertex][label=$1'$]  (3) at (4,4) {};
\draw[->] (0) -- (1) {};
\draw[->] (0) -- (2) {};
\draw[->,blue] (0) -- (3) {};
\draw[->] (1) -- (3) {};
\draw[->] (2) -- (3) {};
\end{tikzpicture}}

 & - \\
\hline
$S_2^{2+1}$ & - 				&
\raisebox{-.5\height}{\begin{tikzpicture}[scale=0.35]
\fill[blue,opacity=0.5] (0,0)-- (4,0)--(4,4)--cycle;
\node[vertex][label=below:$0$]  (0) at (0,0) {};
\node[vertex][label=below:$1$]  (1) at (4,0) {};
\node[vertex][label=$0'$]  (2) at (0,4) {};
\node[vertex][label=$1'$]  (3) at (4,4) {};
\draw[->] (0) -- (1) {};
\draw[->] (0) -- (2) {};
\draw[->,blue] (0) -- (3) {};
\draw[->] (1) -- (3) {};
\draw[->] (2) -- (3) {};
\end{tikzpicture}}
\raisebox{-.5\height}{\begin{tikzpicture}[scale=0.35]
\fill[blue,opacity=0.5] (0,0)-- (0,4)--(4,4)--cycle;
\node[vertex][label=below:$0$]  (0) at (0,0) {};
\node[vertex][label=below:$1$]  (1) at (4,0) {};
\node[vertex][label=$0'$]  (2) at (0,4) {};
\node[vertex][label=$1'$]  (3) at (4,4) {};
\draw[->] (0) -- (1) {};
\draw[->] (0) -- (2) {};
\draw[->,blue] (0) -- (3) {};
\draw[->] (1) -- (3) {};
\draw[->] (2) -- (3) {};
\end{tikzpicture}}
&
\raisebox{-.5\height}{\begin{tikzpicture}[scale=0.35]
\fill[blue,opacity=0.5] (0,0)-- (2,-1)--(4,0)--cycle;
\node[vertex][label=below:$0$]  (0) at (0,0) {};
\node[vertex][label=below:$1$]  (1) at (2,-1) {};
\node[vertex][label=below:$2$]  (2) at (4,0) {};
\node[vertex][label=$0'$]  (3) at (0,4) {};
\node[vertex][label=$1'$]  (4) at (2,3) {};
\node[vertex][label=$2'$]  (5) at (4,4) {};
\draw[->,blue] (0) -- (2) node[midway, above] {};
\draw[->,densely dotted,color=gray] (0) -- (5)  {};
\draw[->] (0) -- (1) node[midway, below] {};
\draw[->] (1) -- (2) node[midway, below] {};
\draw[->] (0) -- (3) node[midway, left] {};
\draw[->] (1) -- (4) node[midway, above right] {};
\draw[->] (2) -- (5) node[midway, right] {};
\draw[->] (3) -- (4) node[midway, below] {};
\draw[->] (4) -- (5) node[midway, below] {};
\draw[->] (3) -- (5) node[midway, above left] {};
\draw[->] (0) -- (4)  {};
\draw[->] (1) -- (5)  {};
\end{tikzpicture}}
\raisebox{-.5\height}{\begin{tikzpicture}[scale=0.35]
\fill[blue,opacity=0.5] (0,0)-- (2,-1)--(4,4)--cycle;
\node[vertex][label=below:$0$]  (0) at (0,0) {};
\node[vertex][label=below:$1$]  (1) at (2,-1) {};
\node[vertex][label=below:$2$]  (2) at (4,0) {};
\node[vertex][label=$0'$]  (3) at (0,4) {};
\node[vertex][label=$1'$]  (4) at (2,3) {};
\node[vertex][label=$2'$]  (5) at (4,4) {};
\draw[->,densely dotted,color=gray] (0) -- (2) node[midway, above] {};
\draw[->,densely dotted,blue] (0) -- (5)  {};
\draw[->] (0) -- (1) node[midway, below] {};
\draw[->] (1) -- (2) node[midway, below] {};
\draw[->] (0) -- (3) node[midway, left] {};
\draw[->] (1) -- (4) node[midway, above right] {};
\draw[->] (2) -- (5) node[midway, right] {};
\draw[->] (3) -- (4) node[midway, below] {};
\draw[->] (4) -- (5) node[midway, below] {};
\draw[->] (3) -- (5) node[midway, above left] {};
\draw[->] (0) -- (4)  {};
\draw[->] (1) -- (5)  {};
\end{tikzpicture}}
\raisebox{-.5\height}{\begin{tikzpicture}[scale=0.35]
\fill[blue,opacity=0.5] (0,0)-- (2,3)--(4,4)--cycle;
\node[vertex][label=below:$0$]  (0) at (0,0) {};
\node[vertex][label=below:$1$]  (1) at (2,-1) {};
\node[vertex][label=below:$2$]  (2) at (4,0) {};
\node[vertex][label=$0'$]  (3) at (0,4) {};
\node[vertex][label=$1'$]  (4) at (2,3) {};
\node[vertex][label=$2'$]  (5) at (4,4) {};
\draw[->,densely dotted,color=gray] (0) -- (2) node[midway, above] {};
\draw[->,densely dotted,blue] (0) -- (5)  {};
\draw[->] (0) -- (1) node[midway, below] {};
\draw[->] (1) -- (2) node[midway, below] {};
\draw[->] (0) -- (3) node[midway, left] {};
\draw[->] (1) -- (4) node[midway, above right] {};
\draw[->] (2) -- (5) node[midway, right] {};
\draw[->] (3) -- (4) node[midway, below] {};
\draw[->] (4) -- (5) node[midway, below] {};
\draw[->] (3) -- (5) node[midway, above left] {};
\draw[->] (0) -- (4)  {};
\draw[->] (1) -- (5)  {};
\end{tikzpicture}}

\\

\hline
$S_3^{2+1}$ & - & -&
\raisebox{-.5\height}{\begin{tikzpicture}[scale=0.35]
\fill[blue,opacity=0.5] (0,0)-- (2,-1)--(4,4)--cycle;
\fill[blue,opacity=0.7] (2,-1)--(4,0)--(4,4)--cycle;
\node[vertex][label=below:$0$]  (0) at (0,0) {};
\node[vertex][label=below:$1$]  (1) at (2,-1) {};
\node[vertex][label=below:$2$]  (2) at (4,0) {};
\node[vertex][label=$0'$]  (3) at (0,4) {};
\node[vertex][label=$1'$]  (4) at (2,3) {};
\node[vertex][label=$2'$]  (5) at (4,4) {};
\draw[->,densely dotted,color=gray] (0) -- (2) node[midway, above] {};
\draw[->,densely dotted,blue] (0) -- (5)  {};
\draw[->] (0) -- (1) node[midway, below] {};
\draw[->] (1) -- (2) node[midway, below] {};
\draw[->] (0) -- (3) node[midway, left] {};
\draw[->] (1) -- (4) node[midway, above right] {};
\draw[->] (2) -- (5) node[midway, right] {};
\draw[->] (3) -- (4) node[midway, below] {};
\draw[->] (4) -- (5) node[midway, below] {};
\draw[->] (3) -- (5) node[midway, above left] {};
\draw[->] (0) -- (4)  {};
\draw[->] (1) -- (5)  {};
\end{tikzpicture}}
\raisebox{-.5\height}{\begin{tikzpicture}[scale=0.35]
\fill[blue,opacity=0.5] (0,0)-- (2,-1)--(2,3)--cycle;
\fill[blue,opacity=0.7] (2,-1)--(2,3)--(4,4)--cycle;
\node[vertex][label=below:$0$]  (0) at (0,0) {};
\node[vertex][label=below:$1$]  (1) at (2,-1) {};
\node[vertex][label=below:$2$]  (2) at (4,0) {};
\node[vertex][label=$0'$]  (3) at (0,4) {};
\node[vertex][label=$1'$]  (4) at (2,3) {};
\node[vertex][label=$2'$]  (5) at (4,4) {};
\draw[->,densely dotted,color=gray] (0) -- (2) node[midway, above] {};
\draw[->,densely dotted,blue] (0) -- (5)  {};
\draw[->] (0) -- (1) node[midway, below] {};
\draw[->] (1) -- (2) node[midway, below] {};
\draw[->] (0) -- (3) node[midway, left] {};
\draw[->] (1) -- (4) node[midway, above right] {};
\draw[->] (2) -- (5) node[midway, right] {};
\draw[->] (3) -- (4) node[midway, below] {};
\draw[->] (4) -- (5) node[midway, below] {};
\draw[->] (3) -- (5) node[midway, above left] {};
\draw[->] (0) -- (4)  {};
\draw[->] (1) -- (5)  {};
\end{tikzpicture}}
\raisebox{-.5\height}{\begin{tikzpicture}[scale=0.35]
\fill[blue,opacity=0.5] (0,0)-- (0,4)--(2,3)--cycle;
\fill[blue,opacity=0.7] (0,4)--(2,3)--(4,4)--cycle;
\fill[blue,opacity=0.8] (0,0)--(2,3)--(4,4)--cycle;
\node[vertex][label=below:$0$]  (0) at (0,0) {};
\node[vertex][label=below:$1$]  (1) at (2,-1) {};
\node[vertex][label=below:$2$]  (2) at (4,0) {};
\node[vertex][label=$0'$]  (3) at (0,4) {};
\node[vertex][label=$1'$]  (4) at (2,3) {};
\node[vertex][label=$2'$]  (5) at (4,4) {};
\draw[->,densely dotted,color=gray] (0) -- (2) node[midway, above] {};
\draw[->,densely dotted,blue] (0) -- (5)  {};
\draw[->] (0) -- (1) node[midway, below] {};
\draw[->] (1) -- (2) node[midway, below] {};
\draw[->] (0) -- (3) node[midway, left] {};
\draw[->] (1) -- (4) node[midway, above right] {};
\draw[->] (2) -- (5) node[midway, right] {};
\draw[->] (3) -- (4) node[midway, below] {};
\draw[->] (4) -- (5) node[midway, below] {};
\draw[->] (3) -- (5) node[midway, above left] {};
\draw[->] (0) -- (4)  {};
\draw[->] (1) -- (5)  {};
\end{tikzpicture}}

\\
\hline
$S_G$& -&
\raisebox{-.5\height}{\begin{tikzpicture}[scale=0.35]
\node[vertex][label=below:$0$]  (0) at (0,0) {};
\node[vertex][label=below:$1$]  (1) at (4,0) {};
\node[vertex][label=$0'$]  (2) at (0,4) {};
\node[vertex][label=$1'$]  (3) at (4,4) {};
\draw[->] (0) -- (1) {};
\draw[->,blue] (0) -- (2) {};
\draw[->,blue] (0) -- (3) {};
\draw[->] (1) -- (3) {};
\draw[->,blue] (2) -- (3) {};
\end{tikzpicture}}

&
\raisebox{-.5\height}{\begin{tikzpicture}[scale=0.35]
\node[vertex][label=below:$0$]  (0) at (0,0) {};
\node[vertex][label=below:$1$]  (1) at (2,-1) {};
\node[vertex][label=below:$2$]  (2) at (4,0) {};
\node[vertex][label=$0'$]  (3) at (0,4) {};
\node[vertex][label=$1'$]  (4) at (2,3) {};
\node[vertex][label=$2'$]  (5) at (4,4) {};
\draw[->,densely dotted,color=gray] (0) -- (2) node[midway, above] {};
\draw[->,densely dotted,blue] (0) -- (5)  {};
\draw[->] (0) -- (1) node[midway, below] {};
\draw[->] (1) -- (2) node[midway, below] {};
\draw[->,blue] (0) -- (3) node[midway, left] {};
\draw[->] (1) -- (4) node[midway, above right] {};
\draw[->] (2) -- (5) node[midway, right] {};
\draw[->] (3) -- (4) node[midway, below] {};
\draw[->] (4) -- (5) node[midway, below] {};
\draw[->,blue] (3) -- (5) node[midway, above left] {};
\draw[->] (0) -- (4)  {};
\draw[->] (1) -- (5)  {};
\end{tikzpicture}}
\\
\hhline{====}
Total &
\raisebox{-.5\height}{\begin{tikzpicture}[scale=0.35]
\node[vertex][label=below:$0$]  (0) at (0,0) {};
\node[vertex][label=$0'$]  (1) at (0,4) {};
\draw[->,blue] (0) -- (1) {};
\end{tikzpicture}}

 & 
\raisebox{-.5\height}{\begin{tikzpicture}[scale=0.35]
\node[vertex][label=below:$0$]  (0) at (0,0) {};
\node[vertex][label=below:$1$]  (1) at (4,0) {};
\node[vertex][label=$0'$]  (2) at (0,4) {};
\node[vertex][label=$1'$]  (3) at (4,4) {};
\draw[->] (0) -- (1) {};
\draw[->,blue] (0) -- (2) {};
\draw[->] (0) -- (3) {};
\draw[->] (1) -- (3) {};
\draw[->] (2) -- (3) {};
\end{tikzpicture}}

  &
\raisebox{-.5\height}{\begin{tikzpicture}[scale=0.35]
\node[vertex][label=below:$0$]  (0) at (0,0) {};
\node[vertex][label=below:$1$]  (1) at (2,-1) {};
\node[vertex][label=below:$2$]  (2) at (4,0) {};
\node[vertex][label=$0'$]  (3) at (0,4) {};
\node[vertex][label=$1'$]  (4) at (2,3) {};
\node[vertex][label=$2'$]  (5) at (4,4) {};
\draw[->,densely dotted,color=gray] (0) -- (2) node[midway, above] {};
\draw[->,densely dotted,color=gray] (0) -- (5)  {};
\draw[->] (0) -- (1) node[midway, below] {};
\draw[->] (1) -- (2) node[midway, below] {};
\draw[->,blue] (0) -- (3) node[midway, left] {};
\draw[->] (1) -- (4) node[midway, above right] {};
\draw[->] (2) -- (5) node[midway, right] {};
\draw[->] (3) -- (4) node[midway, below] {};
\draw[->] (4) -- (5) node[midway, below] {};
\draw[->] (3) -- (5) node[midway, above left] {};
\draw[->] (0) -- (4)  {};
\draw[->] (1) -- (5)  {};
\end{tikzpicture}}
  \\
\hline
Reduced to &
\raisebox{-.5\height}{\begin{tikzpicture}[scale=0.35]
\node[vertex,blue][label=below:$0$]  (0) at (0,0) {};
\end{tikzpicture}}
 &
\raisebox{-.5\height}{\begin{tikzpicture}[scale=0.35]
\node[vertex,blue][label=below:$0$]  (0) at (0,0) {};
\node[vertex][label=below:$1$]  (1) at (4,0) {};
\draw[->] (0) -- (1) node[midway, below] {};
\end{tikzpicture}}
  &
\raisebox{-.5\height}{\begin{tikzpicture}[scale=0.35]
\node[vertex,blue][label=below:$0$]  (0) at (0,0) {};
\node[vertex][label=below:$1$]  (1) at (2,-1) {};
\node[vertex][label=below:$2$]  (2) at (4,0) {};
\draw[->] (0) -- (1) node[midway, below] {};
\draw[->] (1) -- (2) node[midway, below] {};
\draw[->] (0) -- (2) node[midway, below] {};

\end{tikzpicture}}
  \\
\hline
\end{tabular}
\label{tab:S3toS2}
\end{table*}
Like the lower dimensional case, there are two choices of $E$ that gives rise to a difference of $\beta$. They correspond to the two possible spin structures on $S^1$. Putting everything together, we have
\begin{align}
\label{equ:gamma2}
\gamma=i_g(\beta \cup_1 \beta)+i_g \beta \cup i_g \beta+i_g\beta \cup_1 \beta + \eta(S^1)\beta.
\end{align}
The first two terms are respectively $i_g Sq^1 \beta$ and $Sq^1 i_g \beta$, so the expression matches equation \eqref{equ:gamma}.
\subsection{3+1D to 2+1D}
The calculations are almost identical to the previous section. However, it turns out that we must do dimensional reduction in a orientation dependent manner. The calculations are very involved so we have moved them to Appendix \ref{app:4to3calculation}.  Here, we will only state the final result. The resulting sign factor left from dimensional reduction is
\begin{align}
\gamma =  i_g(\beta \cup_2 \beta)+i_g\beta \cup_1 i_g \beta+i_g\beta \cup_2 \beta + \eta(S^1)\beta + \epsilon,
\end{align}
where $\epsilon$ is the cochain
\begin{align}
\epsilon(a,b,c)=& (\beta(g,a,bc)+\beta(a,g,bc))i_g \beta(b,c) \nonumber\\
&+\beta(g,ab,c)  i_g \beta(a,b) + d\mu(a,b,c);\\
\mu(a,b)=&\beta(g,a,b)i_g\beta(a,b).
\end{align}
This expression matches equation \eqref{equ:gamma}. What is left to check is equation \eqref{equ:checkepsilon}. This can be verified using various cocycle conditions of $\beta$. The procedure is outlined in Appendix \ref{app:4to3calculation}.

\section{Topological Invariants for Bosonic SPT Phases}\label{invariantsbosons}
We are now ready to tackle the following question: given a $n$-cocycle, which phase does it correspond to? That is, which equivalence class does it belong to in the cohomology group $\mathcal H^n(G,U(1))$? One way to answer this question is to compare the partition functions for all possible manifolds with flat connections. However, we will show that for finite abelian unitary symmetry group, we only need certain closed orientable manifolds equipped with a non-trivial flat connection. These are our topological invariants. We remark that if the flat connection is trivial, then the partition function contains no information, since it is always unity.\\
Let us elaborate on the word topological. In this sense, it means that the partition function is invariant under coboundary transformations of the cocycle. That is, the partition function is the same value for cocycles that describe the same phase. Indeed, this is why we must demand that our invariants come from closed manifolds, which have no boundary.\\
Recall that a flat connection defines a map from the fundamental group $\pi_1 (\mathcal{M})$ to $G$. That is, it tells us which group elements are assigned to the non-contractible loops of the manifold. Since $G$ is abelian, it is sufficient to consider manifolds with abelian fundamental group. For example, in 1+1D, a genus-$n$ torus can always be triangulated as $n$ genus-1 tori if $G$ is abelian. Thus, there is no extra information to extract from these manifolds.\\
In the following sections, we will demonstrate that the manifolds in 1+1D and 2+1D with abelian fundamental group are enough to distinguish all the cocycles. In higher dimensions, we will need to use the dimensional reduction techniques developed in section \ref{dimredbosons} in order to calculate the invariants.\\
As a reminder, we will use the notation $N^{ijk...} = \text{lcm}(N_i,N_j,N_k,...)$ and $N_{ijk...} = \text{gcd}(N_i,N_j,N_k,...)$ to denote the least common multiple and greatest common divisor respectively. The group is represented as $G=\prod_{i=1}^K \mathbb{Z}_{N_i}$ under addition, and the generator of each subgroup $\mathbb{Z}_{N_i}$ is denoted $e_i$.
\subsection{1+1D}
\begin{figure}[H]
\centering
\includegraphics{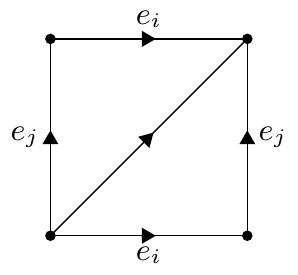}
\caption{Visualization of the partition function $Z^{ij}$, which is a triangulation of $T^2$. Opposite edges of the square are identified.}
\label{fig:torus}
\end{figure}
The only 2-manifold with the mentioned properties is $T^2$ (note that the partition function on $S^2$ is always one.) Let us define $Z^{ij}$ as the partition function on the torus with holonomies $e_i$ and $e_j$ around the two loops of the torus. The visualization is shown in Figure \ref{fig:torus}. Explicitly, the partition function is
\begin{align}
Z^{ij} = i_{e_i}i_{e_j}\omega_2 =\frac{\omega_2(e_i,e_j)}{\omega_2(e_j,e_i)}.
\end{align}
Next, we will consider the following ``canonical'' cocycles
\begin{align}
\label{equ:cocycles2}
\omega_2(a,b)= \exp 2\pi i \sum_{ij} \frac{P_{ij}}{N_{ij}} a_ib_j,
\end{align}
where $P_{ij}$ is an integer tensor. One can check that the expression is indeed a 2-cocycle. For fixed $i$ and $j$, the partition function is then
\begin{align}
Z^{ij} = \exp \frac{2\pi i}{N_{ij}} (P_{ij}-P_{ji}).
\end{align}

We will now proceed through a counting argument, as has been introduced in Ref. \onlinecite{WangLevin2015}. For a fixed $i<j$ , the partition function can take $N_{ij}$ values for all the different possible choices of the integer $P_{ij}$. Hence, for all $i<j$, the partition functions can take \underline{\smash{at least}} $\prod_{i<j}N_{ij}$ different values. On the other hand, we also know from the K\"unneth formula \cite{Spanier1994} that the second cohomology group is
\begin{equation}
\mathcal H^2(G,U(1)) \cong \prod_{i<j} \mathbb Z_{N_{ij}}.
\end{equation}
This implies that among all the SPT phases within the group cohomology classification, there are \underline{\smash{at most}} $\prod_{i<j}N_{ij}$ different values that the invariants can take. We have just shown that the number of different values must be equal to the number of SPT phases. As a consequence, these invariants are able to distinguish all the bosonic SPT phases described by the group cohomology model in 1+1D.

It is worth pointing out that if we instead place multiples of the generators on the torus, then the partition function will always have lower resolution. For example, if $N_i=2$ and $N_j=4$, then replacing $e_j$ with $2e_j$ will give a partition function of one for any cocycle. Thus, the invariants we define must have connections that send generators of the fundamental group to generators of $G$.
\subsection{2+1D}
It is known\cite{aschenbrenner} that the only closed orientable 3-manifolds with abelian fundamental groups are lens spaces $L(p;q)$, $T^3$, and $S^2 \times S^1$. However, $S^2$ does not admit a non-trivial flat connection, which means that any partition function defined on $S^2 \times S^1$ will give unity. As a result, lens spaces and $T^3$ are the only manifolds that we have to use.\\
To have a non-trivial flat connection, we want to put these group elements around the non-contractible loops of the manifolds. The non-contractible loop for the lens space is a link around the ``great circle'' (the perimeter of the base of the bipyramid) as shown in Figure \ref{fig:lens} (also, see Appendix \ref{app:lens} for a visualization and properties of lens spaces). For $T^3$, these are simply the three different circles which form the edges of the box in Figure \ref{fig:lens}. Thus, let us define the three following partition functions:
\begin{figure}[h!]
\centering
\includegraphics{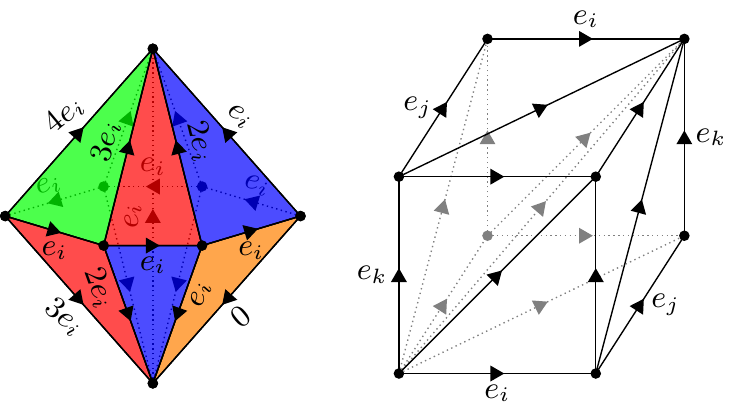}
\caption{Visualization of the partition functions $Z_i$ and $Z_{ijk}$, corresponding to the lens space $L(N_i;1)$ (drawn here for $N_i=6$) and $T^3$, respectively. The lens space is constructed by gluing faces with identical colors together, while $T^3$ is constructed by gluing opposite sides of the box.}
\label{fig:lens}
\end{figure}
\begin{enumerate}
\itemsep0em 
\item $Z_i$ is the partition function on $L(N_i;1)$ with $e_i$ around the great circle.
\item $Z_{ij}$ is the partition function on $L(N^{ij};1)$ with $e_i+e_j$ around the great circle.
\item $Z_{ijk}$ is the partition function on $T^3$ with $e_i$, $e_j$, and $e_k$ around three homotopically inequivalent loops.
\end{enumerate}
We will now show that these invariants are able to distinguish the different elements in
\begin{align}
\mathcal H^3(G,U(1)&) \cong \prod_{i} \mathbb{Z}_{N_i}\prod_{i<j} \mathbb{Z}_{N_{ij}}\prod_{i<j<k} \mathbb{Z}_{N_{ijk}}.
\end{align}
From the given triangulations, we can see that the partition functions are
\begin{subequations}
\begin{align}
\label{equ:Z_i}
Z_i &=\prod_{n=1}^{N_i} \omega_3(e_i,ne_i,e_i),\\
\label{equ:Z_ij}
Z_{ij} &= \prod_{n=1}^{N^{ij}} \omega_3(e_i+e_j,n(e_i+e_j),e_i+e_j),\\
\label{equ:Z_ijk}
Z_{ijk} &=i_{e_i}i_{e_j}i_{e_k}\omega_3 \nonumber\\
&=\frac{\omega_3(e_i,e_j,e_k)\omega_3(e_j,e_k,e_i)\omega_3(e_k,e_i,e_j)}{\omega_3(e_j,e_i,e_k)\omega_3(e_i,e_k,e_j)\omega_3(e_k,e_j,e_i)}.
\end{align}
\end{subequations}
Next, consider the following canonical cocycles:
\begin{align}
\label{equ:canonicalcocycles}
\omega_3(a,b,c)=& \exp 2\pi i  \sum_{ij} \frac{P_{ij}}{N_i N_j}a_i (b_j+c_j-[b_j+c_j]) \nonumber \\
 &\cdot \exp 2\pi i \sum_{ijk} \frac{Q_{ijk}}{N_{ijk}} a_ib_jc_k,
\end{align}
where $P_{ij}$ and $Q_{ijk}$ are integer tensors and $[b_j+c_j]$ means $b_j+c_j$ modulo $N_j$. Without loss of generality, let us assume for the counting argument that $Q_{ijk}=0$ if any of the two indices are identical. The value of the partition functions are then
\begin{subequations}
\begin{align}
Z_i &=\exp \frac{2\pi i}{N_i} P_{ii},\\
Z_{ij} &= \exp 2\pi i N^{ij} \left [ \frac{P_{ii}}{N_i^2} + \frac{P_{ij}+P_{ji}}{N_iN_j} +  \frac{P_{jj}}{N_j^2} \right],\\
Z_{ijk} &=\exp \frac{2\pi i}{N_{ijk}} \sum_{\hat p} \text{sgn} (\hat p) Q_{\hat p(i) \hat p(j) \hat p(k)},
\end{align}
\end{subequations}
where $\hat p$ is the permutation of the indices $i,j,k$ and the sign of $\hat p$ is $\pm 1$ depending on its parity. We can see that the partition functions $Z_i$ and $Z_{ijk}$ can take $N_i$ and $N_{ijk}$ different values for different choices of $P_{ii}$ and $\sum_{\hat p} \text{sgn} (\hat p) Q_{\hat p(i) \hat p(j) \hat p(k)}$, respectively. Furthermore, the quantity
 \begin{align}
\frac{Z_{ij}}{Z_i^{N^{ij}/N_i} Z_j^{N^{ij}/N_j}} &= \exp \frac{2\pi i}{N_{ij}} (P_{ij}+P_{ji})
\end{align}
can take $N_{ij}$ different values for different choices of $P_{ij}+P_{ji}$. Taking all possible values for $i$, $j$, and $k$, the invariants given are able to distinguish all elements of $\mathcal H^3(G,U(1))$.

A few comments are in order. First, the invariant defined on the lens space $L(N_i;q)$ (for any $q$ such that $q$ and $N_i$ are coprime) is also a valid topological invariant. The corresponding invariant for the given canonical cocycle would be
\begin{equation}
\prod_{n=1}^{N_i}  \omega_3(e_i,ne_i,q e_i) = \exp \frac{2\pi i}{N_i} qP_{ii}.
\end{equation}
This invariant can still take $N_i$ different values. However, we have chosen $q$ to be $1$ for simplicity.\\
Second, one can also define invariants on $L(N_i;1)$ that place multiples of the generator (say $p e_i$ for an integer $p$) around the great circle as long as $p$ and $N_i$ are coprime. The corresponding invariant for the given canonical cocycle is
\begin{equation}
\prod_{n=1}^{N_i}  \omega_3(pe_i,npe_i,pe_i) = \exp \frac{2\pi i}{N_i} p^2P_{ii},.
\end{equation}
Analogously, we have chosen $p=1$ for simplicity.\\
\subsection{3+1D}
 It is a known fact\cite{aschenbrenner} that any abelian group can be realized as the fundamental group of some $n$-manifold for $n>3$. Thus, at first, it is not obvious which 4-manifolds we should use. However, if we use dimensional reduction to reduce the theory down to 2+1D and evaluate the partition functions given in the previous subsection, it turns out that they give us enough information to distinguish all the SPT phases classified by the cohomology group
 \begin{align}
\mathcal H^4(G,U(1)&) \cong \prod_{i<j} \mathbb{Z}_{N_{ij}}^2\prod_{i<j<k} \mathbb{Z}_{N_{ijk}}^2\prod_{i<j<k<l} \mathbb{Z}_{N_{ijkl}}.
\end{align}
Going down one dimension via dimensional reduction simply means that we take a product of our 3-manifolds in the previous subsection with a circle. Therefore, let us define the following partition functions:
\begin{enumerate}
\itemsep0em 
\item $Z_{i,l}$ is the partition function on $L(N_i;1) \times S^1$ with $e_i$ around the great circle of $L(N_i;1)$ and $e_l$ around $S^1$.
\item $Z_{ij,l}$ is the partition function on $L(N^{ij};1)\times S^1$ with $e_i+e_j$ around the great circle of $L(N^{ij};1)$ and $e_l$ around $S^1$.
\item $Z_{ijk,l}$ is the partition function on $T^4$ with $e_i$, $e_j$, $e_k$, and $e_l$ around four homotopically inequivalent loops.
\end{enumerate}
The partition functions can be obtained by replacing $\omega_3$ in equations \eqref{equ:Z_i} - \eqref{equ:Z_ijk} with $i_{e_l}\omega_4$: 
\begin{subequations}
\begin{align}
\label{equ:Z_il}
Z_{i,l} &=\prod_{n=1}^{N_i} i_{e_l}\omega_4(e_i,ne_i,e_i),\\
\label{equ:Z_ijl}
Z_{ij,l} &= \prod_{n=1}^{N^{ij}} i_{e_l}\omega_4(e_i+e_j,n(e_i+e_j),e_i+e_j),\\
\label{equ:Z_ijkl}
Z_{ijk,l} &=i_{e_i}i_{e_j}i_{e_k}i_{e_l}\omega_4.
\end{align}
\end{subequations}
We will now evaluate the partition functions using the canonical cocycles
\begin{align}
\omega_4(a,b,c,d)&= \exp 2\pi i \left [\sum_{ijk} \frac{P_{ijk}}{N_{ij} N_k}a_i b_j (c_k+d_k-[c_k+d_k]) \right. \nonumber\\
 & +\left.  \sum_{ijkl} \frac{Q_{ijkl}}{N_{ijkl}} a_ib_jc_kd_l\right].
\end{align}
For simplicity, we make an extra assumption that $Q_{ijkl}$ is zero if $i,j,k$ are not all distinct. The topological invariants are
\begin{subequations}
\begin{align}
Z_{i,l} &=\exp \frac{2\pi i}{N_{il}} (P_{ili}-P_{lii}),\\
Z_{ij,l} &= \exp 2\pi i N^{ij} \left [ \frac{P_{ili}-P_{lii}}{N_{il}^2} + \frac{P_{ilj}-P_{lij}}{N_{il}N_j} +(i \leftrightarrow j) \right], \\
Z_{ijk,l} &=\exp \frac{2\pi i}{N_{ijkl}} \sum_{\hat p} \text{sgn}(\hat p)Q_{\hat p(i) \hat p(j) \hat p(k) \hat p(l)},
\end{align}
\end{subequations}
where $\hat p$ is a permutation on four indices. We refer the reader to Ref. \onlinecite{WangLevin2015} for the completeness of the invariants $Z_{i,l}$, $Z_{ijk,l}$, and the invariant
\begin{equation}
\frac{Z_{ij,l}}{Z_{i,l}^{N^{ij}/N_{il}} Z_{j,l}^{N^{ij}/N_{jl}}} = \exp 2\pi i N^{ij} \left [ \frac{P_{ilj}-P_{lij}}{N_{il}N_j} +\frac{P_{jli}-P_{lji}}{N_{jl}N_i} \right].
\end{equation}
\begin{widetext}
\subsection{4+1D}
We will now show an application of the general dimensional reduction for classifying 4+1D SPT phases. The cohomology group is
\begin{equation}
\mathcal H^5(G,U(1)) \cong \prod_i \mathbb Z_{N_i} \prod_{i<j} \mathbb Z_{N_{ij}}^2 \prod_{i<j<k} \mathbb Z_{N_{ijk}}^4 \prod_{i<j<k<l} \mathbb Z_{N_{ijkl}}^3 \prod_{i<j<k<l<m} \mathbb Z_{N_{ijklm}}.
\end{equation}
First, we define the following partition functions:
\begin{enumerate}
\itemsep0em 
\item $Z_{i,l,m}$ is the partition function on $L(N_i;1) \times T^2$ with $e_i$ around the great circle of $L(N_i;1)$ and $e_l$, $e_m$ around the two non-contractible loops of $T^2$.
\item $Z_{ij,l,m}$ is the partition function on $L(N^{ij};1)\times T^2$ with $e_i+e_j$ around the great circle of $L(N^{ij};1)$ and $e_l$, $e_m$ around the two non-contractible loops of $T^2$.
\item $Z_{ijk,l,m}$ is the partition function on $T^5$ with $e_i$, $e_j$, $e_k$, $e_l$, and $e_m$ around 5 homotopically inequivalent loops.
\end{enumerate}
The partition functions can be evaluated by replacing $\omega_4$ in the 3+1D partition functions \eqref{equ:Z_il} - \eqref{equ:Z_ijkl} with $i_{e_m}\omega_5$:
\begin{subequations}
\begin{align}
Z_{i,l,m} &=\prod_{n=1}^{N_i} i_{e_l}i_{e_m}\omega_5(e_i,ne_i,e_i),\\
Z_{ij,l,m} &= \prod_{n=1}^{N^{ij}} i_{e_l}i_{e_m}\omega_5(e_i+e_j,n(e_i+e_j),e_i+e_j),\\
Z_{ijk,l,m} &=i_{e_i}i_{e_j}i_{e_k}i_{e_l}i_{e_m}\omega_5.
\end{align}
\end{subequations}
However, these partition functions are not enough to distinguish the elements of $\mathcal H^5(G,U(1))$. To see this, consider the following canonical cocycles
\begin{align}
\label{equ:omega5}
\omega_5(a,b,c,d,e) = \exp2\pi i  &\left[\sum_{ijk} \frac{P_{ijk}}{N_iN_jN_k} a_i (b_j+c_j-[b_j+c_j])(d_k+e_k-[d_k+e_k]) \right. \nonumber\\
&\left.+\sum_{ijkl} \frac{Q_{ijkl}}{N_{ijk}N_l} a_ib_jc_k(d_l+e_l-[d_l+e_l])+\sum_{ijklm} \frac{R_{ijklm}}{N_{ijklm}} a_ib_jc_kd_le_m\right].
\end{align}
with the extra assumption that $R_{ijklm}$ is zero unless $i,j,k,l,m$ are all different. First, let us do dimensional reduction on a circle with holonomy $g$. The resulting 4-cocycle is
\begin{align}
i_{g}\omega_5(a,b,c,d) = \exp2\pi i  &\left[\sum_{ijm} \frac{P_{mij}}{N_iN_jN_m}  (a_i+b_i-[a_i+b_i])(c_j+d_j-[c_j+d_j])g_m \right. \nonumber\\
&\left.+\sum_{ijkm} \frac{Q_{mijk}-Q_{imjk}+Q_{ijmk}}{N_{mij}N_k} a_ib_j(c_k+d_k-[c_k+d_k])g_m \right. \nonumber \\
&\left.+\sum_{ijklm} \frac{R_{mijkl}-R_{imjkl}+R_{ijmkl}-R_{ijkml}+R_{ijklm}}{N_{ijklm}} a_ib_jc_kd_lg_m\right].
\end{align}
However, the cocycle on the first line is a coboundary of the cochain
\begin{equation}
\mu_{g}(a,b,c)=\exp 2\pi i \sum_{ijm} \frac{P_{mij}}{N_iN_jN_m}  a_i(b_j+c_j-[b_j+c_j])g_m.
\end{equation}
Hence, the invariants we have defined (and in general the evaluation of $i_{g}\omega_5$ on any closed 4-manifold) will not be able to detect cocycles with different values of $P_{ijk}$. Evaluating the partition functions gives
\begin{subequations}
\begin{align}
Z_{i,l,m}&=\exp \frac{2\pi i}{N_{ilm}} \sum_{\hat p}\text{sgn}(\hat p)Q_{\hat p(i)\hat p(l)\hat p(m)i},\\
Z_{ij,l,m}&= \exp 2\pi i N^{ij} \sum_{\hat p}\text{sgn}(\hat p) \left [ \frac{Q_{\hat p(i)\hat p(l)\hat p(m)i}}{N_{ilm}N_{il}} + \frac{Q_{\hat p(i)\hat p(l)\hat p(m)j}}{N_{ilm}N_{j}} + (i \leftrightarrow j)  \right],\\
Z_{ijk,l,m}&=\exp \frac{2\pi i}{N_{ijklm}} \sum_{\hat p} \text{sgn}(\hat p)R_{\hat p(i) \hat p(j) \hat p(k) \hat p(l) \hat p(m)},
\end{align}
\end{subequations}
where $\hat p$ is a three index permutation for the first two equations, and a five index permutation for the last equation. One can show that the $\prod_{i<j<k} \mathbb Z_{N_{ijk}}^3 \mathbb \prod_{i<j<k<l} \mathbb Z_{N_{ijkl}}^3 \prod_{i<j<k<l<m} \mathbb Z_{N_{ijklm}}$ part of the cohomology group can be distinguished by the above partition functions (we refer to Ref. \onlinecite{ChengTantiWang2017} for further details.)

To distinguish the remaining part of the cohomology group, let us assume that we have distinguished the previous part of the cocycle. That is, we have a choice of the tensors $Q_{ijkl}$ and $R_{ijklm}$ that gives the same values for the three invariants above. If we now divide our cocycle by another canonical cocycle with $P_{ijk}=0$ and the values $Q_{ijkl}$ and $R_{ijklm}$ given previously, the resulting cocycle will only depend on $P_{ijk}$. Let us call this residual part of the cocycle (i.e., the first line of equation \eqref{equ:omega5}) $\omega_5^I$. We can now use the general dimensional reduction via the slant product of $\omega_5^I$ and the chain $(g,h)$, given in equation \eqref{equ:slantgh5}. This gives,
\begin{equation}
i_{(g,h)}\omega_5^I(a,b,c) = \exp 2\pi i \sum_{ijk} \frac{P_{ijk}+P_{ikj}}{N_iN_jN_k} a_j(b_k+c_k-[b_k+c_k])(g_k+h_k-[g_k+h_k]).
\end{equation}
Hence, the 3-cocycle in equation \eqref{equ:cocyclegh}, (which we will call $\alpha_{(g,h)}$) is given by
\begin{align}
\label{equ:alphagh}
\alpha_{(g,h)}(a,b,c)&=i_{(g,h)}\omega_5^I(a,b,c) \left ( \frac{\mu_{g}(a,b,c)\mu_{h}(a,b,c)}{ \mu_{gh}(a,b,c)} \right )\\
&=\exp 2\pi i \sum_{ijk} \frac{P_{ijk}+P_{ikj}+P_{kij}}{N_iN_jN_k} a_i(b_j+c_j-[b_j+c_j])(g_k+h_k-[g_k+h_k]).\nonumber
\end{align}
Note that there is a slight ambiguity in defining the cocycle $\alpha_{(g,h)}$. We made use of the fact that there exists a cochain $\mu_{g}$ such that $d\mu_{g} = i_{g}\omega_5^I$. However, the solution is not unique, since $\mu_g \xi_g$, where $\xi_g \in \mathcal Z^3(G,U(1))$, is also a solution. Consequently, we must make sure that the invariants we define do not depend on this choice of $\xi_g$.

To do this, let us first consider the equivalence class of the cocycle $\alpha_{(g,h)}$ denoted $[\alpha_{(g,h)}] \in \mathcal H^3(G,U(1))$. If we treat $[\alpha_{(\cdot,\cdot)}]$ as a function that depends on $g$ and $h$, then we can define it as a cochain valued in $\mathcal H^3(G,U(1))$. That is, $[\alpha] \in \mathcal C^2(G,\mathcal H^3(G,U(1)))$. For these cochains, we now need to define a coboundary operator. A sensible choice is to use the boundary operator $\partial$. Given a cochain $[\alpha] \in \mathcal C^{m-1}(G,\mathcal H^3(G,U(1)))$, its coboundary is denoted as $[\alpha_{\partial}] \in \mathcal C^{m}(G,\mathcal H^3(G,U(1)))$ defined naturally following equation \eqref{equ:boundarylinks} as
\begin{align}
\left [\alpha_{\partial(g_1, ... ,g_m)} \right ] = &\left [\alpha_{(g_2, ... ,g_m)}\alpha_{(g_1, ... ,g_{m-1})}^{(-1)^m} \prod_{i=1}^{m-1} \alpha_{(g_1, ...,g_{i-1}, g_ig_{i+1},g_{i+1},... ,g_m)}^{(-1)^{i}}\right ].
\end{align}
We can now check that $[\alpha_{(g,h)}]$ given by equation \eqref{equ:alphagh} is actually a cocycle under the $\partial$ coboundary operator. First, notice that $[\mu] \in \mathcal C^1(G,\mathcal H^3(G,U(1)))$ and that the combination
\begin{equation}
\left [\frac{\mu_{g}\mu_{h}}{ \mu_{gh}} \right ] = \left [ \mu_{\partial (g,h)} \right ]
\end{equation}
is a coboundary under $\partial$. Hence, we can write
\begin{equation}
\left [\alpha_{(g,h)} \right ] =  \left [i_{(g,h)}\omega_5 \mu_{\partial(g,h)} \right ].
\end{equation}
Its coboundary is thus given by 
\begin{align}
\left [\alpha_{\partial(g,h,k)} \right ] &=  \left [i_{\partial(g,h,k)}\omega_5 \mu_{\partial^2(g,h,k)} \right ]\\
&=\left[di_{(g,h,k)}\omega_5/i_{(g,h,k)}d\omega_5 \right ] =   \left[di_{(g,h,k)}\omega_5 \right ],
\end{align}
where we simplified the expression using the commutation relation \eqref{equ:commutator} and the facts that $\omega_5$ is a cocycle and $\partial^2=0$. The equivalence class of a coboundary by $d$ is the identity element in $\mathcal H^3(G,U(1))$ and therefore, $[\alpha] \in \mathcal Z^2(G,\mathcal H^3(G,U(1)))$.
Now, consider $[\alpha]$ again when $\mu_g \rightarrow \mu_g \xi_g$. We can see that
\begin{equation}
\left [\alpha_{(g,h)} \right] \rightarrow \left [\alpha_{(g,h)} \frac{\xi_{g}\xi_{h}}{\xi_{gh}} \right ] = \left [\alpha_{(g,h)} \xi_{\partial(g,h)} \right].
\end{equation}
Since $[\alpha]$ is changed by a coboundary of $\partial$, the quantities we define will only be invariants if they do not change under these coboundary transformations. That is, the invariants must be elements of $\mathcal H^2(G,\mathcal H^3(G,U(1)))$.

Let us now proceed to obtain such invariants. First, let us evaluate $\alpha_{(g,h)}$ on the 3-manifolds we used for the 2+1D invariants. We get,
\begin{subequations}
\begin{align}
Z_{i}(g,h) &=\exp \frac{2\pi i}{N_{i}} \sum_{k} \frac{P_{iik}+P_{iki}+P_{kii}}{N_k}(g_k+h_k-[g_k+h_k]),\\
Z_{ij}(g,h) &= \exp 2\pi i N^{ij} \sum_{k} \left [ \frac{P_{iik}+P_{iki}+P_{kii}}{N_i^2N_k} + \frac{P_{ijk}+P_{jki}+P_{kij}}{N_iN_jN_k} + (i \leftrightarrow j) \right] \cdot (g_k+h_k-[g_k+h_k]),\\
Z_{ijk}(g,h) &=0.
\end{align}
\end{subequations}
These partition functions are now elements of $\mathcal H^3(G,U(1))$. We will discard $Z_{ijk}(g,h)$ since there is no information to extract. Now, let us define the following invariants of $\mathcal H^2(G,\mathcal H^3(G,U(1)))$:
\begin{subequations}
\begin{align}
\mathcal I_{ij} &= \prod_{n=1}^{N^{ij}} Z_{i}(e_j,ne_j),\\
\mathcal I_{ijk} &= \prod_{n=1}^{\text{lcm}(N_{ij},N_k)} Z_{ij}(e_k,ne_k).
\end{align}
\end{subequations}
One can check that these expressions are invariant under coboundary transformations. The explicit expression for these invariants are
\begin{subequations}
\begin{align}
\mathcal I_{ij} &= \exp \frac{2\pi i}{N_{ij}}  (P_{iij}+P_{iji}+P_{jii}),\\
\mathcal I_{ijk} &= \exp  \frac{2\pi i}{N_{ijk}}\left [\frac{N_j}{N_i}(P_{iik}+P_{iki}+P_{kii}) +  P_{ijk}+P_{jki}+P_{kij} + (i \leftrightarrow j) \right ].
\end{align}
\end{subequations}
Now, $\mathcal I_{ij}$ and $\mathcal I_{ji}$ can each take $\prod_{i <j} N_{ij}$ independent values, while the invariant
\begin{align}
\frac{\mathcal I_{ijk}}{\mathcal I_{ik}^{\text{lcm} (N_{ik},N_j)/N_i}  \mathcal I_{jk}^{\text{lcm} (N_{jk},N_i)/N_j} } = \exp \frac{2 \pi i}{N_{ijk}} (P_{ijk}+P_{jki}+P_{kij}+P_{ikj}+P_{jik}+P_{kji})
\end{align}
can take $N_{ijk}$ different values. Thus, these invariants can distinguish the $\prod_{i<j} \mathbb Z_{N_{ij}}^2 \prod_{i<j<k} \mathbb Z_{N_{ijk}}$ part of the cohomology group.

The last invariant we need to introduce is the partition function on the generalized lens space (see Appendix \ref{app:lens} for details). We define $\mathcal I_{i}$ as the partition function on $L(N_i;1,1,1)$ with the flat connection given by wrapping $e_i$ around the ``great circle''. The explicit expression is
\begin{align}
\mathcal I_{i} &= \prod_{n,m=1}^{N_i}\omega_5(e_i,ne_i,e_i,me_i,e_i) = \exp \frac{2\pi i}{N_i}P_{iii}.
\end{align}
This invariant can take $\prod_i N_i$ different values, and so it classifies the $\prod_i \mathbb Z_{N_i}$ part in the cohomology group.

To recapitulate, we have defined six topological invariants $Z_{i,l,m}$, $Z_{ij,l,m}$, $Z_{ijk,l,m}$, $\mathcal I_{i}$, $\mathcal I_{ij}$, and $\mathcal I_{ijk}$. The latter two come from the general dimensional reduction using the slant product with a 2-chain $(g,h)$. With this, we were able to distinguish all the cohomology classes of $\mathcal H^5(G,U(1))$. We note that there are applications of these invariants beyond bosonic SPT phases. Namely, they can be used to determine the obstruction class of the Steenrod square for 3+1D fermionic SPT phases. We refer to Ref. \onlinecite{ChengTantiWang2017} for further discussions.

We have seen that, with the help of dimensional reduction, we can define invariants that distinguish all the possible different SPT phases described in the group cohomology model. However, we were not able to determine the manifolds whose partition functions (or product of partition functions) are the invariants $\mathcal I_{ij}$, $\mathcal I_{ijk}$. It would be interesting to see if there is a general formalism to generate these invariants. In particular, we believe that the manifolds required in a given dimension are the generators of the oriented cobordism group with an extra constraint that we only consider cobordisms that preserve the flat connections.
\end{widetext}

\section{Topological Invariants for Gu-Wen Fermionic SPT Phases}\label{invariantsfermions}
In this section, we will extend our results from the previous section by showing that the Gu-Wen SPT phases can be distinguished by evaluating the Gu-Wen partition function on the same manifolds as the bosonic case (equipped with spin structures). Let us first recall that the group is now $G^f = \mathbb Z_2^f \times G$. We will denote the subgroup $\mathbb Z_2^f$ with subscript $0$, which means that $N_0=2$. Accordingly, $N^{0i}=\text{lcm} (2,N_i)$, $N_{0i}=\text{gcd} (2,N_i)$, etc.

We can extract the equivalence class of the cocycle $\beta$ by exploiting the spin structure of the manifolds. The Gu-Wen partition function has a dependence on the spin structure from $E(\beta)$. Hence, for two partition functions given on the same manifold with two different spin structures, represented by chains $E$ and $E'$, their ratio is
\begin{equation}
\label{equ:extractbeta}
\frac{\mathcal Z(\eta)}{\mathcal Z(\eta')}=\frac{E(\beta)}{E'(\beta)}.
\end{equation}
For inequivalent spin structures, the difference between the sets $E$ and $E'$ must be a cycle that is not a boundary. That is, a representative of a non-trivial element of $H_{n-1}(\mathcal M,\mathbb Z_2)$ (see Appendix \ref{app:w_2} for further discussion). We remark that for physical systems, we only have to distinguish the equivalence class in $B\mathcal H^{n-1}(G,\mathbb Z_2)$. It has been shown that for finite abelian groups, there is no obstruction for 1+1D and 2+1D, while there are certain conditions for the cocycle to be obstruction-free in 3+1D\cite{ChengTantiWang2017}. Thus, this consideration is more general in 3+1D.
\subsection{1+1D}
\begin{figure}[H]
\centering
\includegraphics{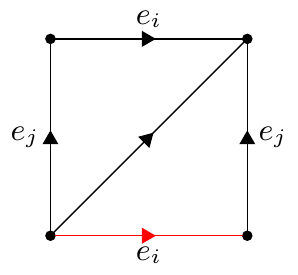}
\caption{A 1-cycle that is not a boundary for $T^2$ (shown in red).}
\label{fig:toruscycle}
\end{figure}
The equivalence class of $\beta_1$ is
\begin{align}
\mathcal H^1(G,\mathbb{Z}_2) \cong \prod_i \mathbb{Z}_{N_{i0}}.
\end{align}
Consider the following canonical cocycles for $\beta_1$
\begin{align}
\label{equ:canonicalbeta1}
\beta_1(a) &= \sum_i \frac{2R_{i}}{N_{0i}} a_i.
\end{align}
For $T^2$, the partition function $\mathcal Z^{ij}(\eta_i,\eta_j)$ will depend on the spin structure on the two circles: $\eta_i$ and $\eta_j$. These two numbers take value $0$ or $1$. A cycle that is not a boundary is the link spanning the $i$ direction, which is shown in red in Figure \ref{fig:toruscycle}. Hence, we see from equation \eqref{equ:extractbeta} that, 
\begin{align}
\frac{\mathcal Z^{ij}(0,1)}{\mathcal Z^{ij}(0,0)}&=\exp \pi i   \beta(e_i)  \nonumber\\
&= \exp \frac{2\pi i}{N_{0i}}R_{i} .
\end{align}
Note that this invariant is independent of the group element placed in the $j$ direction. Therefore, in the case where $G=\mathbb Z_{N_1}$, we can replace $e_j$ with the identity element. For each $i$, the invariant can take one value for $N_i$ odd, and two values for $N_i$ even. Hence, the total number of invariants is $\prod_{i} N_{0i}$. Therefore, the invariants can distinguish all the equivalence classes of $\beta_1$.\\
Now that we have determined $\beta_1$, we can consider the case where we fix $\beta_1$, and attempt to distinguish the different elements of $\mathcal H^2(G,U(1))$. That is, the different cocycles one can add to $\omega_2$ satisfying the Gu-Wen equation \eqref{equ:domega}. They can be obtained from the partition function $\mathcal Z^{ij}(0)$, like the bosonic case. The Grassmann integral and spin structure terms can indeed be evaluated explicitly. However, they only depend on $\beta_1$, which is fixed. Since there are $\left |\mathcal H^2(G,U(1))\right |$ inequivalent cocycles we can add, this means that the partition function can take $\left |\mathcal H^2(G,U(1))\right |$ values. This shows that we can also distinguish the inequivalent $\omega$'s. For this reason, these manifolds can distinguish all Gu-Wen phases in 1+1D.

\subsection{2+1D}
\begin{figure}[H]
\centering
\includegraphics{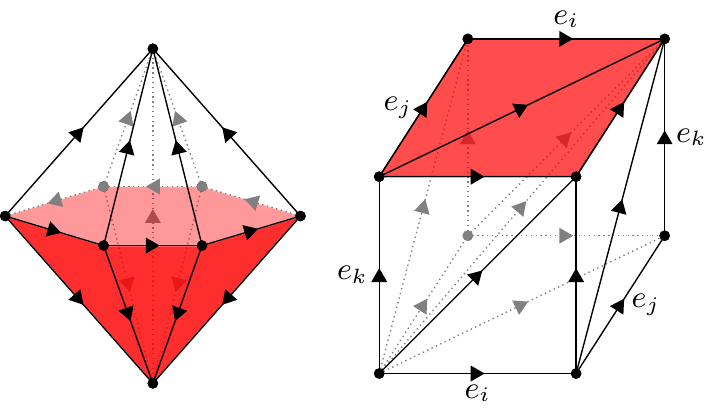}
\caption{2-cycles that are not boundaries for the lens space $L(N^{0i};1)$ and $T^3$ (shown in red).}
\label{fig:lensspin}
\end{figure}
The equivalence class of $\beta_2$ is
\begin{align}
\mathcal H^2(G,\mathbb{Z}_2) \cong \prod_i \mathbb{Z}_{N_{i0}}\prod_{i<j} \mathbb{Z}_{N_{ij0}}.
\end{align}
We will use the following canonical cocycles for $\beta_2$
\begin{align}
\label{equ:canonicalbeta2}
\beta_2(a,b) &= \sum_i \frac{R_i}{N_{i}}(a_i+b_i-[a_i+b_i])+\sum_{ij} \frac{2R_{ij}}{N_{0ij}} a_i b_j,
\end{align}
where we will assume $R_{ii}=0$ to simplify the counting argument.\\
The lens space $L(p;q)$ has two possible spin structures for $p$ even (denoted by $\eta=0,1$), while there is only the trivial spin structure for $p$ odd (denoted by $\eta=0$) \cite{franc1987spin}. On the other hand, $T^3$ has eight possible spin structures, denoted by $\eta_i,\eta_j,\eta_k$ taking values 0 or 1 depending on whether the boundary conditions are periodic or antiperiodic around each circle.

Let us start by placing $e_i$ around the great circle of the lens space $L(N^{0i};1)$ and call this partition function $\mathcal Z_{0i}(\eta)$, with $\eta$ the choice of spin structure. Note that $N^{0i}$ is always even, so the lens space admits two inequivalent spin structures. A 2-cycle that is not a boundary is shown in red in Figure \ref{fig:lensspin}. It has no boundary because the great circle has an even number of links, which cancel out since they are all identical. However, it is not a boundary, as one can see from the figure. The expression we have is then
\begin{align}
\frac{\mathcal Z_{0i}(1)}{\mathcal Z_{0i}(0)}=&\exp \pi i \sum_{n=1}^{N^{0i}} \beta(e_i,ne_i) = \exp \pi i \frac{N^{0i}}{N_i} R_i\nonumber \\
\label{equ:spinlens}
=&  \exp  \frac{2\pi i}{N_{0i}} R_i.
\end{align}
Note that $\mathcal Z_{0i}(0)=(\mathcal Z_{i}(0))^{2/N_{0i}}$. The set of all these invariants for all $i$ can take $\prod_i N_{0i}$ different values.\\
Next, we will consider $T^3$. A 2-cycle that is not a boundary is the cycle spanning the $i$-$j$ plane. Accordingly,
\begin{align}
\frac{\mathcal Z_{ijk}(0,0,1)}{\mathcal Z_{ijk}(0,0,0)}&=\exp \pi i \left (  \beta(e_i,e_j) + \beta(e_j,e_i) \right) \nonumber\\
\label{equ:spinT3}
&= \exp \frac{2\pi i}{N_{ij0}} (R_{ij}+R_{ji}).
\end{align}
Note that like $T^2$ in 1+1D, this value is independent of the element we place in the $k$ direction. The total set of invariants for all $i<j$ can take $\prod_{i<j} N_{ij0}$ different values. Hence, the set of all the invariants can take $\prod_i N_{i0} \prod_{i<j} N_{ij0}$ different values, which is equal to $\left | \mathcal H^2(G,\mathbb{Z}_2) \right |$. Therefore, the invariants can distinguish all the elements of $\mathcal H^2(G,\mathbb{Z}_2)$. Fixing $\beta$, we can distinguish elements in $\mathcal H^3(G,U(1))$ using the same argument as the 1+1D case.

\subsection{3+1D}
We need to classify the cohomology group elements
\begin{equation}
\mathcal H^3(G,\mathbb{Z}_2) \cong \prod_{i} \mathbb Z_{N_{0i}} \prod_{i<j} \mathbb Z_{N_{0ij}}^2 \prod_{i<j<k} \mathbb Z_{N_{0ijk}}
\end{equation}
by assuming the canonical cocycles of the form
\begin{align}
\beta_3(a,b,c) =& \sum_{ij} \frac{2R_{ij}}{N_{0i}N_j}a_i(b_j+c_j-[b_j+c_j])\nonumber\\
&+\sum_{ijk} \frac{2R_{ijk}}{N_{0ijk}} a_i b_j c_k.
\end{align}
First, consider the partition function on $L(N_i;1)\times S^1$, $\mathcal Z_{i,j}(\eta_i,\eta_j)$, where $\eta_i$ and $\eta_j$ are the choices of spin structure on the lens space and $S^1$, respectively. If we change the spin structure on $S^1$, the ratio of the partition functions is equal to evaluating $\beta_3$ on the lens space. Hence, the invariant is
\begin{align}
\frac{\mathcal Z_{i,j}(0,1)}{\mathcal Z_{i,j}(0,0)} &= \exp \pi i \sum_{n=1}^{N_i} \beta(e_i,ne_i,e_i)= \exp  \frac{2\pi i}{N_{0i}}  R_{ii},
\end{align}
which takes $\prod_{i} N_{0i}$ values. Next, if we want to flip the spin structure on the lens space, we need to consider the partition function on $L(N^{0i};1)\times S^1$, $\mathcal Z_{0i,j}(\eta_i,\eta_j)$. We can obtain the corresponding invariant by first performing dimensional reduction over $S^1$ with flux $e_j$ and then evaluating the invariant from the 2+1D case. This gives
\begin{align}
\frac{\mathcal Z_{0i,j}(1,0)}{\mathcal Z_{0i,j}(0,0)} &= \exp \pi i  \sum_{n=1}^{N^{0i}} i_{e_j}\beta(e_i,ne_i)= \exp  \frac{4 \pi i}{N_{0i}N_{0j}}R_{ji}.
\end{align}
This invariant can take two values if $N_i$ and $N_j$ are both even. Otherwise, it takes one value. Swapping $i$ and $j$ for $i \ne j$ gives us another set of invariants, so all together, this set of invariants takes $\prod_{i<j} N_{0ij}^2$ values. Finally, changing the spin structure on one of the circles in $T^4$ gives
\begin{align}
\frac{\mathcal Z_{ijk,l}(0,0,0,1)}{\mathcal Z_{ijk,l}(0,0,0,0)}&= \exp  \frac{2 \pi i}{N_{0ijk}} \sum_{\hat p}R_{\hat p(i)\hat p(j)\hat p(k)},
\end{align}
which takes $\prod_{i<j<k} N_{0ijk}$ values. Thus, these invariants can distinguish elements in $\mathcal H^3(G,\mathbb{Z}_2)$. Fixing $\beta$, the elements in $\mathcal H^4(G,U(1))$ can then be distinguished.

We have shown from 1+1D to 3+1D that all Gu-Wen phases can be distinguished on the given closed manifolds equipped with spin structure. We were not able to show this for 4+1D because we do not know all the 5-manifolds to use. However, we believe that these manifolds can also be realized in general as the generators of the spin cobordism group with an extra condition that the cobordisms must preserve the flat connections.

A few comments are in order. First, in higher dimensions, it is known that the oriented and spin cobordisms group differ, so it is possible that these manifolds will be different as we study SPT phases in higher dimensions. Second, as we have only considered fermionic SPT phases described by the Gu-Wen model, it remains to be shown if these manifolds are enough to distinguish all fermionic SPT phases in general. Some considerations for general fermionic SPT phases for certain symmetry groups have been affirmative. For example, the complete $\mathbb Z_8$ classification of SPT phases with symmetry group $G^f= \mathbb Z_2^f \times \mathbb Z_2$ in 2+1D\cite{GuLevin2014,Kapustinetal2015} can be distinguished by evaluating the partition function on $\mathbb R P^3$, which is the lens space $L(2;1)$\cite{PutrovWangYau2017}. In the next section, we will further support this claim by establishing the partition functions we defined as phases obtained from braiding statistics of excitations of the gauged theories.

\section{Relation to Braiding Statistics}\label{braidingrelation}
In this section, we will establish a relation between the invariants we have constructed to the invariants obtained from braiding statistics. These invariants are obtained from braiding excitations of the gauged SPT phases in 2+1D and above \cite{LevinGu2012}, and are not limited to systems with a TQFT description. Like the invariants we defined, these invariants are also able to distinguish SPT phases with any finite abelian unitary symmetry group in 2+1D and 3+1D\cite{WangLevin2015}. For bosonic SPT phases, we will show that there is a one-to-one correspondence between these invariants, and recast the partition functions we studied in terms of braiding statistics. For Gu-Wen fermionic SPT phases, we will focus the study on 2+1D invariants.

\subsection{Braidings in 2+1D Bosonic SPT Phases}
First, let us review the concept of topological invariants from braiding statistics in 2+1D. Ref. \onlinecite{LevinGu2012} first showed that the two inequivalent SPT phases for the symmetry group $G=\mathbb Z_2$ in 2+1D can be gauged so that it is topologically ordered. Unlike the SPT partition function, the group elements on the links are now summed over so that they are dynamical. That is, the partition function sums over all possible flat connections and is called the Dijkgraaf-Witten partition function \cite{DijkgraafWitten}. Since the theory is topologically ordered, one can study the braidings of flux excitations, which give rise to different phase factors for different SPT phases. This was later generalized by Ref. \onlinecite{WangLevin2015} for general finite abelian unitary symmetry groups. The phases from the braiding procedures they introduced are invariant under coboundary transformations of the cocycle and are able to distinguish all the SPT phases described by the group cohomology model. Let us reproduce those definitions here: let $\xi_i$ represent the flux excitations corresponding respectively to generators $e_i$. The invariants are phases obtained from the following braidings:
\begin{enumerate}
\itemsep0em 
\item $\Theta_i$ = $2 \pi N_i s_{\xi_i}$ where $s_{\xi_i}$ is the topological spin of $\xi_i$
\item $\Theta_{ij}$ is obtained from braiding $\xi_i$ around $\xi_j$ $N^{ij}$ times
\item $\Theta_{ijk}$ is obtained from braiding $\xi_i$ respectively around $\xi_j$, $\xi_k$, followed by $\xi_j$, $\xi_k$ again, but in the reverse direction.
\end{enumerate}
Alternatively, we know that for abelian anyons, $\Theta_i$ can also be thought of as the exchange statistics of $\xi_i$. That is, the phase obtained from swapping positions of two $\xi_i$ particles.

By comparing the expressions from Ref. \onlinecite{WangLevin2015} to those of the partition functions \eqref{equ:Z_i}-\eqref{equ:Z_ijk}, one can see that the two invariants are related by
\begin{subequations}
\begin{align}
\exp i\Theta_{i} &= Z_i,\\
\label{equ:relationij}
\exp i\Theta_{ij} &= \frac{Z_{ij}}{Z_i^{N^{ij}/N_i} Z_j^{N^{ij}/N_i}},\\
\exp i\Theta_{ijk} &= 1/Z_{ijk}.
\end{align}
\end{subequations}

One can ask if there is an interpretation of $Z_{ij}$ in terms of braiding. Since $Z_{ij}$ is a partition function on a lens space like $Z_i$, we should expect it to be the topological spin of some particle. Indeed, an educated guess would be that the particle is $\xi_i \times \xi_j$, the fusion of particles $\xi_i$ and $\xi_j$. We will now show that it is indeed the case.\\
Suppose we attach $\xi_j'$ to $\xi_i$ and $\xi_i'$ to $\xi_j$ (the primes are to label them as different particles, but they carry the same flux.) Consider exchanging the two groups of particles $N_{ij}$ times. This exchange would give rise to three different phases.
\begin{enumerate}
\itemsep0em 
\item The exchange of $\xi_i$ and $\xi_i'$ $N^{ij}$ times gives $\frac{N^{ij}}{N_i}\Theta_i$.
\item The exchange of $\xi_j$ and $\xi_j'$ $N^{ij}$ times gives $\frac{N^{ij}}{N_j}\Theta_j$.
\item The exchange of $\xi_i$ and $\xi_j'$ and the exchange of $\xi_j$ and $\xi_i'$ $N^{ij}$ times. This is equal to the braiding of $\xi_i$ around $\xi_j$ $N^{ij}$ times, which gives $\Theta_{ij}$.
\end{enumerate}
The combination of these three phases is exactly $Z_{ij}$ from equation \eqref{equ:relationij}. Hence, this partition function corresponds to the exchange of two $\xi_i \times \xi_j$ particles $N^{ij}$ times. Since the order of the group element $e_i+e_j$ is $N^{ij}$, we indeed have
\begin{equation}
Z_{ij}= \exp 2\pi i N^{ij} s_{\xi_i \times \xi_j}.
\end{equation}
This agrees with the result obtained by Ref. \onlinecite{Zaletel2014}.

We remark that for 2+1D TQFTs, the partition function on some closed orientable 3-manifold is equal to the partition function of $S^3$ with some defect link, where surgery on the link gives the 3-manifold\cite{WangWenYau2016,PutrovWangYau2017}. These links correspond exactly to those obtained from closing up the worldlines of the particles. Closing up a single vortex worldline gives a trivial knot, on which surgery gives lens spaces, while $T^3$ corresponds to surgery on a Borromean ring.

\subsection{Braidings in 3+1D Bosonic SPT Phases}
\begin{figure}
\includegraphics{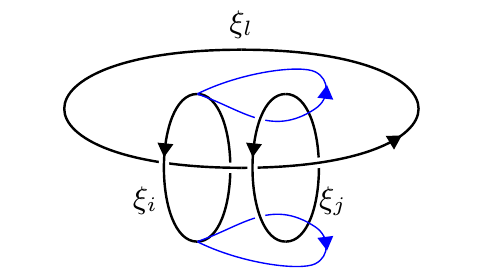}
\caption{The mutual braiding of loop excitations $\xi_i$, $\xi_j$ threaded by another loop $\xi_l$. The trajectory of two points on $\xi_i$ swept out as it braids around $\xi_j$ is shown in blue.}
\label{fig:3loop}
\end{figure}

In 3+1D, the partition functions are related to loop-braiding statistics. In particular, they are related to the so-called 3-loop braidings\cite{WangLevin2014,WangLevin2015}, shown in Figure \ref{fig:3loop}. Let $\xi_i$ now be loop excitations corresponding to group elements $e_i$. The invariants can be defined as the following:
\begin{enumerate}
\itemsep0em 
\item $\Theta_{i,l}$ is obtained from exchanging of two $\xi_i$ loops $N_i$ times, while threaded through by $\xi_l$.
\item $\Theta_{ij,l}$ is obtained from braiding loops $\xi_i$ around $\xi_j$ $N^{ij}$ times, while threaded through by $\xi_l$.
\item $\Theta_{ijk,l}$ is obtained from braiding $\xi_i$ around $\xi_j$, $\xi_k$, then $\xi_j$, $\xi_k$ again in the reverse direction, all while threaded through by $\xi_l$.
\end{enumerate}

Comparing to the explicit expressions in Ref. \onlinecite{WangLevin2015}, the partition functions and these braidings statistics are related via
\begin{subequations}
\begin{align}
\exp i\Theta_{i,l} &= Z_{i,l},\\
\exp i\Theta_{ij,l} &= \frac{Z_{ij,l}}{Z_{i,l}^{N^{ij}/N_{il}} Z_{j,l}^{N^{ij}/N_{jl}}},\\
\exp i\Theta_{ijk,l} &= 1/Z_{ijk,l}.
\end{align}
\end{subequations}
Similarly to the 2+1D case, we can interpret $Z_{ij,l}$ as the phase obtained by exchanging two $\xi_i \times \xi_j$ loops $N^{ij}$ times while being threaded by $\xi_l$. We can also see that the slant product on one group element corresponds to the threaded loop in the 3-loop braiding process. The latter fact is discussed in detail in Refs. \onlinecite{WangLevin2014,WangLevin2015}. Furthermore, the manifolds should correspond to surgery on $S^4$ using a surface link created by closing up the worldsheets of the loop excitations.

\subsection{Braidings in 2+1D Fermionic SPT Phases}
The braiding statistics of gauged interacting fermionic SPT phases in 2+1D has been studied in general in Ref. \onlinecite{WangLinGu2017}. These considerations include phases that are beyond those described by the Gu-Wen model. The 3+1D braidings statistics are much more complicated and have only been partially studied\cite{ChengTantiWang2017}, so we will only make a few remarks towards the end. 

First, let us review the notation used. The group considered in Ref. \onlinecite{WangLinGu2017} is $G^f =\mathbb Z_{2m}^f \times G$ for some positive integer $m$. Here, we restrict to the case where $m=1$. That is, $G^f =\mathbb Z_2^f \times G$.  We will use the index $0$ to denote the subgroup $\mathbb Z_2^f$. In this notation, $N_0=2$ and all the other bosonic particles are labeled by indices $i,j,k,... = 1,...,K$. In addition to the excitations $\xi_i$, we also have the excitation corresponding to gauging the fermionic parity operator called $\xi_0$. For simplicity, we will use Greek indices to include the particle $\xi_0$. That is $\mu,\nu,\lambda,...=0,1,...,K$. The definitions of the braiding invariants are slightly modified in the fermionic case. For Gu-Wen phases,
\begin{enumerate}
\itemsep0em 
\item $\Theta_0 = 2\pi N_i s_{\xi_0}$.
\item $\Theta_i = 2\pi \frac{2N_i}{N_{0i}} s_{\xi_i}$.
\item $\Theta_{\mu\nu}$ is obtained from braiding $\xi_\mu$ around $\xi_\nu$ $N^{\mu\nu}$ times.
\item $\Theta_{\mu\nu\lambda}$ is obtained from braiding $\xi_\mu$ respectively around $\xi_\nu$, $\xi_\lambda$, followed by $\xi_\nu$, $\xi_\lambda$ again, but in the reverse direction.
\end{enumerate}

We will now establish a relation between the partition functions of the Gu-Wen model with inputs $(\omega_3,\beta_2)$ and the invariants from braiding statistics as follows: First, let us define the following invariants:
\begin{enumerate}
\itemsep0em 
\item $\mathcal Z_i(\eta)$ is the partition function on $L(N_i;1)$ with spin structure $\eta=0,1$.
\item $\mathcal Z_{0i}(\eta)$ is the partition function on $L(N^{0i};1)$ with spin structure $\eta=0,1$.
\item $\mathcal Z_{ij}(\eta)$ is the partition function on $L(N^{ij};1)$ with spin structure $\eta=0,1$.
\item $\mathcal Z_{ijk}(\eta_i,\eta_j,\eta_k)$ is the partition function on $T^3$ with spin structure $\eta_i,\eta_j,\eta_k=0,1$ on each circle.
\end{enumerate}
Then, we claim that the following relations hold:
\begin{subequations}
\begin{align}
\label{equ:thetai}
\exp i\Theta_{i}  &=\mathcal Z_i(0),\\
\label{equ:theta0i}
\exp i\Theta_{0i} &=\frac{\mathcal Z_{0i}(1)}{\mathcal Z_{0i}(0)},\\
\label{equ:thetaij}
\exp i\Theta_{ij} &= \frac{\mathcal Z_{ij}(0)}{\mathcal Z_{i}(0)^{\frac{N^{ij}}{N_i}} \mathcal Z_{j}(0)^{\frac{N^{ij}}{N_j}}},\\
\label{equ:theta0ij}
\exp i\Theta_{0ij} &=\frac{\mathcal Z_{ijk}(0,0,1)}{\mathcal Z_{ijk}(0,0,0)},\\
\label{equ:thetaijk}
\exp i\Theta_{ijk}& = 1/\mathcal Z_{ijk}(0,0,0).
\end{align}
\end{subequations}
Note that in equation \eqref{equ:theta0i}, $N^{0i}$ is even and so $L(N^{0i};1)$ always admits two spin structures. Also, although $\Theta_{0ij}$ depends on the partition function $\mathcal Z_{ijk}$, the final answer will not depend on $k$, as in the bosonic case.\\
Equations \eqref{equ:thetai},  \eqref{equ:thetaij}, and  \eqref{equ:thetaijk} are equivalent to their bosonic counterparts, while equations \eqref{equ:theta0i}, \eqref{equ:theta0ij} are invariants that we have calculated previously in equations \eqref{equ:spinlens} and \eqref{equ:spinT3} to distinguish elements of $\mathcal H^2(G,\mathbb Z_2)$ in the Gu-Wen model.

To support our claim, we will show that if the braiding statistics are defined as such, then all the constraints from braiding statistics are satisfied. The general constraints can be found in Ref. \onlinecite{WangLinGu2017}. Here, we specialize the constraints to the case where the group is $G^f= \mathbb Z_2^f \times G$ and eliminate the dependence of $\Theta_0$, $\Theta_{00}$, $\Theta_{000}$, and $\Theta_{00i}$ \footnote{We cannot establish a relation of these braidings with these partition functions because they turn out to be all zero in the Gu-Wen model. However, we will conjecture the relations towards the end.}.
\begin{subequations}
\begin{align}
\label{equ:constraint1}
N_{0i}\Theta_{0i}&=\frac{N^{0i}(N^{0i}-1)}{2} \Theta_{0ii},\\
\label{equ:constraint2}
N_{0ij}\Theta_{0ij}&=0, \\
\label{equ:constraint3}
N_{0i}\Theta_{iii}&=0,\\
\label{equ:constraint4}
\Theta_{iij}&=\Theta_{0ij},\\
\label{equ:constraint5}
N_i\Theta_i &=\begin{cases}
 \Theta_{0i}-\frac{N_i^2(N_i-1)}{4} \Theta_{iii} ;&\text{$N_i$ even}\\
0;&\text{$N_i$ odd}
\end{cases},\\
\label{equ:constraint7}
N_{ij}\Theta_{ij}&=\frac{N^{ij}(N^{ij}-1)}{2} \Theta_{0ij}, &\\
\label{equ:constraint8}
\Theta_{ii}&=\begin{cases}
2\Theta_i + \frac{N_i (N_i-1)}{2}\Theta_{iii};& \text{$N_i$ even}\\
\Theta_i;& \text{$N_i$ odd}
\end{cases},\\
\label{equ:constraint9}
N_{ijk}\Theta_{ijk} &=0. &
\end{align}
\end{subequations}
There is also the constraint that $\Theta_{\mu\nu}$ and $\Theta_{\mu\nu\lambda}$ is invariant under cyclic permutations of the indices, but these are automatically satisfied from the definitions of the partition functions. We can show that constraints \eqref{equ:constraint1} - \eqref{equ:constraint5} are satisfied explicitly. Unfortunately, we cannot derive an explicit expression of constraints \eqref{equ:constraint7} - \eqref{equ:constraint9} but we will instead show for certain groups that they are satisfied.

\begingroup
To begin, let us use the canonical cocycle for $\beta_2$ in equation \eqref{equ:canonicalbeta2}. However, we will not assume that $R_{ii}$ is zero in contrast to the previous section.\\
The expressions for $\Theta_{0i}$ and $\Theta_{0ij}$ have been worked out previously:
\allowdisplaybreaks
\begin{align}
\label{equ:theta0iform}
\Theta_{0i} &= \frac{2 \pi}{N_{0i}} R_i, &&\\
\Theta_{0ii} &= 0, &&\\
\label{equ:theta0ijform}
\Theta_{0ij} &= \frac{2\pi}{N_{ij0}} (R_{ij}+R_{ji}) && \text{for $i \ne j$}.
\end{align}
Inserting these expressions into constraints \eqref{equ:constraint1} and \eqref{equ:constraint2}, we see that they are satisfied.\\
Next, we obtain the expression for $\Theta_{ijk}$. This can be obtained by dimensional reduction. First, we can reduce $(\omega,\beta)$ along the $k$ direction of the torus using equation \eqref{equ:gamma2}:
 \begin{align*}
 \left ((-1)^{i_{e_k} (\beta \cup_1 \beta) +i_{e_k} \beta \cup i_{e_k} \beta +i_{e_k}\beta \cup_1 \beta+\eta_k \beta}i_{e_k}\omega , i_{e_k} \beta \right ).
 \end{align*}
Then, we reduce along the $j$ direction using equation \eqref{equ:gamma1} to get
\endgroup
\begin{widetext}
 \begin{align*}
&\left ((-1)^{i_{e_j}[i_{e_k} (\beta \cup_1 \beta) +i_{e_k} \beta \cup i_{e_k} \beta +i_{e_k}\beta \cup_1 \beta+\eta_k\beta] + i_{e_j}i_{e_k} \beta \cup i_{e_k} \beta+\eta_ji_{e_k}\beta }i_{e_j}i_{e_k}\omega , i_{e_j}i_{e_k} \beta \right )\\
&=\left ((-1)^{i_{e_j}i_{e_k}( \beta \cup_1 \beta) +i_{e_j}(i_{e_k}\beta \cup_1 \beta) + i_{e_j}i_{e_k} \beta \cup i_{e_k} \beta+\eta_k i_{e_j}\beta+\eta_j i_{e_k}\beta } i_{e_j}i_{e_k}\omega , i_{e_j}i_{e_k} \beta \right ).
  \end{align*}
  Note that we discarded $i_{e_j} (i_{e_k} \beta \cup i_{e_k} \beta)$ since it is zero \footnote{by symmetry, or using the fact that it is a coboundary of $i_{e_j}i_{e_k}\beta \cup_1 i_{e_k}\beta$, which is zero}. Evaluating this on the circle with $e_i$ then gives the partition function
 \begin{align}
\mathcal Z_{ijk}(\eta_i,\eta_j,\eta_k)=(-1)^{i_{e_i}i_{e_j}i_{e_k}( \beta \cup_1 \beta) +i_{e_i}i_{e_j}(i_{e_k}\beta \cup_1 \beta)+i_{e_j}i_{e_k} \beta i_{e_i}i_{e_k}\beta+\eta_ki_{e_i}i_{e_j}\beta+\eta_j i_{e_i}i_{e_k}\beta + \eta_ii_{e_j}i_{e_k}\beta } i_{e_i}i_{e_j}i_{e_k}\omega
  \end{align}
  \end{widetext}
If we set all spin structures on the circles to be trivial, we can read off the Grassmann integral for the torus, which we call $\sigma_{ijk}(\beta)$
 \begin{align}
\sigma_{ijk}(\beta)&= (-1)^{i_{e_i}i_{e_j}i_{e_k}( \beta \cup_1 \beta) +i_{e_i}i_{e_j}(i_{e_k}\beta \cup_1 \beta)+i_{e_j}i_{e_k} \beta i_{e_i}i_{e_k}\beta}.
\end{align}
It turns out that $\sigma_{ijk}$ is symmetric under cyclic permutations of $i$, $j$, and $k$, though it is not manifest from the cup-$i$ product expression. Evaluating this using the canonical cocycles of $\beta$, we obtain three different cases:
\begin{align}
\sigma_{iii}(\beta)=& 1, &&\\
\sigma_{iij}(\beta)=& \exp \frac{2\pi i}{N_{ij0}} (R_{ij}+R_{ji}); \ \ \ \ \ \ \ \ i \ne j,\\
    \sigma_{ijk}(\beta) =& \exp \frac{4\pi i}{N_{ik0}N_{jk0}} (R_{ik}+R_{ki})(R_{jk}+R_{kj})\nonumber\\
     & \cdot (\text{cyclic permutations}); \ \ \ i \ne j \ne k.
  \end{align}

Let us consider some special cases. For $\Theta_{iii}$, the Grassmann integral is one and $i_{e_i}i_{e_i}i_{e_i}\omega=1$ from the explicit expression. Therefore,
\begin{equation}
\label{equ:thetaiiiform}
\Theta_{iii}=0,
\end{equation}
and so constraint \eqref{equ:constraint3} is satisfied. For $\Theta_{iij}$, the bosonic integral is also one for the same reason. Therefore,
\begin{align}
\Theta_{iij}&=\frac{2\pi}{N_{ij0}} (R_{ij}+R_{ji}).
  \end{align}
This, along with equation \eqref{equ:theta0ijform}, satisfies constraint \eqref{equ:constraint4}.

Next, we will evaluate $N_i\Theta_i$ in constraint \eqref{equ:constraint5}, which is equivalent to calculating $\mathcal Z_i(0)^{N_i}$. Hence, we need to evaluate each of the three terms in the partition function to the power of $N_i$ .
First, for the bosonic integral, we have
\begin{align}
Z_i^{N_i}&=\prod_{n=1}^{N_i} \omega(e_i,ne_i,e_i)^{N_i} \nonumber\\
&=\prod_{n,m=1}^{N_i} d\omega(e_i,ne_i,e_i,me_i)\nonumber\\
&= \exp \pi i \sum_{n,m=1}^{N_i} \beta(e_i,ne_i) \beta(e_i,me_i) \nonumber \\
&=\exp \pi i R_i.
\end{align}
In the second line, we used the property of the coboundary operator, and substituted for $\beta$ using the Gu-Wen equation in the third line.

\begin{figure}[h!]
\centering
\includegraphics{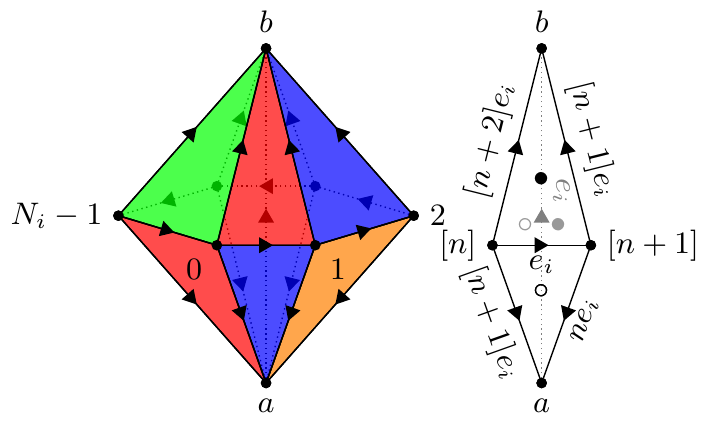}
\caption{Global ordering of vertices on the lens space $L(N_i;1)$. The Grassmann integrand for each tetrahedron is shown on the right, with $[n]$ defined modulo $N_i$.}
\label{fig:grassmannlens}
\end{figure}

 Next, let us evaluate the Grassmann integral. We first assign a global ordering of vertices $0,1,...,N_i,a,b$ to the lens space as shown in Figure \ref{fig:grassmannlens}. Writing down only the integrand, the Grassmann integral on the lens space is
\begin{align}
\sigma_i(\beta)= \prod_{n=N_i-1}^{0} [\theta_{[n+1]ab}^{\beta(ne_i,e_i)}\theta_{[n][n+1]b}^{\beta(e_i,[n+1]e_i)}\bar\theta_{[n]ab}^{\beta([n+1]e_i,e_i)}\bar\theta_{[n][n+1]a}^{\beta(e_i,ne_i)}],
\end{align}
where $[n]$ is defined modulo $N_i$. Notice that we have ordered the tetrahedra in descending order of $n$. That is, from east to west in the figure. First, we want to pair up the triangles that are adjacent in the figure. We can do so by swapping $\bar\theta_{[n]ab}^{\beta([n+1]e_i,e_i)}$ (on each left triangle in the tetrahedra) with $\bar\theta_{[n][n+1]a}^{\beta(e_i,ne_i)}$ so that we can pair it up with $\theta_{[n]ab}^{\beta([n+1]e_i,e_i)}$ (on each right triangle in the tetrahedra to the left), which is the first variable in the square bracket to the right. Doing so gives us a sign factor of $(-1)^{\sum_{n=1}^{N_i} \beta([n+1]e_i,e_i)\beta(e_i,ne_i)}$. Note that we also need to move the last term $\bar\theta_{0ab}^{\beta(e_i,e_i)}$ to the front so that it is paired up with $\theta_{0ab}^{\beta(e_i,e_i)}$. This costs an extra $(-1)^{\beta(e_i,e_i)}$. We can now integrate out all the adjacent pairs and we are left with
\begin{align*}
 \prod_{n=N_i-1}^{0} (\theta_{[n][n+1]b}^{\beta(e_i,[n+1]e_i)}\bar\theta_{[n][n+1]a}^{\beta(e_i,ne_i)}).
\end{align*}
Next, we swap the positions in each bracket, which costs $(-1)^{\sum_{n=1}^{N_i} \beta(e_i,[n+1]e_i)\beta(e_i,ne_i)}$ and move $\theta_{01b}^{\beta(e_i,e_i)}$ to the front, which costs $(-1)^{\beta(e_i,e_i)}$. We can now regroup the terms as
\begin{align*}
\prod_{n=N_i-1}^{0} (\theta_{[n+1][n+2]b}^{\beta(e_i,ne_i)}\bar\theta_{[n][n+1]a}^{\beta(e_i,ne_i)}).
\end{align*}
Recall that for lens spaces, the triangles $([n][n+1]a)$ and $([n+1][n+2]b)$ are identical. Therefore, we can integrate everything out upon swapping each pair in the brackets. This costs $(-1)^{\sum_{n=1}^{N_i} \beta(e_i,ne_i)\beta(e_i,ne_i)}$. Putting all sign factors together, the Grassmann integral gives
\begin{align}
&(-1)^{\sum_{n=1}^{N_i}\beta(e_i,ne_i) \left [\beta([n+1]e_i,e_i) + \beta(e_i,[n+1]e_i)+ \beta(e_i,ne_i) \right]} \nonumber\\
=&\exp \pi i \sum_{n=1}^{N_i} \beta(e_i,ne_i)\beta(ne_i,e_i)=\exp \pi i R_i.
\end{align}
\begin{figure}[H]
\centering
\includegraphics{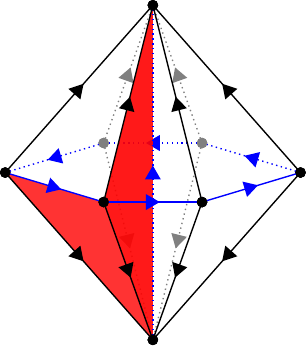}
\caption{The set $S$ shown in blue for the lens space. The set $E$ for the trivial spin structure is shown in red. One can obtain $E$ for the non-trivial spin structure by adding the 2-cycle shown in Figure \ref{fig:lensspin}.}
\label{fig:Striviallens}
\end{figure}
Finally, we calculate the spin structure term $E_i(\beta)$. Using the rules given in Table \ref{tab:m} or equation \eqref{equ:S2_1D}, we can obtain the set $S$ as the links around the great circle and the one going up vertically from $a$ to $b$. This is shown in blue in Figure \ref{fig:Striviallens}. It is a boundary of the area shaded in red for the trivial spin structure. Hence, the spin structure term is
\begin{equation}
\label{equ:Eibeta}
E_i(\beta)=(-1)^{\beta(e_i,ne_i)+\beta(ne_i,e_i)}=1.
\end{equation}
Together, we have
\begin{align}
\mathcal Z_i(0)^{N_i}  &= Z_i^{N_i} \sigma_i(\beta)^{N_i} = \exp \left (\pi i(N_i+1) R_i \right).
\end{align}
As a result,
\begin{align}
N_i\Theta_i   &= \pi(N_i+1) R_i.
\end{align}
This equation, along with equations \eqref{equ:theta0iform} and \eqref{equ:thetaiiiform}, satisfies the constraint \eqref{equ:constraint5}.

We were not able to obtain a closed form for $\Theta_{i}$, $\Theta_{ij}$ or $\Theta_{ijk}$. This is because we need an explicit expression for the cochain $\omega$ that satisfies the Gu-Wen equation \eqref{equ:domega} for any symmetry group $G$, which isn't possible in general. For this reason, we need to solve the equation explicitly for a given group $G$ by first writing them as linear modular equations and solving them using the Smith decomposition. We show explicitly in Appendix \ref{app:smith} that constraints \eqref{equ:constraint7} - \eqref{equ:constraint9} are satisfied for some example groups, namely, $\mathbb Z_2 \times \mathbb Z_4$, $\mathbb Z_2 \times \mathbb Z_6$, and $\mathbb Z_2 \times \mathbb Z_2 \times \mathbb Z_2$. In general, we believe that these constraints are also satisfied for any finite abelian unitary group $G$.

Similarly to the bosonic case, one can ask for the interpretation of the partition functions with non-trivial spin structure in terms of braiding. Assuming that the relations between braiding statistics and partition functions that we have established are true, flipping the spin structure should correspond to the braiding with the original particle fused with $\xi_0$. For example, equations \eqref{equ:theta0ij} and \eqref{equ:thetaijk} together give
\begin{align}
 \mathcal Z_{ijk}(0,0,1)^{-1} =\exp i \left (\Theta_{ijk}+\Theta_{0ij} \right ).
\end{align}
Since $\Theta_{0ij}=\Theta_{ij0}$, this partition function corresponds to the braiding of $\Theta_{ijk}$ with $\xi_k$ replaced by $\xi_k \times \xi_0$. As another example, combining equations \eqref{equ:thetai} and \eqref{equ:theta0i} gives
\begin{align}
\mathcal Z_{0i}(1)= \exp i \left (\Theta_{0i}+\frac{N^{0i}}{N_i}\Theta_{i}\right ).
\end{align}
This partition function can be related to the topological spin of $\xi_i \times \xi_0$ in an identical analysis as that of $Z_{ij}$ in the bosonic case upon the condition that $\Theta_{0}=0$.

With an understanding of the interpretation of the partition functions, one could now ask if there are partition functions that correspond to $\Theta_0$, $\Theta_{00}$, $\Theta_{000}$, and $\Theta_{00i}$ that we have eliminiated earlier. First, let us generalize the definitions of the partition functions in the following way:
\begin{enumerate}
\itemsep0em 
\item $\mathcal Z_\mu(\eta)$ is the partition function on $L(N_\mu;1)$ with spin structure $\eta=0,1$.
\item $\mathcal Z_{\mu\nu}(\eta)$ is the partition function on $L(N^{\mu\nu};1)$ with spin structure $\eta=0,1$.
\item $\mathcal Z_{\mu\nu\lambda}(\eta_\mu,\eta_\nu,\eta_\lambda)$ is the partition function on $T^3$ with spin structure $\eta_\mu,\eta_\nu,\eta_\lambda=0,1$.
\end{enumerate}
Furthermore, assume the convention that $e_0=0$, the identity element in $G^f$. Then, we conjecture that the following relations between braiding statistics and partition functions:
\begin{subequations}
\begin{align}
\exp i\Theta_{\mu}  &=\mathcal Z_\mu(\delta_{\mu,0}),\\
\exp i\Theta_{\mu\nu} &= \frac{\mathcal Z_{\mu\nu}(\delta_{(\mu+\nu),0})}{\mathcal Z_{\mu}(\delta_{\mu,0})^{\frac{N^{\mu\nu}}{N_\mu}} \mathcal Z_{\nu}(\delta_{\nu,0})^{\frac{N^{\mu\nu}}{N_\nu}}},\\
\exp i\Theta_{\mu\nu\lambda}& = 1/\mathcal Z_{\mu\nu\lambda}(\delta_{\mu,0},\delta_{\nu,0},\delta_{\lambda,0}).
\end{align}
\end{subequations}
Here, $\delta_{\mu,0}$ is the Kronecker delta, which flips the spin structure only when $\mu=0$.

One can check that the given equations reproduce the relations \eqref{equ:thetai} - \eqref{equ:thetaijk}. Furthermore, $\Theta_0$,$\Theta_{00}$,$\Theta_{000}$, and $\Theta_{00i}$ are all zero for the Gu-Wen partition function, which agrees with Ref. \onlinecite{WangLinGu2017}. This is also why we have eliminated the dependence of these phases from our constraints. For general fermionic SPT phases, these quantities are not necessarily zero, so it would be interesting to check whether the conjectured relations hold in general or not. It would also be interesting to examine if the partition function with a $\xi_0$ particle defect in a spin-TQFT can be related to the partition function of the manifold obtained by surgery along that knot with the spin structure also flipped.

We now remark on extending this correspondence to loop-braiding statistics for 3+1D fermionic SPT phases. The abelian braiding statistics have been studied in Ref. \onlinecite{ChengTantiWang2017}. It would be interesting to check whether the extension of this hypothesis applies to 3+1D. That is, whether we can associate partition functions to braiding statistics in a systematic way by checking whether all the braiding constraints are satisfied. Furthermore, these partition functions could possibly offer us hints on the form of the non-abelian loop braiding constraints.

\section{Conclusion}
In summary, we have explored dimensional reduction as a method of studying SPT phases by reducing them to a lower dimension. A one-dimension reduction corresponds to compactifying over a circle. In addition, the general dimensional reduction using the general notion of the slant product provides further information of the SPT phases that would have otherwise been lost in the regular dimensional reduction, as we have shown in distinguishing the equivalence classes of $\mathcal H^5(G,U(1))$. In addition, we derived the dimensional reduction procedure over a circle for Gu-Wen SPT phases. We leave the general dimensional reduction for Gu-Wen phases for future studies.

With the help of dimensional reduction, we were able to construct topological invariants defined as the partition function on closed oriented manifolds  equipped with certain flat connections that were able to distinguish all the bosonic SPT phases in the group cohomology model. For Gu-Wen fermionic SPT phases, the manifolds must be equipped with a spin structure. We then used the spin structure dependence of the manifold to extract the cocycle $\beta$. In 1+1D, we only need the torus. In 2+1D, we need lens spaces and the 3-torus. The manifolds in 3+1D are, by virtue of dimensional reduction, the product of the 3-manifolds mentioned and a circle. We were not able to identify all the manifolds in 4+1D, but we believe they can be realized as generators of the cobordism group with the restriction that the cobordism preserves the flat connection.

Finally, we showed that for bosonic SPT phases, these invariants are equivalent to those obtained from braiding statistics of the gauged theories. For Gu-Wen SPT phases, we believe they are also equivalent and have provided evidence to support our claim. In the latter case, it would be interesting if the equivalence can be established more rigorously via surgery on spin manifolds.

Some future directions would be to see whether the manifolds mentioned are enough to classify fermionic SPT phases that are beyond the classification of the Gu-Wen model such as charge $2m$-superconductors\cite{Wang2016} or phases with the so-called Majorana edge modes\cite{FidkowskiKitaev2011,Bhardwajetal2017,KapustinThorngren2017,WangGu2017}. It would also be interesting to see if this framework is applicable to SPT phases with non-unitary symmetries such as time-reversal on non-orientable manifolds\cite{Barkeshlietal2016,Bhardwaj2017}, including those that are beyond the group cohomology classification\cite{VishwanathSenthil2013,Kapustin2014}, or point-group symmetries\cite{Shiozakietal2017,Shapourianetal2017,Songetal2017}.

\begin{acknowledgements}
I owe a huge debt of gratitude to Davide Gaiotto for suggesting this project, and for his sharp insights that have continuously guided me through the numerous challenges of this work. I am also very grateful to Chenjie Wang for his patience in explaining to me many physical concepts and detailed explanations of his works. Furthermore, I would like to thank them both for their supervision and encouragement during my Master's program at Perimeter Institute, and for careful readings and suggestions on this manuscript. I would also like to thank Zheng-Cheng Gu, Theo Johnson-Freyd, Surya Raghavendran, and Yehao Zhou for fruitful discussions, Meng Cheng and Chenjie Wang for collaboration on a related work, and Jingxiang Wu for valuable comments. This work was supported by the Perimeter Institute for Theoretical Physics and by the Marsland family through an Honorary PSI Scholarship Award.  Research at Perimeter Institute is supported by the Government of Canada through the Department of Innovation, Science and Economic Development Canada and by the Province of Ontario through the Ministry of Research, Innovation and Science.
\end{acknowledgements}

\appendix

\section{$\mathbb Z_2$ Homology and the Second Stiefel-Whitney Class} \label{app:w_2}
First, let us discuss the $\mathbb Z_2$ cohomology. Given a simplicial complex of a manifold $\mathcal M$, an $n$-cochain is defined as a function from an $n$-simplex in $\mathcal M$ to $\mathbb Z_2$. The set of all functions forms a group $C^n(\mathcal M,\mathbb Z_2)$. Similarly to group cohomology\footnote{Note that we use regular symbols for $C$, $Z$, $B$, and $H$, in contrast to group cohomology, where we use calligraphic symbols $\mathcal C$, $\mathcal Z$, $\mathcal B$, and $\mathcal H$.}, we can define $n$-cocycles and $n$-coboundaries using a coboundary operator $d$. The form of the coboundary operator is not important for this discussion. These two objects form groups $Z^n(\mathcal M,\mathbb Z_2)$ and $B^n(\mathcal M,\mathbb Z_2)$ respectively. The $n^{th}$ cohomology group of this manifold is
\begin{equation}
 H^n(\mathcal M,\mathbb Z_2) = Z^n(\mathcal M,\mathbb Z_2)/ B^n(\mathcal M,\mathbb Z_2).
\end{equation}
The second Stiefel-Whitney class $[w_2]$ is a certain element of $H^2(\mathcal M,\mathbb Z_2)$, which we represent via a 2-cocycle $w_2$. On spin manifolds, a unique property is that $w_2$ is a coboundary. As a result, there exists a 1-cochain $\eta$ such that $d\eta=w_2$. This cochain is not unique, since it can be added by any 1-cocycle. Hence, the different $\eta$'s are classified by the group $H^1(\mathcal M,\mathbb Z_2)$. The choice of $\eta$ reflects the choice of spin structure on the manifold

A more visual way to view the second Stiefel-Whitney class is to consider its Poincar\'e dual. Consider a $\mathbb Z_2$ simplicial complex in the manifold $\mathcal M$. An $n$-chain is a collection of $n$-simplices from the simplicial complex. $n$-chains form a group $C_n(\mathcal M, \mathbb Z_2)$, where they add modulo 2. Note that for $\mathbb Z_2$, we can ignore the orientation of the simplices. The boundary operator $\partial$ takes $n$-chains to $(n-1)$-chains. The definition of $\partial$ acting on an $n$-simplex is
\begin{align}
\partial(012...n) &= \sum_{i=0}^{n} \left (012...\hat{i}...(n-1)n \right) (\text{mod 2}).
\end{align}
This is identical to equation \eqref{equ:boundary}, but with addition instead of multiplication.
Next, we define $n$-cycles as $n$-chains that have no boundary. They form a subgroup $Z_n(\mathcal M, \mathbb Z_2)$. Furthermore, $n$-boundaries are $n$-cycles who are themselves boundaries. These form a subgroup $B_n(\mathcal M, \mathbb Z_2)$. The $n^{th}$ homology group is then the quotient group
\begin{equation}
 H_n(\mathcal M,\mathbb Z_2) = Z_n(\mathcal M,\mathbb Z_2)/ B_n(\mathcal M,\mathbb Z_2).
\end{equation}
On an $n$-manifold, the cocycle $w_2$ is dual to an $(n-2)$-cycle $S$. We will call $S$ a chain-representative of $w_2$. On spin manifolds, $S$ is an $(n-2)$-boundary. Thus, there exists an $(n-1)$-chain $E$ such that $\partial E =S$. Similarly, $E$ is the chain-representative of $\eta$. The choice of $E$ is not unique, since one can always add a cycle to $E$. Furthermore, from the cocycle condition of $\beta$, the spin structure term $\int_E \beta$ in the Gu-Wen partition function is invariant under changing $E$ by a boundary. Hence, the equivalence class of the set $E$ reflects the choice of spin structure on $\mathcal M$. Now, consider two inequivalent choices $E$ and $E'$. Their difference must be a cycle that is not a boundary. This means that their difference is a representative of a non-trivial element of $H_{n-1}(\mathcal M,\mathbb Z_2)$, which is isomorphic to $H^{1}(\mathcal M,\mathbb Z_2)$.

Let us now show that the explicit expression of $S$ in different dimensions are given by equations \eqref{equ:S0_1D} - \eqref{equ:S3_1D} and prove that they are equivalent to definitions of Ref. \onlinecite{GuWen2014} given in Table \ref{tab:m}. A chain representative of $w_2$ for any triangulation of a manifold has been given explicitly in Ref. \onlinecite{goldstein1976}. Here, we will reproduce a simplified but equivalent definition of the chain representative.\\
Within a triangulation of an $n$-manifold, consider a $p$-dimensional subsimplex $s=(v_0v_1...v_p)$ which is contained in a $k$-dimensional subsimplex $t=(012...k)$ (implicitly, $p \le k \le n$). $s$ is defined to be \underline{\smash{regular}} in $t$ if
\begin{enumerate}
\itemsep0em 
\item $v_0=0$.
\item $v_{i+1} =v_i +1$ for $i$ odd.
\item $v_p=k$ for odd $p$.
\end{enumerate}
Let $\partial_p(t_k)$ denote all the regular subsimplices of dimension $p$ in $t_k$, then the chain-representative for the $2^\text{nd}$  Stiefel-Whitney class $w_2$ of a triangulated $n$-manifold is given (mod 2) by
\begin{align}
\label{equ:Sdef}
S&=  \sum_{k= n-2}t_k +  \sum_{k= n-1}\partial_{(n-2)}(t_k) + \sum_{k = n}\partial_{(n-2)}(t_k).
\end{align}
We will call the three terms on the right hand side the sets (or equivalently chains) $S_1$, $S_2$ and $S_3$, respectively. One can see that $S_1$ is already present in the definition of Ref. \onlinecite{GuWen2014}. As a result, we only need to check that the sets $S_2$ and $S_3$ together gives the results listed in Table \ref{tab:m}.

In 0+1D, $S$ is obviously empty, since there are no objects of dimension $-1$.

\begin{figure}[h!]
\centering
\includegraphics{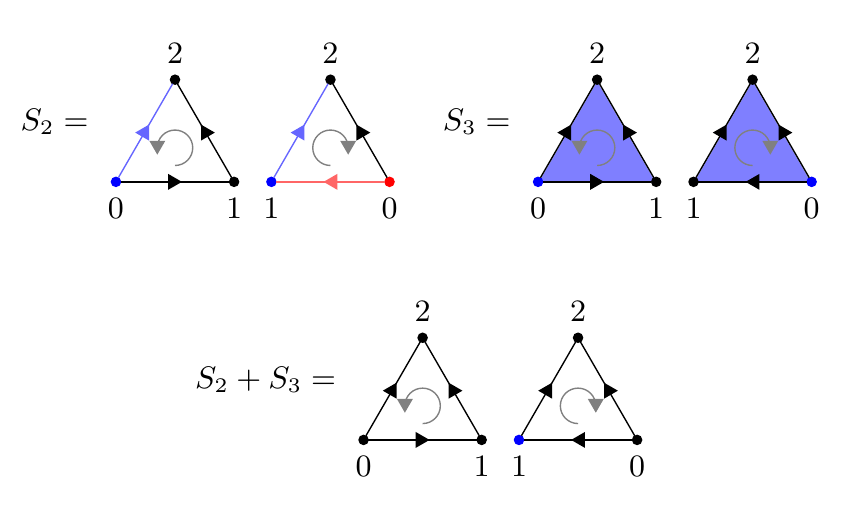}
\caption{Assignment of the sets $S_2$ and $S_3$ in 1+1D depending on orientation of the simplex to give the choice listed in Table \ref{tab:m}. $(+)$ is shown on the left, while $(-)$ is shown on the right. The different colors indicate from which subsimplex the contribution was from.}
\label{fig:2Dassignment}
\end{figure}

In 1+1D, $p=0$. We only need $v_0=0$ so the rules give
\begin{align}
 S=&\{\text{all 0-simplices}\} + \{\text{(0) in any 1-simplex}\}\nonumber\\
 \label{equ:Sdef1_1D}
  &+\{\text{(0) in any 2-simplex}\},
 \end{align}
 or pictorially,
 \begin{align}
 \label{equ:S1_1Dpic_app}
 S_1=&
 \raisebox{-.5\height}{\begin{tikzpicture}[scale=0.5]
\node[vertex,blue][label=below:$0$]  (0) at (0,0) {};
\end{tikzpicture}}, &
 S_2=&
\raisebox{-.5\height}{\begin{tikzpicture}[scale=0.5]
\node[vertex,blue][label=below:$0$]  (0) at (0,0) {};
\node[vertex][label=below:$1$]  (1) at (2,0) {};
\draw[->,lblue] (0) -- (1) node[midway, below] {};
\end{tikzpicture}} , &
S_3=&
   \raisebox{-.5\height}{\begin{tikzpicture}
   \begin{scope}[scale=0.6]
\filldraw[ fill=blue,   fill opacity=0.5,draw=white](0,0)--(2,0)--(1,3.46/2)--cycle;
\node[vertex,blue][label=below:$0$] (0) at (0,0) {};
\node[vertex][label=below:] (1) at (2,0) {};
\node[vertex][label=above:] (2) at (1,3.46/2) {};
\draw[->] (0) -- (1) node[midway, below] {}  node[midway, above] {};
\draw[] (1) -- (2) node[midway, left] {} node[midway, right] {};
\draw[->] (0) -- (2) node[midway, right] {} node[midway, left] {} ;
\end{scope}
\end{tikzpicture}},
 \end{align}
where the elements in each set are the blue vertices. Note that for $S_3$, one of the arrows is omitted to convey the fact that the triangle is orientation independent.
The vertices in $S_2$ are attached to links, each of which is always a side of two adjacent triangles. If we assign those to the triangle to the right of the link, then a $(+)$ triangle would get the vertex $(0)$, while a $(-)$ triangle would get $(1)$ and $(0)$. For each triangle, $S_3$ gives $(0)$ for both orientations, so $S_2+S_3$ only has  $(1)$ left from each $(-)$ triangle. This is illustrated in Figure \ref{fig:2Dassignment}.

Note that the triangle on the right of a certain 1-simplex is $(+)$ if that simplex is missing an odd number and $(-)$ if it is missing an even number. We will use this fact in higher dimensions.

\begin{figure}[h!]
\centering
\includegraphics{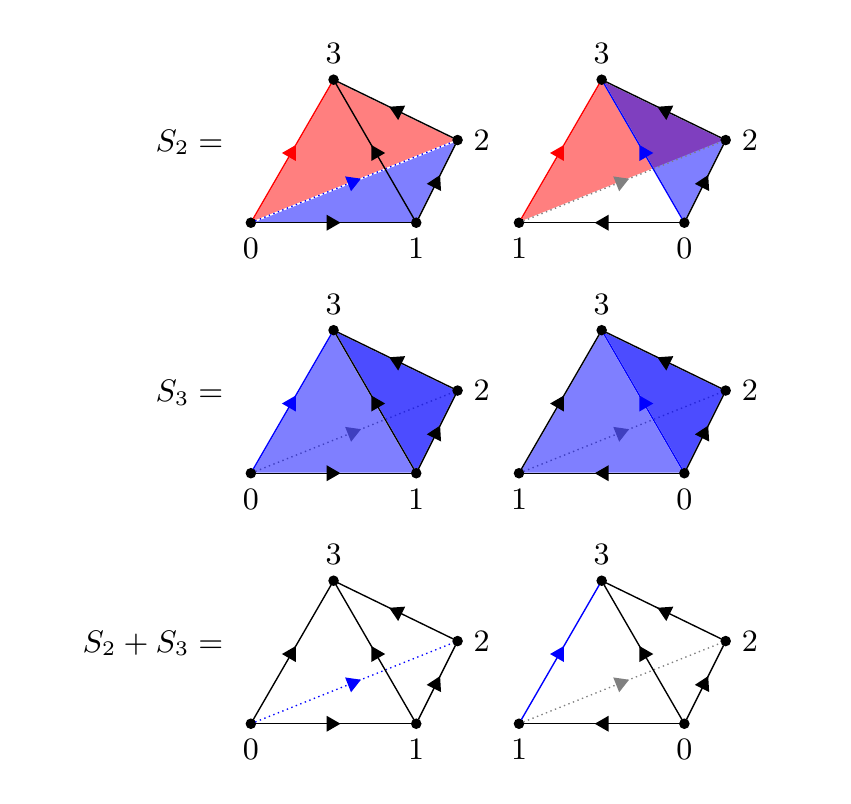}
\caption{Assignment of the sets $S_2$ and $S_3$ in 2+1D. $(+)$ is shown on the left, while $(-)$ is shown on the right. The colors indicate from which subsimplex the contribution was from.}
\label{fig:2_1Dassignment}
\end{figure}

In 2+1D, $p=1$. We need $v_0=0$ and $v_1=k$, so the rules give
\begin{align}
S=&\{\text{all 1-simplices}\} + \{\text{(02) in any 2-simplex}\} \nonumber\\
\label{equ:Sdef2_1D}
 &+ \{\text{(03) in any 3-simplex}\}
 \end{align}
Or pictorially,
 \begin{align}
 \label{equ:S2_1Dpic_app}
 S_1=&
\raisebox{-.5\height}{\begin{tikzpicture}[scale=0.5]
\node[vertex][label=below:$0$]  (0) at (0,0) {};
\node[vertex][label=below:$1$]  (1) at (2,0) {};
\draw[->,blue] (0) -- (1) node[midway, below] {};
\end{tikzpicture}} , &
S_2=&
   \raisebox{-.5\height}{\begin{tikzpicture}
   \begin{scope}[scale=0.6]
\filldraw[ fill=blue,   fill opacity=0.5,draw=white](0,0)--(2,0)--(1,3.46/2)--cycle;
\node[vertex][label=below:$0$] (0) at (0,0) {};
\node[vertex] (1) at (2,0) {};
\node[vertex][label=above:$2$] (2) at (1,3.46/2) {};
\draw[->] (0) -- (1) node[midway, below] {}  node[midway, above] {};
\draw[->] (1) -- (2) node[midway, left] {} node[midway, right] {};
\draw[->,blue] (0) -- (2) node[midway, right] {} node[midway, left] {} ;
\end{scope}
\end{tikzpicture}}, &
S_3=&
   \raisebox{-.5\height}{\begin{tikzpicture}
   \begin{scope}[scale=0.6]
\filldraw[ fill=blue,   fill opacity=0.5,draw=white](0,0)--(2,0)--(1,3.46/2)--cycle;
\filldraw[ fill=blue,   fill opacity=0.7,draw=white](2.5,1)--(2,0)--(1,3.46/2)--cycle;
\node[vertex][label=below:$0$] (0) at (0,0) {};
\node[vertex][label=below:] (1) at (2,0) {};
\node[vertex][label=above:] (2) at (2.5,1) {};
\node[vertex][label=above:$3$] (3) at (1,3.46/2) {};
\draw[->] (0) -- (1) node[midway, below] {}  node[midway, above] {};
\draw (1) -- (2) node[midway, left] {} node[midway, right] {};
\draw[->,densely dotted,gray] (0) -- (2) node[midway, right] {} node[midway, left] {} ;
\draw[->,blue] (0) -- (3);
\draw[->] (1) -- (3);
\draw[->] (2) -- (3);
\end{scope}
\end{tikzpicture}},
 \end{align}
where the elements in each set are the blue links. The links in $S_2$ are attached to triangles, each of which is always a side of two adjacent tetrahedra. We can assign the link depending on the orientation of those triangles with respect to the tetrahedra. This corresponds to assigning the link to $(+)$ if the triangle it is attached to is missing an odd number, and to $(-)$ if the triangle is missing an even number, as shown in Figure \ref{fig:2_1Dassignment}. From $S_2$, we assign $(02)$ from $(012)$ and $(03)$ from $(013)$ to $(+)$, and assign $(03)$ from $(023)$ and $(13)$ from $(123)$ to $(-)$. $S_3$ gives $(03)$ for both orientations, so the final result is $(02)$ from each tetrahedron in $(+)$ and $(13)$ from each tetrahedron in $(-)$.

In 3+1D, $p=2$. We need $v_0=0$, and $v_1$ and $v_2$ are adjacent so the rules give
\begin{align}
S^{3+1}=&\{\text{all 2-simplices}\} + \{\text{(012),(023) in any 3-simplex}\}\nonumber\\
 &+ \{\text{(012),(023),(034) in any 4-simplex}\}.
 \end{align}
The triangles in $S_2$ are attached to tetrahedra, each of which is always a side of two adjacent 4-simplices. We can assign a triangle to $(+)$ if the corresponding tetrahedron is missing an even number, and to $(-)$ if it is missing an odd number. Note that this assignment is opposite from the previous cases. From $S_2$, we assign $(012),(023)$ from $(0123)$, $(013),(034)$ from $(0134)$, and $(123),(134)$ from $(1234)$ to $(+)$. For $(-)$ we assign $(023),(034)$ from $(0234)$ and $(012),(024)$ from $(0124)$.  From $S_3$, we get $(012),(023),(034)$ for both orientations. Hence, we are left with $(013),(134),(123)$ for $(+)$ and $(024)$ for $(-)$.

We remark that the choice made in Gu-Wen is not unique. This comes from the fact that one can assign the regular subsimplices to the opposite orientation. That is, we can swap the assignments of the $(+)$ and $(-)$ triangles, which is equivalent to swapping the entries in the columns of Table \ref{tab:m}. A simple reasoning that this assignment is also valid comes from the fact that the original assignments satisfy the mirrored Pachner moves of those listed in Ref. \onlinecite{GuWen2014}. Accordingly, the new assignment must satisfy the Pachner moves listed in Ref. \onlinecite{GuWen2014} as well. As an example, we can see that we had to swap the assignments in the 3+1D case to match those given in Table \ref{tab:m}.

\section{Interpretation of $m$ in the Gu-Wen Model}\label{app:m}
In the original Gu-Wen model \cite{GuWen2014}, the function $m$ was defined to take in variables from the vertices. To distinguish it from our definition, let us call it $\tilde m$, a function from $G^{n-2}$ to $M$. Similarly, $\beta$ was defined to take in group elements from the vertices. We will call the corresponding function in that model $n$. $\tilde m$ and $n$ are related via $d\tilde m=n$. However, $\tilde m$ is not homogeneous and is therefore not a cochain. This prevents $n$ from being a coboundary.

To generalize the Gu-Wen partition function to include manifolds with non-trivial flat connections, we can relate $n$ to $\beta$ in our model in the same way that $\nu$ is related to $\omega$:
\begin{equation}
 \beta_{n-1}(g_1,...,g_{n-1})=n_{n-1}(1,g_1,g_1g_2,...,g_1g_2g_3 \cdots g_{n-1}).
 \end{equation}
On the other hand, there is actually no function $m$ that would satisfy $dm = \beta$, since $\beta$ is not necessarily a coboundary! As a result, $m$ in our model is merely a theoretical tool to help with the calculations, and can be completely removed in the end of the calculation on spin manifolds, on which we can rewrite $\int_{S} m = \int_{E} \beta$ where $\partial E = S$.

Indeed, one can define the partition function on spin-manifolds in a way that completely does not depend on $m$ as follows: the spin structure term is $(-1)^{\int_{E} \beta}$ where $\partial E$ is a chain representative of the $2^\text{nd}$ Stiefel-Whitney class $[w_2]$. Different choices of $E$ correspond to different spin structures. However, it is harder to perform the dimensional reduction under this definition. 

\onecolumngrid
\section{Cup-$i$ Products and Steenrod Squares}\label{app:cup}
The cup product is a bilinear map $ \cup: \mathcal C^m(G,M) \times \mathcal  C^n(G,M) \rightarrow \mathcal C^{m+n}(G,M)$ defined as
\begin{align}
 \alpha_m \cup \beta_n (g_1,...,g_{m+n}) =  \alpha_m(g_1,...,g_m)\beta_n(g_{m+1},...,g_{m+n}).
 \end{align}
It is also an operation in the level of cohomology. In the case where $M=\mathbb{Z}_2$, a generalization called the cup-$i$ product\cite{Steenrod1947} $\cup_i: \mathcal C^m(G,\mathbb{Z}_2) \times \mathcal C^n(G,\mathbb{Z}_2) \rightarrow \mathcal C^{m+n-i}(G,\mathbb{Z}_2)$ can be defined for all $i \ge 0$ with $\cup_0 \equiv \cup$. However, they are not operations in cohomology in general. Using multiplication as the group operation, the first two cup-$i$ products are
\begin{align}
 \alpha_m \cup_1 \beta_n (g_1,...,g_{m+n-1}) = \sum_{j=0}^{m-1}&  \alpha_m\left (g_1,...,g_j, \prod_{i=1}^ng_{j+i} , g_{j+m+1},...,g_{m+n-1}\right)\beta_n(g_{j+1},...,g_{j+n}),\\
 \alpha_m \cup_2 \beta_n (g_1,...,g_{m+n-2}) = \sum_{j=0}^{m-1}\sum_{k=1}^{n-1} &  \alpha_m\left (g_1,...,g_j,\prod_{i=1}^kg_{j+i}, g_{j+k+1},...,g_{m+k-1}\right)\nonumber \\
 &\cdot \beta_n \left (g_{j+1},...,g_{j+k},\prod_{i=1}^{m-1-j}g_{j+k+i},g_{m+k},...,g_{m+n-2}\right).
 \end{align}
Note that compared to the general expression in Ref. \onlinecite{Steenrod1947}, we have ignored all negative signs because of using $M=\mathbb{Z}_2$. These definitions are used for calculations. For manipulations, there is a useful relation between cup-$i$ products. The coboundary operator applied to a cup product obeys the Leibniz rule. For cup-$i$ products, the failure is amounted by
\begin{align}
\label{equ:dcup}
d(\alpha_m \cup_{i+1} \beta_n)  = d \alpha_m\cup_{i+1} \beta_n+ \alpha_m \cup_{i+1}d\beta_n +\alpha_m \cup_i \beta_n + \beta_n \cup_i \alpha_m.
 \end{align}
Here, it is implied that $\alpha_m\cup_i\beta_n$ is zero for $i<0$.\\
For the regular cup product, one can also check that the slant product acting on the cup product obeys a Leibniz rule
\begin{gather}
\label{equ:icup}
i_g (\alpha_m\cup \beta_n)  = i_g \alpha_m \cup\beta_n +\alpha_m \cup i_g \beta_n.
\end{gather}
However, we are not aware of a relation for $i_g$ acting on cup-$i$ products in general.

The cup-$i$ products are not operations in cohomology in general. However, the $m^\text{th}$ Steenrod square defined as
\begin{align}
Sq^m (\beta_n) = \beta_n \cup_{n-m} \beta_n
 \end{align}
for $m=0,...,n$ is an operation in cohomology, that is, cocycles are sent to cocycles, and coboundaries are sent to coboundaries. This claim can be checked using equation \eqref{equ:dcup}.\\ 
From the definition, we can see that $i_g$ commutes with the $m^\textit{th}$ Steenrod square only when $m \le n-1$ for $\beta_{n-1}$. However, we show in section \ref{dimredfermions} that the commutator $[i_g,Sq^2]$ is always a coboundary up to spacetime dimension 4. We believe that in general, $i_g$ and $Sq^2$ should commute in cohomology since it is necessary for dimensional reduction to hold (specifically, for equation \eqref{equ:reduced4} to hold.)

The explicit expressions of cup-$i$ products used in this paper are
\begin{align}
\alpha_1 \cup_1 \beta_2 (a,b)&= \alpha_1(ab)\beta_2(a,b),\\
\alpha_2 \cup_1 \beta_2 (a,b,c)&= \alpha_2(ab,c)\beta_2(a,b)+\alpha_2(a,bc)\beta_2(b,c),\\
\alpha_2 \cup_2 \beta_3 (a,b,c)&= \alpha_2(a,b)\beta_3(a,b,c)+\alpha_2(ab,c)\beta_2(a,b,c),\\
\alpha_3 \cup_2 \beta_3 (a,b,c,d)&= \alpha_3(a,b,c)\beta_3(a,bc,d)+\alpha_3(ab,c,d)\beta_3(a,b,cd)+(\alpha_3(a,b,c)+\alpha_3(a,bc,d))\beta_3(b,c,d).
\end{align}

\section{Grassmann Integral Manipulation}\label{app:grassmann}
The notation used in this paper is that anything in $\left [\quad \right]$ is Grassmann even. In other words, $[\theta_1^{\beta_1}\cdots  \theta_n^{\beta_n}]$ satisfies $\sum_{i=1}^n \beta_i \equiv 0$ (mod 2) i.e., it is a linear combination of cocycle conditions of $\beta$. Using this, the following manipulations are equivalent to collecting signs from swapping Grassmann numbers and then applying cocycle conditions:
\begin{align}
\theta_0^{\beta_0}[\theta_1^{\beta_1}\theta_2^{\beta_2}\cdots  \theta_n^{\beta_n}] &= [\theta_1^{\beta_1}\theta_2^{\beta_2}\cdots  \theta_n^{\beta_n}] \theta_0^{\beta_0},\\
[\theta_1^{\beta_1}\theta_2^{\beta_2}\cdots  \theta_n^{\beta_n}] &= (-1)^{\beta_1} [\theta_2^{\beta_2}\cdots  \theta_n^{\beta_n}\theta_1^{\beta_1}],\\
[\theta_1^{\beta_1}\theta_2^{\beta_2}\cdots  \theta_m^{\beta_m} \bar\theta_0^{\beta_0}][\theta_0^{\beta_0}\tilde\theta_1^{\tilde\beta_1}\tilde\theta_2^{\tilde\beta_2}\cdots  \tilde\theta_n^{\tilde\beta_n}] &= [\theta_1^{\beta_1}\theta_2^{\beta_2}\cdots  \theta_m^{\beta_m} \tilde\theta_1^{\tilde\beta_1}\tilde\theta_2^{\tilde\beta_2}\cdots  \tilde\theta_n^{\tilde\beta_n}].
\end{align}

\section{Calculation of 3+1D to 2+1D Dimensional Reduction for Gu-Wen SPT Phases}\label{app:4to3calculation}
First, we imagine a 4D prism with two tetrahedra as caps. For the positive orientation, we have the integrand
\begin{align*}
& [\theta_{1233'}^{\beta(b,c,g)}\theta_{0133'}^{\beta(a,bc,g)}\theta_{0123}^{\beta(a,b,c)}\bar\theta_{0233'}^{\beta(ab,c,g)}\bar\theta_{0123'}^{\beta(a,b,cg)}][\theta_{0123'}^{\beta(a,b,cg)}\theta_{022'3'}^{\beta(ab,g,c)}\bar\theta_{0122'}^{\beta(a,b,g)}\bar\theta_{012'3'}^{\beta(a,bg,c)}\bar\theta_{122'3'}^{\beta(b,g,c)}] \nonumber\\
& [\theta_{11'2'3'}^{\beta(g,b,c)}\theta_{012'3'}^{\beta(a,bg,c)}\theta_{011'2'}^{\beta(a,g,b)}\bar\theta_{01'2'3'}^{\beta(ag,b,c)}\bar\theta_{011'3'}^{\beta(a,g,bc)}][\theta_{00'1'3'}^{\beta(g,a,bc)} \theta_{01'2'3'}^{\beta(ag,b,c)}\bar\theta_{00'1'2'}^{\beta(g,a,b)}\bar\theta_{00'2'3'}^{\beta(g,ab,c)}\bar\theta_{0'1'2'3'}^{\beta(a,b,c)}].
\end{align*}
After permutations and integrating out the unnecessary variables, one can reduce the integrand to
\begin{gather*}
         [ (\theta_{11'2'3'}^{\beta(g,b,c)}\theta_{1233'}^{\beta(b,c,g)}\bar\theta_{122'3'}^{\beta(b,g,c)}) (\theta_{00'1'3'}^{\beta(g,a,bc)}\theta_{0133'}^{\beta(a,bc,g)}\bar\theta_{011'3'}^{\beta(a,g,bc)}) (\theta_{022'3'}^{\beta(ab,g,c)}\bar\theta_{0233'}^{\beta(ab,c,g)}\bar\theta_{00'2'3'}^{\beta(g,ab,c)} ) (  \theta_{011'2'}^{\beta(a,g,b)}\bar\theta_{0122'}^{\beta(a,b,g)}  \bar\theta_{00'1'2'}^{\beta(g,a,b)} )   ]
\end{gather*}
with sign factor
\begin{gather*}
i_g(\beta \cup_2 \beta)(a,b,c)+i_g\beta \cup_1 i_g \beta (a,b,c)+i_g \beta \cup_2 \beta(a,b,c)+ \beta(a,b,c)+\beta(g,ab,c)\beta(g,a,b)+\beta(g,ab,c)\beta(ab,c,g) \nonumber\\
+\beta(ab,g,c)+\beta(ab,g,c)\beta(g,ab,c)+\beta(a,b,g)(\beta(g,a,b)+\beta(g,ab,c))+\beta(a,g,b)+ \beta(a,g,b)(\beta(g,a,b)+\beta(g,ab,c))\nonumber\\
+i_g\beta(b,c) (\beta(a,g,bc)+\beta(g,a,bc))+\beta(g,b,c)(\beta(b,c,g)+\beta(b,g,c))+\beta(g,a,bc)(\beta(a,bc,g)+\beta(a,g,bc)).
 \end{gather*}
The $(-)$ simplex has the integrand reversed, and the roles of $\theta,\bar\theta$ swapped, but will give same sign factor. Next, we can rename the variables in each bracket (which corresponds to each face of the tetrahedron), and gain an extra factor. For example, the first bracket $(\theta_{11'2'3'}^{\beta(g,b,c)}\theta_{1233'}^{\beta(b,c,g)}\bar\theta_{122'3'}^{\beta(b,g,c)})$ can be reduced to $(-1)^{\beta(b,g,c)} \theta_{123}^{i_g\beta(b,c)}$ as shown in Table \ref{tab:reducegrassmann3to2}.
\begin{table}[h!]
\caption{Reducing $(\theta_{11'2'3'}^{\beta(g,b,c)}\theta_{1233'}^{\beta(b,c,g)}\bar\theta_{122'3'}^{\beta(b,g,c)})$ to $(-1)^{\beta(b,g,c)} \theta_{123}^{i_g\beta(b,c)}$}
\begin{tabularx}{\textwidth}{|c|c|c|X|}
\hline
$\beta(g,b,c)$ &$\beta(b,c,g)$ &$\beta(b,g,c)$ & \\
\hline
0&0&0&  Put back $\theta_{123}^{i_g\beta(b,c)}$.\\
\hline
1&0&0& Rename $\theta_{11'2'3'}^{\beta(g,b,c)}= \theta_{123}^{\beta(g,b,c)}=\theta_{123}^{i_g\beta(b,c)}$. Put back $(-1)^{\beta(b,g,c)}=1$.   \\
\hline
0&1&0& Rename $\theta_{1233'}^{\beta(b,c,g)}= \theta_{123}^{\beta(b,c,g)}=\theta_{123}^{i_g\beta(b,c)}$ . Put back $(-1)^{\beta(b,g,c)}=1$.\\
\hline
0&0&1&   Rename $\bar\theta_{122'3'}^{\beta(b,g,c)})= \bar\theta_{123}^{\beta(b,g,c)})=\bar\theta_{123}^{i_g\beta(b,c)}$ and swap with $\theta_{123}^{i_g\beta(b,c)}$ to get sign $(-1)^{\beta(b,g,c)}$.\\
\hline
1&1&0&   Integrate out $\theta_{11'2'3'}^{\beta(g,b,c)}\theta_{1233'}^{\beta(b,c,g)}$ and $\bar\theta_{1233'}^{\beta(b,c,g)}\bar\theta_{11'2'3'}^{\beta(g,b,c)}$ with no sign. Put back $\theta_{123}^{i_g\beta(b,c)}$ and $(-1)^{\beta(b,g,c)}$.\\
\hline
1&0&1&  Integrate out $\theta_{11'2'3'}^{\beta(g,b,c)}\bar\theta_{122'3'}^{\beta(b,g,c)}$ and $\theta_{122'3'}^{\beta(b,g,c)}\bar\theta_{11'2'3'}^{\beta(g,b,c)}$ to get $(-1)^{\beta(b,g,c)}$. Put back $\theta_{123}^{i_g\beta(b,c)}$.\\
\hline
0&1&1&  Integrate out $\theta_{1233'}^{\beta(b,c,g)}\bar\theta_{122'3'}^{\beta(b,g,c)}$ and $\theta_{122'3'}^{\beta(b,g,c)}\bar\theta_{1233'}^{\beta(b,c,g)}$ to get $(-1)^{\beta(b,g,c)}$. Put back $\theta_{123}^{i_g\beta(b,c)}$. \\
\hline
1&1&1&  Integrate out $\theta_{1233'}^{\beta(b,c,g)}\bar\theta_{122'3'}^{\beta(b,g,c)}$ and $\theta_{122'3'}^{\beta(b,g,c)}\bar\theta_{1233'}^{\beta(b,c,g)}$ to get $(-1)^{\beta(b,g,c)}$. Rename $\theta_{11'2'3'}^{\beta(g,b,c)}= \theta_{123}^{\beta(g,b,c)}=\theta_{123}^{i_g\beta(b,c)}$.  \\
\hline
\end{tabularx}
\label{tab:reducegrassmann3to2}
\end{table}

Unfortunately, the method we used to assign $S^{3+1}$ into $S^{2+1}$ fails because $S^{3+1}$ cannot be nicely grouped into $S_2^{2+1}$ and $S_3^{2+1}$ separately. As a result, we must do the calculation for the two orientations separately using the definition of $S$ in Table \ref{tab:m}.\\
First, we must assign the factors obtained from Table \ref{tab:reducegrassmann3to2} from each triangle to only one tetrahedron. Since each triangle bounds two adjacent tetrahedra, we need a local rule to assign the sign factor to only one of them. The rule we will use is to assign the sign factor depending on the number that is missing from the Grassmann variable. That is, the number on the vertex opposite to each face.\\
For $(+)$, we will assign the factors coming from triangles that are missing an \underline{\smash{odd number}}. That is, we assign $(-1)^{\beta(g,ab,c)+\beta(ab,c,g)}$ coming from $(023)$ and $(-1)^{\beta(g,a,b)+\beta(a,b,g)}$ coming from $(012)$.\\
For $(-)$, we will assign the factors coming from triangles that are missing an \underline{\smash{even number}}. That is, we assign $(-1)^{\beta(b,c,g)+\beta(g,b,c)}$ coming from $(123)$ and $(-1)^{\beta(a,bc,g)+\beta(g,a,bc)}$ coming from $(013)$.\\
Now, let us reduce the set $S^{3+1}$ to $S^{2+1}$. The triangles in $S_1^{3+1}$ that reduce nicely to links in $S_1^{2+1}$ are the ones directly above those links. For example, $(001')+(00'1')$ reduces to $(01)$ under the slant product. The remaining triangles are  $(012),(013),(023),(123),(012'),(013'),(023'),(123'),(01'2'),(01'3'),(02'3'),(12'3')$. This is the first contribution to $\Delta S$.

 The second contribution to $\Delta S$ is from the mismatch of $S_2^{3+1}+S_3^{3+1}$ and $S_2^{2+1}+S_3^{2+1}$. First, the rules in Table \ref{tab:m} give $S_2^{2+1}+S_3^{2+1}$ as $(02)$ per $(+)$ tetrahedron and $(13)$ per $(-)$ tetrahedron. If we lift this up to 3+1D using the slant product (equation \eqref{equ:lift}), we would get $(022')+(00'2')$ for $(+)$ and $(133')+(11'3')$ for $(-)$.\\
 To simplify the calculation, we will swap the assignments for $(+)$ and $(-)$ in Table \ref{tab:m}, which is also a valid assignment (as discussed in Appendix \ref{app:w_2}.) This means that to obtain $S_2^{3+1}+S_3^{3+1}$, we will instead assign $(024)$ from each $(+)$ tetrahedron in the prism, and $(013),(134),(123)$ from each $(-)$ tetrahedron in the prism. This gives
\begin{equation}
S_2^{3+1}+S_3^{3+1} = \begin{cases}
(00'2')+(0'2'3')+(0'1'2')+(01'3')+(012')+(12'3')+(122')+(023') & (+)\\
(01'3')+(012')+(12'3')+(11'2')+(023')+(013)+(133')+(123) & (-)
\end{cases}
\end{equation}
per prism. Subtracting off the lift of $S_2^{2+1}+S_2^{3+1}$, we obtain the second contribution to $\Delta S$ per prism:
\begin{equation*}
\begin{cases}
(022')+(0'2'3')+(0'1'2')+(01'3')+(012')+(12'3')+(122')+(023') & (+)\\
(01'3')+(012')+(12'3')+(11'2')+(023')+(013)+(11'3')+(123) & (-)
\end{cases}.
\end{equation*}

The next step is to note that the triangles from the first contribution are always at the boundary of two adjacent prisms. Consequently, we need a local rule to assign these triangles to only one of the prisms. These two assignments will reflect the two different possible spin structures on the compactified circle.
\begin{enumerate}
\item For $(+)$, assign those missing 0 or 2: $(013),(123),(013'),(123'),(01'3'),(12'3')$ and for $(-)$, assign those missing 1 or 3:  $(012),(023),(012'),(023'),(01'2'),(02'3')$. Combining this with the second contribution,
\begin{equation}
\Delta S  = \begin{cases}
(013)+(123')+(0'2'3')+(012)+(022')+(012')+(122')+(023')+(013')+(123') & (+)\\
(01'3')+(12'3')+(11'2')+(013)+(11'3')+(123)+(012)+(023)+(01'2')+(02'3') & (-)
\end{cases}
\end{equation}
per prism. This is the boundary of
\begin{equation}
\Delta E  = \begin{cases}
(0122')+(0233')+(0133')+(1233') & (+)\\
(00'1'2')+(00'1'3')+(00'2'3')+(11'2'3') & (-)
\end{cases}
\end{equation}
per prism. Thus, we get
\begin{equation}
\int_{\Delta E} \beta = \begin{cases}
\beta(a,b,g)+\beta(ab,c,g)+\beta(a,bc,g)+\beta(b,c,g) & (+)\\
\beta(g,a,b)+\beta(g,ab,c)+\beta(g,a,bc)+\beta(g,b,c) & (-)
\end{cases}
\end{equation}
per prism. If we add these to the terms from the Grassmann integral that we assigned differently, we will get
\begin{equation*}
\begin{cases}
\beta(a,b,g)+\beta(ab,c,g)+\beta(a,bc,g)+\beta(b,c,g)+\beta(ab,g,c)+\beta(a,g,b) & (+)\\
\beta(g,a,b)+\beta(g,ab,c)+\beta(g,a,bc)+\beta(g,b,c)+\beta(b,g,c)+\beta(a,g,bc) & (-)
\end{cases}
\end{equation*}
per prism. These two expressions are actually equal, since their sum is $di_g\beta(a,b,c)$, which is zero.
\item  We swap the assignments of $(012),(013),(023),(123)$. That is, we assign $(012), (023), (013'),  (123'), (01'3'),\allowbreak (12'3')$ to $(+)$ and $(012),(023),(012'),(023'),(01'2'),(02'3')$ to $(-)$. Repeating the calculation, we get
\begin{equation}
\int_{\Delta E} \beta = \begin{cases}
\beta(a,b,g)+\beta(ab,c,g)+\beta(a,bc,g)+\beta(b,c,g)+\beta(a,b,c) & (+)\\
\beta(g,a,b)+\beta(g,ab,c)+\beta(g,a,bc)+\beta(g,b,c)+\beta(a,b,c) & (-)
\end{cases}
\end{equation}
per prism. This means that we get an extra factor of $\beta(a,b,c)$ in $\gamma(a,b,c)$ for changing the spin structure.
\end{enumerate}
Combining this term with the sign factor from the Grassmann integral, the cochain $\gamma$ takes the form
\begin{align}
\gamma(a,b,c)=&i_g(\beta \cup_2 \beta)(a,b,c)+i_g \beta \cup_2 \beta(a,b,c)+ \eta(S^1)\beta(a,b,c)+i_g\beta \cup_1 i_g \beta (a,b,c) +\beta(g,ab,c)\beta(g,a,b) \nonumber\\
&+\beta(g,ab,c)\beta(ab,c,g)+\beta(ab,g,c)\beta(g,ab,c)+\beta(a,b,g)(\beta(g,a,b)+\beta(g,ab,c)) \nonumber\\
& +\beta(a,g,b)(\beta(g,a,b)+\beta(g,ab,c))+i_g\beta(b,c) (\beta(a,g,bc)+\beta(g,a,bc))+\beta(g,b,c)(\beta(b,c,g)+\beta(b,g,c)) \nonumber\\
&+\beta(g,a,bc)(\beta(a,bc,g)+\beta(a,g,bc))+\beta(a,b,g)+\beta(ab,c,g)+\beta(a,bc,g)+\beta(b,c,g)+\beta(ab,g,c) 
\end{align}
After some reorganization, one can write the cochain as
\begin{align}
\gamma=&i_g(\beta \cup_2 \beta)+i_g\beta \cup_1 i_g \beta+i_g\beta \cup_2 \beta + \eta(S^1)\beta  + \epsilon
\end{align}
where $\epsilon$ is the cochain
\begin{align}
\epsilon(a,b,c)&=\beta(g,ab,c)  i_g \beta(a,b)+ (\beta(g,a,bc)+\beta(a,g,bc))i_g \beta(b,c) + d\mu(a,b,c);\\
\mu(a,b) &=\beta(g,a,b) i_g\beta(a,b).
\end{align}
Finally, we need to check that modulo 2,
\begin{equation}
d\epsilon+i_g\beta \cup_1 \beta + \beta \cup_1 i_g\beta + Sq^2(i_g \beta) + i_g Sq^2 (\beta)=0.
\end{equation}
The easiest way to do so is to simplify the expression via a computer program by using repeated applications of the following cocycle conditions:
\begin{align}
\beta(ab,c,d)=& \beta(a,b,c)+\beta(a,bc,d)+\beta(a,b,cd)+\beta(b,c,d)\\
\beta(g,ab,c)=&\beta(g,a,b)+\beta(g,a,bc)+\beta(a,b,c)+\beta(a,g,b)+\beta(a,gb,c)+\beta(a,g,bc)+\beta(g,b,c)
\end{align}
Note that $g$ is treated separately from the other variables. Doing the given substitutions forces either $g$ or a product of group elements to be sent to the right. Thus, the final expressions are limited to a subset of $\beta$'s that will eventually cancel modulo 2.
\vspace{20pt}
\twocolumngrid
\section{Lens Spaces}\label{app:lens}
\begin{figure}[h!]
\centering
\includegraphics{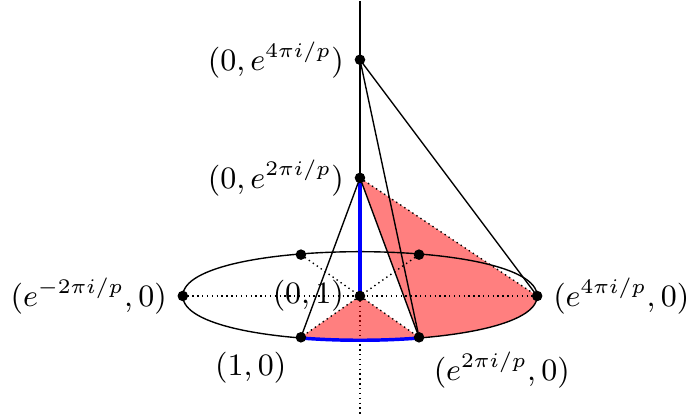}
\caption{The lens space $L(6;1)$ visualized as a quotient of $S^3$ using stereographic projection. The generator of the fundamental group is shown in blue. The two blue segments are homotopic.}
\label{fig:lensvisual}
\end{figure}

A lens space $L(p;q)$ where $p$ and $q$ are coprime, is a 3-manifold created by taking the quotient of $S^3$ by $\mathbb{Z}_p$. Viewing $S^3$ as the unit sphere in $\mathbb{C}^2$, the action is generated by
\begin{equation}
(z_1,z_2) \rightarrow (z_1 e^{2\pi i /p},z_2 e^{2\pi i q/p} ).
\end{equation}
One can visualize this using sterographic projection onto $\mathbb{R}^3$, shown in Figure \ref{fig:lensvisual}. The circle corresponds to $|z_1|=1$ and the vertical line corresponds to $|z_2|=1$, both of which are great circles in $S^3$. We will focus on the case where $q=1$. The example drawn is for $p=6$. The shade in the plane is moved under the action to the shade on the right, and the two are identified under the quotient. Thus, the 3-sphere is broken down into the lens space. Alternatively, one can also think of gluing the top hemisphere of a 3-ball to the bottom hemisphere with a twist of $2\pi q/p$ radians, as shown in Figure \ref{fig:lens}. The gluing of triangles of the same color corresponds to the identification of the shaded area in Figure \ref{fig:lensvisual}.

The fundamental group of the lens space is generated by the non-contractible loop shown in blue going along the ``great circle'' $|z_1|=1$ from $(1,0)$ to $(e^{2\pi i/p},0)$ (note that the two end points are identified.) As a result, we need $p$ such loops placed around the great circle so that it can be contracted to a point. Hence, the fundamental group of the lens space isomorphic to $\mathbb{Z}_p$. One might notice that the loop going up along $|z_2|=1$ from $(0,1)$ to  $(0,e^{2\pi i/p})$ is also non-contractible. However, these two loops are homotopic. Therefore, a flat connection on a lens space will always place the same group element on these two loops. This fact can also be worked out from the flatness condition of the triangulation in Figure \ref{fig:lens}.

Lens spaces can be generalized to higher odd dimensions. In $2n-1$ dimensions, the generalized lens space $L(p;q_1,...,q_n)$, where $q_1,...,q_n$ are (not necessarily pairwise) coprime, is the quotient of $S^{2n-1}$ (viewed as the unit sphere in $\mathbb C^n$) by $\mathbb{Z}_{p}$ via the action
\begin{equation}
(z_1,...,z_n) \rightarrow (z_1 e^{2\pi i q_1/p},...,z_n e^{2\pi i q_n/p} ).
\end{equation}
We remark that in 3D, $L(p;q)$ means $L(p;1,q)$ in the generalized notation.\\
The fundamental group of $L(p;q_1,...,q_n)$ is isomorphic to $\mathbb{Z}_{p}$. For the 5D invariant defined on $L(N_i;1,1,1)$, we can triangulate the manifold using three great circles that do not intersect: $|z_1|=1$, $|z_2|=1$, and $|z_3|=1$, each of which is broken up into $N_i$ segments. The connection we use places a generator of the fundamental group $e_i$ around these three great circles. Therefore, the expression of the partition function $\mathcal I_i$ is 
\begin{equation}
\mathcal I_i =\prod_{n,m=1}^{N_i}\omega_5(e_i,ne_i,e_i,me_i,e_i),
\end{equation}
where the $1^\text{st}$, $3^\text{rd}$, and $5^\text{th}$ slots correspond to the three great circles where we place $e_i$.

\section{Solving the Gu-Wen Equation Using the Smith Normal Form}\label{app:smith}
\allowdisplaybreaks
First, let us rewrite $\omega_n$ as a $\mathbb{R}/\mathbb{Z}$-valued function. That is, taking the natural log, dividing by $2\pi i$ and choosing the branch such that $\omega_n$ is valued in $[0,1)$. The Gu-Wen equation \eqref{equ:domega} now takes the form
\begin{equation}
d\omega_n(g_1,...,g_{n+1}) = \frac{1}{2}Sq^2 \beta_{n-1}(g_1,...,g_{n+1}) \ \ \ \text{(mod 1)}.
\end{equation}
In the remaining discussion, all equations are implicitly defined modulo integers. For a group $G$, there are $|G|^{n+1}$ such linear equations for all possible values of $g_1,...,g_{n+1} \in G$. Let us write $\boldsymbol{\omega}$ as the vector of $\omega_n(g_1,...,g_n)$ and $\boldsymbol{B}$ as the vector of $\frac{1}{2}Sq^2 \beta_{n-1}(g_1,...,g_{n+1})$. We then have
\begin{equation}
\label{equ:matrixequation}
 \boldsymbol{M\omega} =\boldsymbol{B} ,
 \end{equation}
where $\boldsymbol{M}$ is some integer matrix of size $|G|^{n+1} \times |G|^{n}$. We can solve for $ \boldsymbol{\omega}$ using the Smith normal form of $\boldsymbol{M}$, whose decomposition is the following: there exists integer matrices, $\boldsymbol{U},\boldsymbol{D},\boldsymbol{V}$ of sizes $|G|^{n+1} \times |G|^{n+1}$, $|G|^{n+1} \times |G|^{n}$, and $|G|^{n} \times |G|^{n}$, respectively such that $\boldsymbol{U},\boldsymbol{V}$ are invertible (and their inverses are also integer matrices), $\boldsymbol{D}$ is non-zero only along the diagonal, and
\begin{equation}
 \boldsymbol{UMV} = \boldsymbol{D}.
  \end{equation}
To solve for $\boldsymbol{\omega}$, apply $\boldsymbol{U}$ to both sides of equation \eqref{equ:matrixequation} and use the decomposition to get
\begin{equation}
 \boldsymbol{D}(\boldsymbol{V}^{-1}\boldsymbol{\omega})  = \boldsymbol{UB} \ \ \ \text{(mod 1)}.
  \end{equation}
Define the vector $\boldsymbol{W} =\boldsymbol{V}^{-1}\boldsymbol{\omega} $. Since $\boldsymbol{D}$ is diagonal, the general solution of  $\boldsymbol{W}$ is\\
\vspace{-10pt}
\begin{equation}
W_i = \begin{cases}
\frac{(\boldsymbol{UB})_i+n_i}{D_{ii}} &; D_{ii} \ne 0\\
0 &; D_{ii} = 0
\end{cases}
\end{equation}
for $i=1,...,|G|^n$ and $n_i \in \mathbb{Z}$. Note that this solution is only valid if $(\boldsymbol{UB})_i =0$ for all $i$ such that $D_{ii}=0$ or $i>|G|^{n}$. We can now solve for $\boldsymbol{\omega}$ using $\boldsymbol{\omega} = \boldsymbol{VW}$.\\
The different choices of $n_i$ give different solutions to the equation, corresponding to adding inequivalent cocycles. thus, the product of all non-zero elements $D_{ii}$ is $|H^n(G,\mathbb R/ \mathbb Z)|$.

Before providing solutions to the Gu-Wen equation, let us first simplify the constraints \eqref{equ:constraint7} -  \eqref{equ:constraint9} so that they only depend on the bosonic integrals $Z_i$,$Z_{ij}$ and $Z_{ijk}$. First, to simplify equation \eqref{equ:constraint7}, note that the spin structure term $E_{ij}(\beta)$ is an evaluation on a lens space, and therefore is unity like $E_i(\beta)$ in equation \eqref{equ:Eibeta} for the trivial spin structure. Next, we will show that the combined Grassmann integrals that contribute to $N_{ij}\Theta_{ij}$ vanishes. Writing it out explicitly,
\begin{widetext}
\begin{gather}
\left (\frac{\sigma_{ij}(\beta)}{\sigma_i(\beta)^{N^{ij}/N_i}  \sigma_j(\beta)^{N^{ij}/N_j}} \right)^{N_{ij}}=\frac{\exp \pi i N_{ij}\sum_{n=1}^{N^{ij}} \beta(e_i+e_j,n(e_i+e_j))\beta(n(e_i+e_j),e_i+e_j)]}{\sigma_i(\beta)^{N_j}  \sigma_j(\beta)^{N_i}} \nonumber\\
= \exp \pi i N_{ij}\left [\sum_{n=1}^{N^{ij}} \left (  \frac{R_i}{N_i}(1+n-[1+n])_i +   \frac{R_j}{N_j}(1+n-[1+n])_j + \frac{2R_{ij}}{N_{0ij}}[n]_j+\frac{2R_{ji}}{N_{0ij}}[n_i]  \right) \cdot \left (i \leftrightarrow j \right) -N_j R_i - N_i R_j \right]\nonumber\\
\label{equ:Grassmannij}
=\exp \pi i N_{ij} \left [ \frac{4N^{ij}R_{ij}R_{ji}} {N_{0ij}^2N_i} \sum_{n=1}^{N_i} n^2  + \frac{4R_{ij}}{N_{0ij^2}} \sum_{n=1}^{N^{ij}}[n]_i[n]_j + \frac{2(R_{ij}+R_{ji})R_i}{N_{0ij}} \left (\frac{N^{ij}}{N_i} +\sum_{n=1}^{N^{ij}/N_i} [nN_i-1]_j \right ) + \left (i \leftrightarrow j \right)  \right]
\end{gather}
\end{widetext}
If either $N_i$ or $N_j$ is odd, all terms will vanish since $N_{0ij}=1$. On the other hand, if they are both even, $N_{ij}$ is even, and so all terms will still vanish. Hence, there is no net contribution from the Grassmann integrals. Inserting the explicit expression of $\Theta_{0ij}$ from equation \eqref{equ:theta0ijform} into constraint \eqref{equ:constraint7}, we can rewrite it as
 \begin{align}
\frac{Z_{ij}^{N_{ij}}}{Z_i^{N_j} Z_j^{N_i}}  &= \begin{cases}
 \exp \pi i (R_{ij}+R_{ji}) & \text{$N_i \equiv N_j \equiv 2$ (mod 4)}\\
1 &  \text{otherwise}
 \end{cases},
\end{align}
where we have used the fact that $\frac{N^{ij}(N^{ij}-1)}{2}$ is odd only for the former case, and $N^{ij}N_{ij}=N_iN_j$.

Next, we simplify the constraint \eqref{equ:constraint8}. The Grassmann integral that contributes to $\Theta_{ii}$ is
\begin{align}
\frac{\sigma_{ii}(\beta)}{\sigma_i(\beta) \sigma_i(\beta)}  =& \exp \pi i\sum_{n=1}^{N_{i}}   \beta(2e_i,2ne_i) \beta(2ne_i,2e_i)   \nonumber\\
=& \exp \pi i\sum_{n=1}^{N_{i}}  \frac{R_i}{N_i}(2+[2n]_i-[2+2n]_i)\nonumber\\
&\cdot \exp \pi i \sum_{n=1}^{N_{i}}\frac{2R_{ii}}{N_{0i}}(2)(2n),
\end{align}
which vanishes. Again, we don't have contributions from the spin structure term for lens spaces. So the constraint can be written as
 \begin{align}
  Z_{ii}&=
 \begin{cases}
Z_i^4 &\text{(for $N_i$ even)}\\
Z_i^3 & \text{(for $N_i$ odd)}
\end{cases}.
\end{align}
Finally, we simplify the constraint \eqref{equ:constraint9}. $\sigma_{ijk}=-1$ only when $N_i$, $N_j$, and $N_k$ are all even. Therefore, $\sigma_{ijk}^{N_{ijk}}=1$, so the constraint reduces to
 \begin{align}
Z_{ijk}^{N_{ijk}}&=1.
\end{align}

We will now solve the Gu-Wen equation for the groups $\mathbb Z_2 \times \mathbb Z_4$, $\mathbb Z_2 \times \mathbb Z_6$, and $\mathbb Z_2 \times \mathbb Z_2 \times \mathbb Z_2$ and evaluate the explicit expressions for the bosonic integrals. We can then check whether the constraints given are satisfied. Note that we do not have to consider odd subgroups because if $G$ has only odd subgroups, then there are no non-trivial cocycles in $\mathcal H^n(G,\mathbb Z_2)$. On the other hand, we have the isomorphism $\mathbb Z_2 \times \mathbb Z_p \cong \mathbb Z_{2p}$ for $p$ odd.

\subsection{$G=\mathbb Z_2 \times \mathbb Z_4$}
The bosonic integrals are
\begin{align}
Z_1&= \exp \pi i \left [n_1 + \frac{R_1 +R_{11}}{2} \right ],\\
Z_2&= \exp \pi i \left [ \frac{n_2}{2} + \frac{R_2}{4} \right],\\
Z_{11} &=1,\\
Z_{22}&=\exp \pi i R_2,\\
Z_{12} &= \exp \pi i \left [n_{12}-\frac{n_2}{2} + \frac{R_2}{4} \right],
\end{align}
where $n_1$,$n_2$ and $n_{12}$ are independent integers. We first note that by varying these integers, this set of partition functions can take $2\times4\times2=16$ different values, which is equal to $| \mathcal H^3(\mathbb Z_2 \times \mathbb Z_4,U(1))|$. One can now check that the conditions are satisfied.
\begin{align}
Z_{11} &= Z_1^4 =1,\\
Z_{22} &= Z_2^4 =1, \\
\frac{Z_{12}^2}{Z_1^4Z_2^2}&= 1.
\end{align}
\subsection{$G=\mathbb Z_2 \times \mathbb Z_6$}
We obtain the bosonic integrals
\begin{align}
Z_1= \exp \pi i &\left [n_1 + \frac{R_1 +R_{11}}{2} \right ],\\
Z_2= \exp \pi i &\left [ n_1 +n_2- \frac{2n_3}{3} -\frac{R_2+R_{22}}{2} \right],\\
Z_{11}=1, \ \ \ \ \  & \\
Z_{22}=\exp \pi i& \frac{2n_3}{3},\\
Z_{12} = \exp \pi i &\left [n_2+\frac{n_3}{3}+ f(R)  \right. \nonumber\\
&\left.-\left(\frac{R_1 + R_{11} + R_{12}}{2} +(1 \leftrightarrow 2)\right)  \right],
\end{align}
where $n_1$,$n_2$ and $n_3$ are independent integers and $f(R)$ is a complicated integer that depends on the $R$'s. By varying $n_1$,$n_2$ and $n_3$, this set of partition functions can take $2\times2\times6=24$ different values, which is equal to $| \mathcal H^3(\mathbb Z_2 \times \mathbb Z_6,U(1))|$. The conditions are satisfied, since
\begin{align}
Z_{11} &= Z_1^4 =1,\\
Z_{22} &= Z_2^4 =\exp \pi i \frac{2n_3}{3}, \\
\frac{Z_{12}^2}{Z_1^6Z_2^2}&= \exp \pi i (R_{12}+R_{21}).
\end{align}

\subsection{$G=\mathbb Z_2 \times \mathbb Z_2 \times \mathbb Z_2$}
We find the following expression of the bosonic integrals:
\begin{align}
Z_i&=\exp  \pi i \left [n_i + \frac{R_i + R_{ii}}{2} \right ]; \ \ \ \ i=1,2,3 \\
Z_{ii} &=1; \ \ \ \ \ \ \ \ \ \ \ \ \ \ \ \ \ \ \ \ \ \ \ \ \ \ \ \ \ \ \ \ \ i=1,2,3\\
Z_{ij}&=\exp  \pi i \left[n_{ij} +\left(\frac{R_i + R_{ii} + R_{ij}}{2} +(i \leftrightarrow j)\right)\right ]; i<j,\\
Z_{123}&=\exp  \pi i n_{123}, 
\end{align}
for independent integers $n_i$, $n_{ij}$, and $n_{123}$. The set of partition functions can take $2^3\times2^3\times2=2^7$ different values, which is equal to $| \mathcal H^3(\mathbb Z_2 \times \mathbb Z_2 \times \mathbb Z_2,U(1))|$. The conditions are indeed satisfied: 
\begin{align}
Z_{ii} &= Z_i^4 =1; & i=1,2,3\\
\frac{Z_{ij}^2}{Z_i^2 Z_j^2} &= \exp \pi i (R_{ij}+ R_{ji}); & i<j\\
Z_{123}^2&= 1. &
\end{align}
\onecolumngrid
\vspace{20pt}
\twocolumngrid

\bibliography{references}

\end{document}